\documentclass[12pt]{spieman}  
\usepackage{amsmath,amsfonts,amssymb,bm}
\usepackage{graphicx}
\usepackage{setspace}
\usepackage{tocloft}
\usepackage{lineno}
\usepackage{ulem}
\usepackage{comment}

\title{Control problem in millimeter-wave adaptive optics}

\author[a*]{Ichiro Jikuya}
\author[b]{Yoichi Tamura}
\affil[a]{Kanazawa University, Faculty of Frontier Engineering, Institute of Science and Engineering, Kakuma, Kanazawa, Ishikawa 920-1192, Japan}
\affil[b]{Nagoya University, Graduate School of Science, Department of Physics, Furo, Chikusa, Nagoya, Aichi 464-8602, Japan}

\newcommand{\RevSakibara}[1]{\textcolor{black}{#1}}
\newcommand{\RevJikuya}[1]{\textcolor{black}{#1}}
\newcommand{\Add}[1]{\textcolor{black}{#1}}

\newcommand{\T}{\mathrm{T}}
\newcommand{\Image}{\mathrm{Image}\!\,\,}
\newcommand{\Ker}{\mathrm{Ker}\!\,\,}

\cftpagenumbersoff{figure}
\cftpagenumbersoff{table} 
\begin{document} 
\maketitle

\begin{abstract}
Millimeter-wave Adaptive Optics (MAO) is essential for high-precision large-aperture submillimeter telescopes, requiring real-time compensation of wavefront errors by capturing them as spatially-discrete Excess Path Length (EPL) fluctuations. This paper presents a unified control-theoretic framework for the EPL compensation problem. We first model the optical drive system as a plant where input commands relate to measured EPL through a first-order system representing mechanical response delay and a measurement matrix characterizing the actuator-to-sensor coupling. We mathematically formulate the control task as an asymptotic disturbance suppression problem, specifically targeting low-frequency disturbances such as thermal and wind-induced deformations.
Second, we propose an Anti-Windup Proportional-Integral (AWPI) control law. By employing a decoupling strategy, the design is reduced to a loop-shaping problem for decoupled scalar sensitivity functions, ensuring both stability margins and asymptotic disturbance suppression of constant-valued disturbances. The anti-windup mechanism is integrated to maintain control continuity during the recovery from saturation, preventing undesirable discontinuities in the drive command. 
Third, we introduce practical operational tools: a manual focus adjustment scheme that allows observer intervention without interfering with the feedback loop, and the cosine similarity index to quantify the suppressibility of specific Zernike modes. Numerical simulations, incorporating a three-axis secondary reflector drive and five-point EPL measurements, demonstrate direction-dependent disturbance rejection and the suppression of von Karman-modeled wind turbulence, validating the effectiveness of the proposed framework for real-world telescope applications.
\end{abstract}


{\noindent \footnotesize\textbf{*}Ichiro Jikuya,  \linkable{jikuya@se.kanazawa-u.ac.jp} }




\section{Introduction}
\label{sect:introduction}

Large-aperture single-dish submillimeter (submm) telescopes are essential for exploring the universe through wide-field imaging and broadband spectroscopy, leveraging the full potential of direct photon detection technologies.\cite{Kawabe16, Klaassen20, Kohno20} In addition to their standalone capabilities, these telescopes offer advantages when integrated into an interferometric network as an individual array element\cite{Akiyama23}. Projects such as the Large Millimeter Telescope Alfonso Serrano (LMT),\cite{Hughes10} the Atacama Large-Aperture Submillimeter Telescope (AtLAST)\cite{Klaassen20, Mroczkowski25}, the Large Submillimeter Telescope (LST)\cite{Kawabe16}, and a planned Chinese submm telescope \cite{Lou20} exemplify this approach. 

Since the 1970s, radio telescopes have successfully mitigated optical performance degradation due to gravitational deformation of the primary reflector (M1) by employing homologous deformation and secondary reflector (M2) tracking at the prime focus\cite{vonHoerner67}. However, further scaling and operation at higher frequencies are now constrained by thermal and wind-induced deformations.\cite{Baars18} For apertures up to $\approx 100$~m, the dominant factors limiting optical performance and causing wavefront degradation are not atmospheric water vapor fluctuations at the world's best site for submm astronomy\cite{Asaki05}, but rather external environmental loads such as heat and wind. This necessitates the development of real-time optical compensation systems or metrology mechanisms.\cite{Mroczkowski25, Reichert24}

Various metrology systems to mitigate them have been proposed and evaluated to date. Although active surface control mechanisms are implemented in many telescopes, methods for measuring wavefront errors and structural deformations remain under active investigation. For AtLAST/LST, promising approaches include the laser antenna surface scanning instrument (LASSI)\cite{Salas20}, mm-wave adaptive optics (MAO)/phase adaptive stabilization system (PASS)\cite{Tamura20, Nakano22, Liang23, Cheng25}, laser interferometry based on the Etalon Absolute Multiline Technology$^{\rm TM}$\cite{Rakich16, Schloerb22, Attoli23, Attoli24}, and temperature-based correction techniques for thermal deformation\cite{Attoli23, Attoli24}. These systems typically sample the wavefront or telescope structure discretely across the aperture plane. Consequently, it becomes necessary to estimate continuous wavefront errors from spatially discrete excess path length (EPL) measurements and to compensate for these errors using multiple independent optical elements or mechanisms.

One of the key challenges in radio wavefront compensation lies in spatially-discrete\Add{, sparse} sampling of the wavefront with a method that is totally different from the optical and near-infrared (NIR) techniques (e.g., Shack--Hartmann sensors). In addition, systematic changes in optical path lengths induced by the elevation-dependent homologous deformation are substantially large and need to be taken into account for wavefront correction.

Another challenge in achieving radio wavefront compensation is correction of the wavefront in the aperture plane or its conjugate pupil, as done with optical and NIR telescopes\cite{Babcock53, Beckers93}. Unlike optical--NIR telescopes, which can form compact exit pupils within the instrument, radio telescopes face difficulties in achieving such configurations since wave optics applies to radio telescopes and instruments and thus requires relatively large optical elements to form an exit pupil. Therefore, compensation must be applied at the entrance pupil, typically involving M1 or M2.

In this context, low-frequency, low-order deformation modes (e.g., tip--tilt, defocus, and coma) can be corrected via alt--azimuthal adjustments of M1 or translational movements of M2. High-frequency, high-order modes are better addressed through active optics applied to M1 or M2 with deformable surfaces. This division of lower- and higher-mode deformations among multiple optical mechanisms is particularly relevant for radio telescopes that rely on the M2 tracking to compensate for focal shifts caused by homologous deformation.

However, the optical components available for lower-order compensation (e.g., M1 and M2) tend to be large and massive, resulting in long control time constants and making their control non-trivial. Ensuring stable and reliable control is mandatory; otherwise, improper actuation may cause mechanical damage. Moreover, depending on the sensor configuration and the available optical elements, certain disturbance modes can be suppressed while others cannot. For example, in the translational control of M1 and M2, disturbances corresponding to phase-error modes such as tip–tilt, defocus, and coma can in principle be suppressed. Achieving this, however, requires sampling discrete points in a manner that captures the characteristic spatial features of each mode. Disturbance modes that cannot be generated by the optical elements or adequately sampled by the sensors fall outside the set of suppressible modes. Despite this, no prior studies have attempted such a control problem, and thus a tailored control method for submillimeter telescopes needs to be established.

This paper presents a formalism for a MAO control system extracting relevant wavefront modes from spatially discrete EPL errors measured by MAO wavefront sensors. 
We focus on EPL measurements from the aperture to the focal plane obtained via aperture-plane interferometry\cite{Tamura20}. 
Although this study focuses on this specific method, the proposed formalism is also applicable to other discrete\Add{/sparse} phase-error measurements and continuous wavefront sensing across the aperture.
Our formalism is scalable to \RevJikuya{an} unlimited number of discrete EPL measurements while we give an example of control performances with a 5-point wavefront sensor.

This paper is organized as follows. Sec.~\ref{sect:assumed_telescope_system} presents the telescope system assumed for the formalism. In Sec.~\ref{sect:control_design} we describe plant modeling in a general control problem and then formulate our control problem with MAO. We also analyze its stability and asymptotic behavior to shape the {open-loop transfer function matrix}. Sec.~\ref{sect:performance_evaluation} is devoted to discussions on its performance and practical application with a real-world telescope, which is followed by numerical simulations that exemplify a controller with a 5-point EPL sensor and M2 translational actuation in Sec.~\ref{sect:numerical_examples}. Finally, Sec.~\ref{sect:conclusions} concludes the paper.


\section{Assumed Telescope System}\label{sect:assumed_telescope_system}

In this section, we describe the assumptions regarding the telescope system used for the formulation presented in this study.


\subsection{Control System}

Here the term ``optical drive system'' refers collectively to all mechanisms that actuate the telescope optics, including deformable mirrors with multiple actuators, M1's active primary surfaces, azimuth--elevation (Az--El) telescope drives, and translational/rotational drives of M2. All optical drive systems are assumed to be controlled by a central control program that issues command values to actuators and measures the resulting displacements via encoders in a coordinate system fixed to the optical components. This integrated system is referred to as the adaptive optics (AO) control system. Note that in the optical and NIR regime, AO refers to the optical system that compensates for wavefront degradation caused by any fast disturbances along the optical path that is typically dominated by atmospheric wavefront distortions\cite{Babcock53, Beckers93}. As the wind and thermal loads dominates the disturbances in a submm single-dish telescope at a dry site, here the term AO refers to any closed loop correction techniques that ``sample the wavefront in the telescope aperture and use this to adjust some of the optical elements to correct for the measured disturbance in real time,'' following the description by Baars \& K\"{a}rcher.\cite{Baars18}
Each actuator within the optical drive system is modeled as a first-order system with a characteristic time constant. 
Although alternative representations, such as a second-order system, could be used, such cases are beyond the scope of this study (see Sec.~\ref{sect:plant_modeling}).

The AO control system operates as a closed-loop system, periodically executing measurement, computation, and actuation processes. It receives command values derived from multiple EPL measurements obtained by the wavefront sensor and applies them to the optical drive system. The period of this loop is referred to as the control period. When implementing the AO control system, it is ensured that all processes are completed within each control period.
In general, a shorter control period leads to better performance. The sampling period of EPL measurements is assumed to be significantly shorter than the control period of the optical drive system, providing a precise characterization of the wavefront before each actuation. 
The control bandwidth is designed to be lower than the natural frequencies of the mechanical components. Among these, the whole telescope structure (particularly M1) has the lowest natural frequency, typically $\gtrsim 1$~Hz.\cite{Mroczkowski25, Imamura25}


%
\begin{figure}[th!]
	\begin{center}
		\includegraphics[width=0.95 \textwidth]{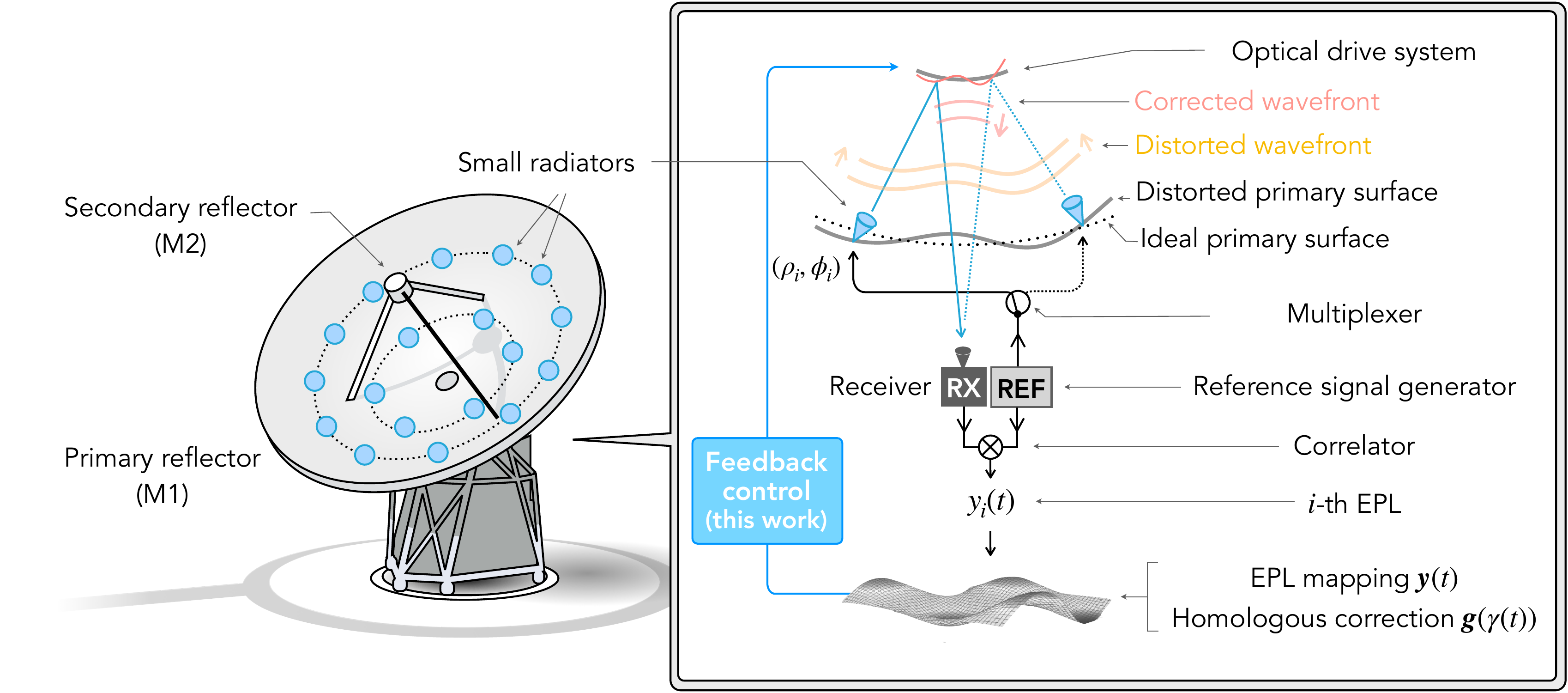}
		\caption{
			Schematic illustration of the wavefront sensor and the control system concept for mm-wave adaptive optics (MAO).\cite{Tamura20} The wavefront sensor uses micro/mm-wave to measure the excess path lengths (EPLs) from the feeds placed on the M1 surface to the receiver feed at the focal point. The reconstructed wavefront deformations are corrected by optical drive systems.
		}\label{fig:FigMAOConcept}
	\end{center}
\end{figure}
%


\subsection{Wavefront Sensor}

The wavefront is spatially sampled by measuring the EPL from each of $m$ discrete points on the aperture plane (e.g., M1) to \Add{a certain point on the focal plane} (e.g., the receiver feed\Add{; hereafter we just use the term, focal point}) as illustrated in Fig.~\ref{fig:FigMAOConcept}. 
The sampling rate is assumed to be sufficiently higher than the natural frequencies of the optical components being controlled. For details on the principle and instrumentation, refer to Tamura et al. (2020).\cite{Tamura20}
The temporal variation of EPL is measured as an offset from a calibrated reference point. Absolute surface measurements are assumed to rely on static techniques such as radio holography, including the phase-retrieval \cite{Morris85} and the out-of-focus holography methods.\cite{Nikolic07}

The relationship between the drive amounts of the optical drive system and the measured EPL variations is described by a measurement matrix, or its pseudoinverse, the control matrix. This matrix is derived by evaluating the ratio of EPL measurements to the actuation range of each optical drive component. Note that the measurement matrix is sometimes referred to as an influence matrix, an interaction matrix, or a response matrix. We use the term measurement matrix throughout this paper. 


\section{Theoretical Framework and Controller Design}\label{sect:control_design}


\subsection{Plant Modeling}\label{sect:plant_modeling}

The dynamics of an optical drive system can be modeled by the first-order differential equation 
\begin{align}
	t_{\mathrm{s}} \, \dot{p}(t) + p(t) = u(t), 
\end{align}
where the real variable $t \in \mathbb{R}$ denotes the time, $t_{\mathrm{s}} > 0$ is the time constant, $u(t) \in \mathbb{R}$ is the drive command,  $p(t) \in \mathbb{R}$ is the drive amount, and $\dot{p}(t)$ is the time derivative of $p(t)$, \textit{i.e.}, $\dot{p}(t) := \frac{dp(t)}{dt}$. 
It is important to note that this equation assumes an optical drive system that controls position in response to a position input command; \RevJikuya{while a system that controls position based on a velocity input command would require a different model incorporating an integrator, the following discussion assumes a position-command-based architecture.} 
An optical drive system that controls position in response to a velocity input command value would necessitate a different model incorporating an integrator; however, the subsequent discussion will be based on the assumption that the system operates using a position input command value to control position. 
The solution of the above equation is given by 
\begin{align}
	p(t) = p(0) e^{- \frac{t}{t_{\mathrm{s}}}} + \RevSakibara{\frac{1}{t_{\mathrm{s}}}}\int_0^t e^{-\frac{t-\tau}{t_{\mathrm{s}}}} u(\tau) d \tau. 
\end{align}
The sudden change in the drive command $u(t)$ can be represented by a Heaviside step function $H(t)$ defined by 
\begin{align}
	H(t) := \left\{ \begin{array}{ll} 0 & t < 0 \\ \frac{1}{2} & t = 0 \\ 1 &  0 < t \end{array} \right. ,
\end{align}
where the symbol $:=$ means definition, \textit{i.e.}, \RevJikuya{$X := Y$ defines $X$ as $Y$, while $Y =: X$ defines $X$ as $Y$.} 
By setting the initial drive amount to be zero $p(0) = 0$ and the drive command to be the Heaviside step function $u(t) = H(t)$, the step response is then given by 
\begin{align}
	p(t) = 1 - e^{- \frac{t}{t_{\mathrm{s}}}}. 
\end{align}
At time $t = t_{\mathrm{s}}$, $p(t_{\mathrm{s}})$ is found to be 63.2\% of the steady state value $p(\infty) = 1$, \textit{i.e.}, 
\begin{align}
	p(t_{\mathrm{s}}) = 1 - e^{-1} \simeq 0.632,
\end{align}
which confirms the definition of $t_{\mathrm{s}}$ as the time constant.
This relationship is highly practical and useful as it is used to perform step response tests on optical drive systems in actual environments and fit an exponential function to the step response to determine the time constant. 
We note that the dynamics of an optical drive system can also be modeled by the first-order differential equation 
\begin{align}
	t_{\mathrm{s}} \, \dot{p}(t) + p(t) = u(t - t_{\mathrm{d}}), 
\end{align}
with the time delay $t_{\mathrm{d}} > 0$. 
\RevJikuya{When the control period $t_{\mathrm{c}}$, the duration of each control cycle, and $t_{\mathrm{d}}$ are sufficiently smaller than $t_{\mathrm{s}}$, the effect of the time delay can be treated as unmodeled dynamics to keep the nominal model mathematically concise.} 
In the control system design described below, we plan to ensure robustness by considering not only the pure time delay but also the signal delay associated with digital control by taking into account the phase margin in the loop shaping design. 
We also note that the dynamics of an optical drive system can also be modeled by the second-order differential equation 
\begin{align}
	\ddot{p}(t) + 2 \zeta \omega_{\mathrm{n}} \, \dot{p}(t) + \omega_{\mathrm{n}}^2 \, p(t) = u(t),
\end{align}
where $\ddot{p}(t)$ is the second time-derivative of $p(t)$, $\zeta > 0$ is the damping ratio and $\omega_{\mathrm{n}} > 0$ is the natural angular frequency. 
When designing an optical drive system, the damping coefficient is set to be large, \textit{i.e.,} $\zeta \simeq 1$, to avoid vibration characteristics,  and then, the step response of the second-order differential equation can be expressed by the linear combination of exponential functions similar to the step response of the first-order differential equation. 
Hence, we neglect the time delay $t_{\mathrm{d}}$ or the second-order time-derivative $\ddot{p}(t)$ and adopt the first-order differential equation in modeling the optical drive system. 

MAO assumes the use of decoupled multiple optical drive systems, where the axes of each drive system are also \RevJikuya{dynamically} decoupled. 
The dynamic characteristics of the entire optical drive system can be expressed by the first order system, 
\textit{i.e.}, the system of first-order differential equations 
\begin{align}
	\bm{T}_{\mathrm{s}} \, \dot{\bm{p}}(t) + \bm{p}(t) = \bm{u}(t), \label{eqn:plant-01}
\end{align}
or equivalently
\begin{align}
	\dot{\bm{p}}(t) = -\bm{T}_{\mathrm{s}}^{-1}  \bm{p}(t) + \bm{T}_{\mathrm{s}}^{-1} \bm{u}(t), \label{eqn:plant-01-2}
\end{align}
where $\bm{T}_{\mathrm{s}} \in \mathbb{R}^{n \times n}$ is the diagonal matrix with the individual time constants on the diagonal elements, $\bm{p}(t) \in \mathbb{R}^n$ is the column vector of drive amounts, $\bm{u}(t) \in \mathbb{R}^n$ is the column vector with the drive commands arranged, the constant $n \in \mathbb{N}$ is the degree of freedom in the entire optical drive system, 
and the superscript $-1$ in $\bm{T}_{\mathrm{s}}^{-1}$ denotes the inverse matrix of $\bm{T}_{\mathrm{s}}$.  
The difference in reaction speed of each optical drive system is reflected in the time constant. 
Despite the limitations on the drive amount and speed, the subsequent control system design is formulated taking into account only the saturation of the drive amount. 
\RevJikuya{This is because, in the telescope instrumentation under consideration, the drive amount limits are more restrictive than the speed limits during typical operation. Neglecting the speed saturation thus allows for a more mathematically tractable formulation while providing a practical anti-windup solution without unnecessary complexity.}

As introduced in Sec.~\ref{sect:assumed_telescope_system}, in MAO the optical path length between the focal point and the M1 surface is measured in real time at multiple locations and controlled in real time to maintain the desired optical path length. 
Ranging radio waves are radiated from multiple radiators on the M1 surface and received at the focal point. 
By calculating the cross power spectrum of the radiated and received radio waves, we find that the phase of the cross power spectrum is affine to the product of the delay time between the radiated and received radio waves and the frequency of the radio waves. 
Therefore, calculating the phase gradient with respect to \RevJikuya{the propagation delay of the radio waves,} which is then converted into the optical path length from the emission point to the focal point. 
Here, assuming there are $m$-radiators, the EPL measurements $\bm{y}(t) \in \mathbb{R}^m$ are obtained in real time, where the constant $m \in \mathbb{N}$ represents the degree of freedom in the EPL measurement system. 

In large radio telescopes, the shape of the M1 surface deforms homologously depending on the elevation angle but keeps its shape paraboloidal or hyperboloidal in Cassegrain or Ritchey--Chr\'{e}tien optics, respectively, and the position of the M2 is optimally adjusted to take this shape deformation into account. 
Thus, the {desired optical path length} $\bm{g}(\gamma(t))$ induced by homologous deformation is determined according to the time-varying elevation angle $\gamma(t) \in \mathbb{R}$. 
Various error factors, such as errors in the telescope design and construction, vibrations associated with telescope pointing control, thermally-induced drift, and wind disturbances, can cause the optical path length to deviate from the desired value. 
Here, we assume that the change in EPLs driven by the optical drive system is sufficiently small compared to the EPLs themselves and linearly approximate them as $\bm{M} \bm{p}(t)$, where $\bm{M} \in \mathbb{R}^{m \times n}$ represents the coefficient matrix of the linear approximation and is called the {measurement matrix}. 
Note that we could also assume that $\bm{M}$ depends on $\gamma(t)$ and model $\bm{M}$ as $\bm{M}(\gamma(t))$, but for simplicity we model it as a constant matrix $\bm{M}$, which matches the trends suggested by our preliminary experiments. 
Real-time {EPL measurements} $\bm{y}(t)$ is expressed by the sum of the desired optical path length $\bm{g}(\gamma(t))$ and the optical path length change $\bm{M} \bm{p}(t)$, with the {error term} represented by $\bm{w}(t)$ encompassing all unmodeled factors as follows:
\begin{align}
	\bm{y}(t) =  \bm{M} \bm{p}(t) + \bm{g}(\gamma(t)) + \bm{w}(t). \label{eqn:plant-02}
\end{align}
When comparing the magnitudes of the terms on the right-hand side, the desired optical path length $\bm{g}(\gamma(t))$ is typically the largest. 
Meanwhile, the drive-induced change $\bm{M} \bm{p}(t)$ is supposed to be of a comparable order to the error term $\bm{w}(t)$, ensuring that the disturbance can be effectively compensated for within the available range of drive amounts. 
The time delay associated with EPL measurement can be an obstacle in control system design, but here we assume that it is sufficiently shorter than the control time, which is defined as the transmission interval of drive commands in the digital control system. 
Needless to say, a short control time is desirable to improve control performance, so it is also desirable that the time delay associated with EPL measurement is short. 

\begin{figure}[th!]
	\begin{center}
		\includegraphics[width=0.75 \textwidth]{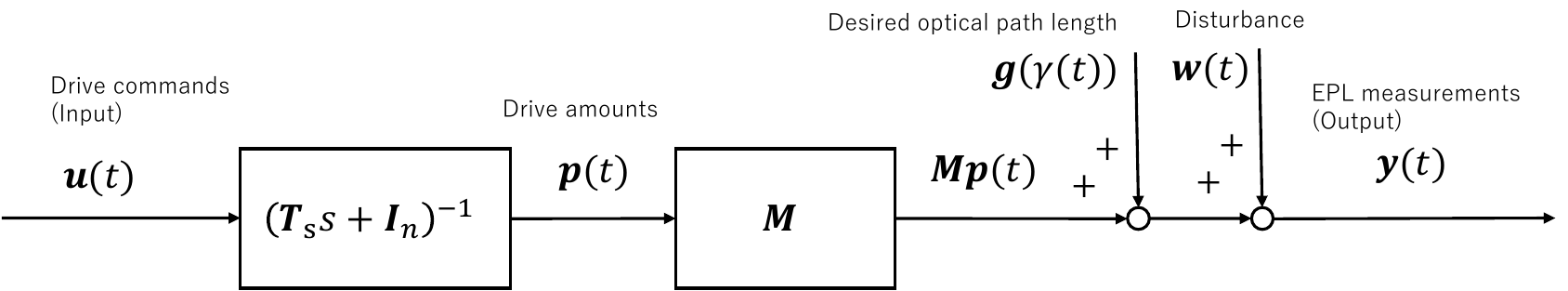}
		\caption{
			Block diagram of the plant model for control in MAO. The plant is described by Eqs.~\eqref{eqn:plant-01} and \eqref{eqn:plant-02}, and it is a mapping from the input $\bm{u}(t)$ to the output $\bm{y}(t)$ corrupted by the desired optical path length $\bm{g}(\gamma(t))$ and the disturbance $\bm{w}(t)$.  The plant dynamics are described by the first-order system in Eq.~\eqref{eqn:plant-01}, whose transfer function is given by Eq.~\eqref{eqn:plant-04}. The symbol $s$ in the transfer function represents the complex frequency that appears in the Laplace transformation in Eq.~\eqref{eqn:LaplaceTrans-01}, \RevJikuya{and $\bm{I}_n$ denotes the $n \times n$ identity matrix.} 
		}\label{fig:FigPlant}
	\end{center}
\end{figure}

Borrowing terminology from control engineering, our ``plant'' model for control in MAO is given by Eqs.~\eqref{eqn:plant-01-2} and \eqref{eqn:plant-02}, where Eq.~\eqref{eqn:plant-01-2} is called the {control equation}, Eq.~\eqref{eqn:plant-02} is called the {measurement equation}. 
In this context, the drive commands $\bm{u}(t)$ are called the {input}, the drive amount $\bm{p}(t)$ are called the {state}, the EPL measurements $\bm{y}(t)$ are called the {output}, 
and the error term $\bm{w}(t)$ is called the {disturbance}. 
The optical drive system corresponds to the {actuator}, and the EPL measurement system serves as the {sensor}. 
If there is a measurement system other than the EPL measurement system that can measure the optical path length, it can also be incorporated into the sensor model. 
As is customary in control engineering, block diagrams are frequently used to visualize the connections between signals and systems, which provides a clear overall understanding. 
The input--output relation of the plant model is visualized using the block diagram as shown in Fig.~\ref{fig:FigPlant}. 

As a supplemental note, to understand the symbol $(\bm{T}_{\mathrm{s}} \, s+ \bm{I}_n )^{-1}$ in Fig.~\ref{fig:FigPlant}, we briefly discuss the usage of Laplace transformation in control engineering. 
The Laplace transform, or more precisely, the one-sided Laplace transform, $\mathcal{L}$ is the integral transformation given by 
\begin{align}
	\mathcal{L}[\bm{p}(t)](s) := \int_{0}^{\infty} \bm{p}(t) e^{-st} dt, \label{eqn:LaplaceTrans-01}
\end{align}
where the complex variable $s \in \mathbb{C}$ denotes the complex frequency. 
The following formulas are known for the time derivative and time integral:
\begin{align}
	 \mathcal{L}\left[ \dot{\bm{p}}(t) \right](s) &= s \, \mathcal{L}[\bm{p}(t)](s) - \bm{p}(0), \label{eqn:LaplaceTrans-02} \\
	 \mathcal{L}\left[ \int_{-\infty}^{t} \bm{p}(\tau) d \tau \right] &=  \frac{1}{s} \left( \mathcal{L}[\bm{p}(t)](s) + \int_{-\infty}^{0} \bm{p}(\tau) d \tau \right). \label{eqn:LaplaceTrans-03}
\end{align}
In other words, the Laplace transform converts the time derivative into multiplication by $s$ and the time integral into division by $s$, or equivalently, multiplication by $1/s$. 
By setting the initial condition to be zero, $\bm{p}(0) = \bm{0}$, calculating the Laplace transformation of Eq.~\eqref{eqn:plant-01} gives 
\begin{align}
	s \, \bm{T}_{\mathrm{s}} \, \bm{p}(s) + \bm{p}(s) = \bm{u}(s), \label{eqn:plant-03}
\end{align}
where $\bm{p}(s) = \mathcal{L}[\bm{p}(t)](s)$ and $\bm{u}(s) = \mathcal{L}[\bm{u}(t)](s)$ are the Laplace transforms of $\bm{p}(t)$ and $\bm{u}(t)$. 
Equation \eqref{eqn:plant-03} can be algebraically converted to 
\begin{align}
	\bm{p}(s) = ( \bm{T}_{\mathrm{s}} \, s + \bm{I}_n )^{-1} \bm{u}(s), \label{eqn:plant-04}
\end{align}
where $\bm{I}_n \in \mathbb{R}^{n \times n}$ is the identity matrix. 
The symbol $(\bm{T}_{\mathrm{s}} \, s+ \bm{I}_n )^{-1}$ in Fig.~\ref{fig:FigPlant} is called the transfer function matrix from $\bm{u}(t)$ to $\bm{p}(t)$ and represents the input-output relation of the optical drive system in the frequency domain. 
The symbol $\bm{M}$ in Fig.~\ref{fig:FigPlant} denotes multiplying signal $\bm{p}(t)$ by matrix $\bm{M}$ to obtain signal $\bm{M} \bm{p}(t)$.
The symbol ``$+$'' in Fig.~\ref{fig:FigPlant} represents summation of signals to obtain the relation in Eq.~\eqref{eqn:plant-02}. 
We note that, as is permitted by convention in control engineering, Fig.~\ref{fig:FigPlant} shows the mixture of the frequency-domain transfer function matrix $( \bm{T}_{\mathrm{s}} \, s + \bm{I}_n )^{-1}$ and the time-domain signals such as $\bm{u}(t)$, $\bm{p}(t)$, etc.

%


\subsection{Fundamental Limitation of Control Problem}

To recognize the fundamental limitation of the control problem in MAO, we temporarily assume that the drive amount $\bm{p}(t)$ can be set instantaneously, \RevJikuya{\textit{i.e.}, $\bm{p}(t) = \bm{u}(t)$ by ignoring the derivative term $\bm{T}_{\mathrm{s}} \dot{\bm{p}}(t)$ in Eq.~\eqref{eqn:plant-01}, and} and first consider the ideal control problem.
Since the desired optical path length $\bm{g}(\gamma(t))$ is not the target of control in Eq.~\eqref{eqn:plant-02}, we define the {residual} $\bm{q}(t)$ as follows:
\begin{align}
	\bm{q}(t) := \bm{y}(t) - \bm{g}(\gamma(t)) = \bm{M} \bm{p}(t) + \bm{w}(t), \label{eqn:q-def}
\end{align}
and the ideal control problem is to suppress $\bm{q}(t)$ using $\bm{p}(t)$.

The solvability condition for this problem depends on the rank condition of $\bm{M}$. 
If $\bm{M}$ is a tall matrix, $m \geq n$, and has column-full rank, $\mathrm{rank}~\bm{M} = n$, 
the disturbance $\bm{w}(t)$ can be canceled in a least-squares approximation by  
\begin{align}
	\bm{p}(t) = - \left( \bm{M}^{\T} \bm{M} \right)^{-1} \bm{M}^{\T} \bm{w}(t), \label{eqn:p-ideal-02}
\end{align}
where the superscript $\T$ in $\bm{M}^{\T}$ denotes the transpose of $\bm{M}$. 
We note that Eq.~\eqref{eqn:p-ideal-02} does not always yield $\bm{q}(t) = \bm{0}$ because only the suppressible part of $\bm{w}(t)$ is canceled by $\bm{M} \bm{p}(t)$. 
For completeness, we recall that the rank, $\mathrm{rank}~\bm{M}$, of the matrix $\bm{M}$ is a number of linearly independent row or column vectors. 

Otherwise, we need to consider the general case where $\bm{M}$ may be rank deficient. 
To this end, we consider the singular value decomposition (SVD) of $\bm{M}$ as follows: 
\begin{align}
	\bm{M} = \bm{U} \bm{\Sigma} \bm{V}^{\T} = 
	\begin{bmatrix} \bm{U}_1 & \bm{U}_2 \end{bmatrix} 
	\begin{bmatrix} \bm{\Sigma}_{11} & \bm{0}_{r \times (n-r)} \\ \bm{0}_{(m-r) \times r} & \bm{0}_{(m-r) \times (n-r)} \end{bmatrix}  
	\begin{bmatrix} \bm{V}_1^{\T} \\ \bm{V}_2^{\T} \end{bmatrix}, \label{eqn:Msvd}
\end{align}
where $\bm{U} \in \mathbb{R}^{m \times m}$ and $\bm{V} \in \mathbb{R}^{n \times n}$ are orthogonal matrices, $\bm{0}$ is the zero matrix whose subscript denotes its matrix or  vector size, $\bm{\Sigma}$ is  the rectangular diagonal matrix with non-negative real numbers on the diagonal, and their sub-matrices are defined in compatible dimensions so that $r := \mathrm{rank}~\bm{M} = \mathrm{rank}~\bm{U}_1 = \mathrm{rank}~\bm{\Sigma}_{11} = \mathrm{rank}~\bm{V}_1$. 
The disturbance $\bm{w}(t)$ can be suppressed in a least-squares approximation by  
\begin{align}
	\bm{p}(t) = - \bm{V}_1 \bm{\Sigma}_{11}^{-1} \bm{U}_1^{\T} \bm{w}(t), \label{eqn:p-ideal-03}
\end{align}
which does not always yield $\bm{q}(t) = \bm{0}$. 
It readily follows that Eq.~\eqref{eqn:p-ideal-03} is a natural extension of Eq.~\eqref{eqn:p-ideal-02}. 
By rearranging the measurement matrix and the drive amount as follows:
\begin{align}
	\tilde{\bm{M}} = \bm{M} \bm{V}_1, \quad 
	\tilde{\bm{p}}(t) = \bm{V}_1^{\T} \bm{p}(t),  
\end{align}
we have 
\begin{align}
	\bm{M} \bm{p}(t) = \tilde{\bm{M}} \tilde{\bm{p}}(t). 
\end{align}
The use of $\tilde{\bm{p}}(t)$ instead of $\bm{p}(t)$ constrains the drive amount $\bm{p}(t)$ to a linear subspace $\Image \bm{V}_1$ rather than allowing full freedom in $\mathbb{R}^n$. 
Therefore, this cancellation problem can be equivalently reduced to the problem of replacing the measurement matrix $\bm{M}$ with a column-full rank measurement matrix $\tilde{\bm{M}}$. 

Hence, without loss of generality, we restrict our attention to the column-full rank $\bm{M}$ and will next consider a realistic control problem where the instantaneous value of the drive amount $\bm{p}(t)$ cannot be specified.


\subsection{Problem Formulation}

In reality, the true values of $\bm{g}(\gamma(t))$ and $\bm{M}$ are unknown. 
We need to determine the estimated residual value $\hat{\bm{q}}(t)$ using the estimated values $\hat{\bm{g}}(\gamma(t))$ of $\bm{g}(\gamma(t))$ as follows: 
\begin{align}
	\hat{\bm{q}}(t) = \bm{y}(t) - \hat{\bm{g}}(\gamma(t)), \label{eqn:hatq} 
\end{align}
and formulate the control problem for $\hat{\bm{q}}(t)$ using the estimated measurement matrix $\hat{\bm{M}}$ of $\bm{M}$. 

However, to avoid unnecessarily complicated discussions and promote intuitive understanding, we assume that $\hat{\bm{g}}(\gamma(t))$ and $\hat{\bm{M}}$ adequately approximate $\bm{g}(\gamma(t))$ and $\bm{M}$ and that $\hat{\bm{q}}(t)$ also adequately approximates $\bm{q}(t)$, and formulate the control problem using the true residual $\bm{q}(t)$ in Eq.~\eqref{eqn:q-def} and true measurement matrix $\bm{M}$. 
This simplification can also be justified by interpreting the approximation errors introduced by $\hat{\bm{g}}(\gamma(t))$ and $\hat{\bm{M}}$ as being included in the disturbance $\bm{w}(t)$. 

Let us decompose the residual $\bm{q}(t)$ in Eq.~\eqref{eqn:q-def} as follows:
\begin{align}
	\bm{q}(t) = \bm{M} \bm{\xi}(t) + \bm{\eta}(t), \label{eqn:q-dec-01} 
\end{align}
where the first term $\bm{M} \bm{\xi}(t)$ belongs to the image of $\bm{M}$ and the second term $\bm{\eta}(t)$ belongs to its orthogonal complement subspace: 
\begin{align}
	\bm{M} \bm{\xi}(t) &\in \Image \bm{M}, \label{eqn:q-dec-02}  \\
	\bm{\eta}(t) &\in \left( \Image \bm{M} \right)^{\perp} = \Ker \bm{M}^{\T}. \label{eqn:q-dec-03} 
\end{align}
Here, $\bm{\xi}(t) \in \mathbb{R}^n$ is defined as the actuation coefficients, which represents the combined effect of the input and the source of the suppressible disturbance. 
For self-completeness, we recall that the definitions of the image, $\Image \bm{X}$, and the kernel, $\Ker \bm{X}$, of a matrix $\bm{X} \in \mathbb{R}^{m \times n}$ are given by 
\begin{align}
	\Image \bm{X} &:= \left\{ \bm{X} \bm{x} \middle|  \bm{x} \in \mathbb{R}^n \right\}, \\
	\Ker \bm{X} &:= \left\{ \bm{x} \in \mathbb{R}^n \middle|  \bm{X} \bm{x} = \bm{0}_m \right\},
\end{align}
and the definition of the orthogonal complement subspace, $\bm{\mathcal{X}}^{\perp}$, of a subspace $\bm{\mathcal{X}} \subset \mathbb{R}^n$ is given by 
\begin{align}
	\bm{\mathcal{X}}^{\perp} &= \left\{ \bm{y} \in \mathbb{R}^n \middle| \bm{y}^{\T} \bm{x} = 0, \forall \bm{x} \in \bm{\mathcal{X}} \right\}. 
\end{align}

We aim to determine the input $\bm{u}(t)$ such that the residual $\bm{q}(t)$ is suppressed. 
Since $\bm{\eta}(t)$ in Eq.~\eqref{eqn:q-dec-01} cannot be suppressed by any choice of $\bm{u}(t)$\RevJikuya{,} the only possible scenario is to suppress $\bm{M} \bm{\xi}(t)$ in Eq.~\eqref{eqn:q-dec-01} by an appropriate choice of $\bm{u}(t)$. 
Due to causality, it is impossible to find $\bm{u}(t)$ that makes $\bm{M} \bm{\xi}(t)$ identically zero for all time. 
A typical alternative approach is to find $\bm{u}(t)$ such that $\bm{M} \bm{\xi}(t)$ asymptotically converges to zero for any initial condition $\bm{p}(0)$. 
Considering the internal model principle underlying servo systems, this approach is intractable for an arbitrary bounded function $\bm{w}(t)$. 
Hence, we restrict our attention to a constant disturbance $\bm{w}$ and seek $\bm{u}(t)$ such that $\bm{M} \bm{\xi}(t)$ asymptotically converges to zero for any initial condition $\bm{p}(0)$ and any constant disturbance \RevSakibara{$\bm{w}$}. 

By premultiplying the Moore-Penrose pseudo-inverse matrix $\bm{M}^{\dagger}$ from the left to Eq.~\eqref{eqn:q-dec-01}, the actuation coefficient vector $\bm{\xi}(t)$ can be recovered from $\bm{q}(t)$ as follows: 
\begin{align}
	\bm{\xi}(t) &= \bm{M}^{\dagger} \bm{q}(t), \label{eqn:xi-recover} \\
	\bm{M}^{\dagger} &:= \left( \bm{M}^{\T} \bm{M} \right)^{-1} \bm{M}^{\T}, \label{eqn:Mdag-def}
\end{align}
where the {Moore-Penrose pseudo-inverse matrix} $\bm{M}^{\dagger}$ is well-defined due to the column-full rank assumption on $\bm{M}$ made without loss of generality. 

Our control problem is defined as the asymptotic disturbance suppression problem. Its goal is to design a controller which is a mapping from $\bm{y}(t)$ to $\bm{u}(t)$ such that the closed-loop system is asymptotically stable and that the following asymptotic convergence properties are satisfied 
\begin{align}
	\lim_{t \rightarrow \infty} \bm{M} \bm{\xi}(t) = \bm{0}_m 
	\quad \Leftrightarrow \quad \lim_{t \rightarrow \infty} \bm{\xi}(t) = \bm{0}_n 
	\quad \Leftrightarrow \quad  \lim_{t \rightarrow \infty} \bm{p}(t) = - \bm{M}^{\dagger}  \bm{w} 
	\label{eqn:ACP-01}
\end{align}
for any initial condition $\bm{p}(0)$ and any constant disturbance $\bm{w}$. 
The asymptotic convergence on the left means that the suppressible component $\bm{M} \bm{\xi}(t)$ of the residual $\bm{q}(t)$ is suppressed. 
\RevJikuya{Physically, this represents the minimization of the optical wavefront errors that fall within the controllable range of the actuators.} 
Monitoring the estimated residual during actual operation is useful for evaluating the deviation of the optical path from the desired one. 
The asymptotic convergence in the middle means that the actuation coefficients $\bm{\xi}(t)$ converge to zero. 
Monitoring the estimated actuation coefficients during actual operation is useful for verifying the effectiveness of the control law. 
The asymptotic convergence on the right means that the suppressible part $\bm{M}^{\dagger}  \bm{w}$ of the disturbance is compensated for by the drive amount $\bm{p}(t)$. 
\RevJikuya{This implies that the actuators have successfully reached the specific positions required to cancel out the static optical disturbances.} 
This relationship represents the expected steady-state behavior for maintaining the desired optical path length within its suppressible subspace. 

Note that while the term ``disturbance rejection'' is commonly used in standard control problems, we intentionally employ the term ``disturbance suppression'' in this paper. 
In typical control configurations \RevSakibara{where} the matrices corresponding to our measurement matrix $\bm{M}$ are wide or square, the objective is to eliminate the disturbance impact entirely. In our case, however, the matrix $\bm{M}$ is tall, which introduces a fundamental geometric constraint, \textit{i.e.}, directionality, on the control action. 
By using ``suppression'', we emphasize that the control action is inherently directional, resulting in the selective attenuation of disturbance components within the image of $\bm{M}$, rather than the omnidirectional elimination implied by ``rejection''.


\subsection{Anti-Windup Proportional-Integral Control Law} 

Considering the internal model principle, the effects of a constant disturbance $\bm{w}$ can be asymptotically eliminated by incorporating an integrator into the controller. 
Because input saturation occurs due to the physical constraints of the optical drive system, a saturation function must be applied to the input inside the controller. 
However, it is well known that we need to avoid integrator windup in integral control, where the integrator windup is a phenomenon such that the internal integrator continues to accumulate the control error, resulting in an excessive buildup of the integral term, even when the physical control input is saturated. This accumulation causes a significant delay in recovering from the saturated state, severely degrading control performance. To address this, the mechanism adjusts the integrator variable by feeding back the difference between the pre-saturation input and the actual saturated output.

Therefore, we aim to solve our control problem using the following {Anti-Windup Proportional-Integral (AWPI or anti-windup PI) controller:}
\begin{align}
	\hat{\bm{q}}(t) &= \bm{y}(t) - \bm{g}(\gamma(t)), \label{eqn:AWPI-00} \\
	\hat{\bm{\xi}}(t) &= \bm{M}^{\dagger} \hat{\bm{q}}(t), \label{eqn:AWPI-0} \\
	\dot{\bm{v}}(t) &= \hat{\bm{\xi}}(t) + \bm{K}_{\mathrm{a}} \left( \tilde{\bm{u}}(t) - \bm{u}(t)   \right), \label{eqn:AWPI-01} \\
	\tilde{\bm{u}}(t) &= - \bm{K}_{\mathrm{i}} \, \bm{v}(t) - \bm{K}_{\mathrm{p}} \, \hat{\bm{\xi}}(t), \label{eqn:AWPI-02} \\
	\bm{u}(t) &= \mathrm{\mathbf{sat}} \left( \tilde{\bm{u}}(t) \right), \label{eqn:AWPI-03}
\end{align}
where $\hat{\bm{q}}(t) \in \mathbb{R}^m$ is an {estimated residual}, $\hat{\bm{\xi}}(t)  \in \mathbb{R}^n$ is an estimated actuation coefficients, $\bm{v}(t) \in \mathbb{R}^n$ is an {integrator variable}, $\tilde{\bm{u}}(t) \in \mathbb{R}^m$ is the unsaturated input, $\bm{K}_{\mathrm{i}} , \bm{K}_{\mathrm{p}}, \bm{K}_{\mathrm{a}} \in \mathbb{R}^{n \times n}$ are diagonal matrices with positive diagonal elements representing the {integral, proportional, and anti-windup gains}, respectively, and $\mathrm{\mathbf{sat}}(\cdot): \mathbb{R}^{n} \rightarrow \mathbb{R}^{n}$ is the {saturation function}.  
The $i$-th element of the saturation function $\mathrm{\mathbf{sat}}(\cdot)$ is defined as  
\begin{align}
	\mathrm{sat}_i \left( x \right) 
	:= \left\{ \begin{array}{ll} \bar{u}_i & \bar{u}_i < x \\ 
	x & - \bar{u}_i \leq x \leq \bar{u}_i \\ 
	- \bar{u}_i & x < - \bar{u}_i  \end{array} 
	\right. , \label{eqn:satdef}
\end{align}
with the constants $\pm \bar{u}_i \in \mathbb{R}$ representing the upper and lower limits of saturation for each $i = 1, \ldots, n$. 
\RevJikuya{We note that while one could mathematically set $\bm{K}_{\mathrm{p}} = \bm{0}_{n \times n}$, we assume $\bm{K}_{\mathrm{p}} \neq \bm{0}_{n \times n}$ in this study. This is because the inclusion of the proportional term is essential for ensuring sufficient phase margins at high frequencies and for enabling the systematic loop-shaping design discussed in Sec.~\ref{sec:LoopShaping}.} 
Similarly, if an anti-windup mechanism is not needed, one can simply set $\bm{K}_{\mathrm{a}} = \bm{0}_{n \times n}$ in Eq.~\eqref{eqn:AWPI-01} and $\tilde{\bm{u}}(t) = \bm{u}(t)$ in Eq.~\eqref{eqn:AWPI-02} such that the pure PI controller in the subsequent subsection is obtained. 
We note that the residual $\bm{q}(t)$ in Eq.~\eqref{eqn:q-def} and the estimated residual in Eq.~\eqref{eqn:AWPI-00} are mathematically identical, but they are intentionally distinguished by using different symbols to account for the discrepancy between the desired optical path length $\bm{g}(\gamma(t))$ and the estimated one $\hat{\bm{g}}(\gamma(t))$ in reality as discussed in Eq.~\eqref{eqn:hatq}. 
We also note that the actuation coefficient $\bm{\xi}(t)$ in Eq.~\eqref{eqn:xi-recover} and the estimated actuation coefficients $\hat{\bm{\xi}}(t)$ in Eq.~\eqref{eqn:AWPI-0} are mathematically identical, but they are intentionally distinguished by using different symbols to account for the discrepancy between the true measurement matrix $\bm{M}$ and the estimated one $\hat{\bm{M}}$ in reality. 
\RevJikuya{In other words, to maintain simplicity in the subsequent control law derivation, we hereafter omit the carets and assume that any such discrepancies are effectively absorbed into the disturbance $\bm{w}(t)$}

\begin{figure}[th!]
	\begin{center}
		\includegraphics[width=\textwidth]{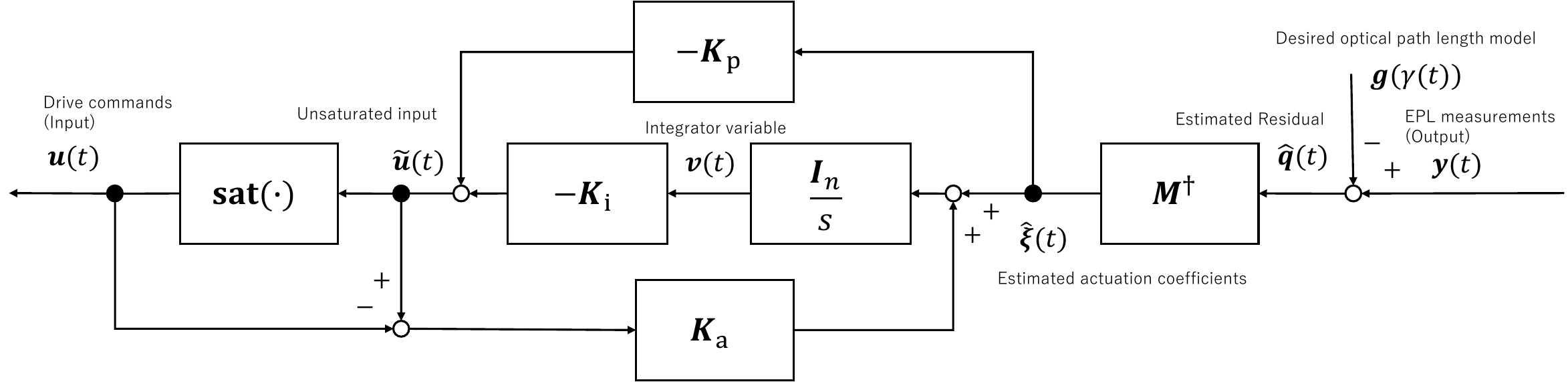}
		\caption{
			Block diagram of the AWPI controller. The controller is described by Eqs.~\eqref{eqn:AWPI-00}-\eqref{eqn:AWPI-03}, incorporating the saturation function $\mathrm{\mathbf{sat}}(\cdot)$ defined in Eq.~\eqref{eqn:satdef}. This controller is a mapping from output $\bm{y}(t)$ to the input $\bm{u}(t)$, and internally performs the correction by first subtracting the desired optical path length $\bm{g}(\gamma(t))$ from $\bm{y}(t)$. The resulting estimated residual $\hat{\bm{q}}(t)$ is then processed by the PI control logic, and the anti-windup mechanism. The core part of the PI controller is the term  $\frac{\bm{I}_n}{s}$ corresponding to $n$ integrators connected in parallel, where the symbol $\frac{1}{s}$ represents the time-integration action and appears in the Laplace transformation of time-integral in Eq.~\eqref{eqn:LaplaceTrans-03}. The design parameters in the controller are the proportional, integral, and anti-windup gains, $\bm{K}_{\mathrm{p}}$,  $\bm{K}_{\mathrm{i}}$,  and $\bm{K}_{\mathrm{a}}$, respectively, where the local feedback represented by $\bm{K}_{\mathrm{a}}$ constitutes the anti-windup mechanism. The design procedure for these gain is described in the main text. 
		}\label{fig:FigAWPI}
	\end{center}
\end{figure}

The input-output relation of the AWPI control can be visualized using the block diagram shown in Fig.~\ref{fig:FigAWPI}. 
The symbols ``$\frac{\bm{I}_n}{s}$'', ``$-\bm{K}_{\mathrm{i}}$'', and ``$- \bm{K}_{\mathrm{p}}$''  in Fig.~\ref{fig:FigAWPI} represents the PI controller, which is a parallel connection of proportional and integral actions in the frequency domain and serves as the core part of the AWPI controller. 
Conceptually, connecting the input $\bm{u}$ and the output $\bm{y}$ in Figs.~\ref{fig:FigPlant} and \ref{fig:FigAWPI} completes a closed loop, thereby creating a feedback control system.

To understand the role of the anti-windup mechanism, we first consider the local feedback loop involving the integrator variable $\bm{v}(t)$ and the anti-windup gain $\bm{K}_{\mathrm{a}}$, which is designed to maintain stability of the local feedback loop. 
Consider a situation where the $i$-th element of the saturation function $\mathrm{\mathbf{sat}}(\cdot)$ saturates in the positive direction. 
In this case, the $i$-th element of $\tilde{\bm{u}}(t)$ is greater than the $i$-th element of $\bm{u}(t)$. 
Since the $i$-th element of $\tilde{\bm{u}}(t) - \bm{u}(t)$ is positive, \RevJikuya{the second term on the right-hand side of Eq.~\eqref{eqn:AWPI-01} acts to increase the $i$-th element of integrator variable $\bm{v}(t)$. Because $\tilde{\bm{u}}(t)$ is determined by $- \bm{K}_{\mathrm{i}} \bm{v}(t)$ in Eq.~\eqref{eqn:AWPI-02}, this increase in the $i$-th element of $\bm{v}(t)$ results in a decrease in the $i$-th element of $\tilde{\bm{u}}(t)$ due to the negative gain $- \bm{K}_{\mathrm{i}}$.} 
A similar, symmetric argument applies when the $i$-th element of the saturation function saturates in the negative direction.
Therefore, we can see that the anti-windup mechanism plays a role in bringing the $i$-th element of $\tilde{\bm{u}}(t)$ closer to $\bm{u}(t)$ regardless of the direction of saturation. 

When selecting an anti-windup gain $\bm{K}_{\mathrm{a}}$, it is important to note that if the gain is too large, the system may be sensitive to high-frequency noise and vibration called the chattering may occur in the input, while if the gain is too small, there is a risk of slow recovery from saturation and the windup phenomena may occur. 
Therefore, selecting an appropriate anti-windup gain $\bm{K}_{\mathrm{a}}$ is crucial. 
One way to do this is to match the time constant $\bm{K}_{\mathrm{i}}^{-1} \bm{K}_{\mathrm{a}}^{-1} $ of the anti-windup mechanism with the integral time $\bm{K}_{\mathrm{i}}^{-1} \bm{K}_{\mathrm{p}}$ of the PI controller. 
In this case, solving $\bm{K}_{\mathrm{i}}^{-1} \bm{K}_{\mathrm{a}}^{-1}  =  \bm{K}_{\mathrm{i}}^{-1} \bm{K}_{\mathrm{p}}$ gives $\bm{K}_{\mathrm{a}} = \bm{K}_{\mathrm{p}}^{-1}$. 
Using this value as a starting point and adjusting the parameters in $\bm{K}_{\mathrm{a}}$ through simulation, one can expect to obtain a gain that balances stability and recovery speed. 
The method for selecting the integral gain $\bm{K}_{\mathrm{i}}$ and the proportional gain $\bm{K}_{\mathrm{p}}$ will be discussed later.

Applying the Euler approximation, the {AWPI controller for digital implementation} can be obtained by time-discretization  as follows:
\begin{align}
    \hat{\bm{\xi}}[k] &= \bm{M}^{\dagger} \left( \bm{y}[k] - \bm{g}(\gamma[k]) \right), \label{eqn:AWPI-04f} \\
    \tilde{\bm{u}}[k] &= - \bm{K}_{\mathrm{i}} \, \bm{v}[k] - \bm{K}_{\mathrm{p}} \, \hat{\bm{\xi}}[k], \label{eqn:AWPI-05f} \\
    \bm{u}[k] &= \mathrm{\mathbf{sat}} \left( \tilde{\bm{u}}[k] \right), \label{eqn:AWPI-06f} \\
    \bm{v}[k+1] &= \bm{v}[k] + t_{\mathrm{c}} \left[ \hat{\bm{\xi}}[k] + \bm{K}_{\mathrm{a}} \left( \tilde{\bm{u}}[k] - \bm{u}[k] \right) \right], \label{eqn:AWPI-07f}
\end{align}
in the forward difference form or 
\begin{align}
	\hat{\bm{\xi}}[k] &= \bm{M}^{\dagger} \left( \bm{y}[k] - \bm{g}(\gamma[k]) \right), \label{eqn:AWPI-04b} \\
	\bm{v}[k] &= \bm{v}[k-1] + t_{\mathrm{c}} \left[ \, \hat{\bm{\xi}}[k] +  \bm{K}_{\mathrm{a}} \left( \tilde{\bm{u}}[k-1] - \bm{u}[k-1]  \right) \right], \label{eqn:AWPI-05b} \\
	\tilde{\bm{u}}[k] &= - \bm{K}_{\mathrm{i}} \, \bm{v}[k] - \bm{K}_{\mathrm{p}} \, \hat{\bm{\xi}}[k], \label{eqn:AWPI-06b} \\
	\bm{u}[k] &= \mathrm{\mathbf{sat}} \left( \tilde{\bm{u}}[k] \right), \label{eqn:AWPI-07b}
\end{align}
in the backward difference form, where $k \in \mathbb{Z}$ is the discrete-time index and $t_{\mathrm{c}} > 0$ is the control period. 

In actual operation, it is important to monitor the status of the control system, and it is recommended to specifically monitor the estimated actuation coefficients $\hat{\bm{\xi}}[k]$ and the unsaturated input $\tilde{\bm{u}}[k]$. 
If the PI controller is functioning properly in the absence of saturation, $\hat{\bm{\xi}}[k]$ is expected to converge to near zero. If the anti-windup mechanism is functioning properly, $\tilde{\bm{u}}[k]$ will not normally saturate, and even if it does saturate, it is designed to return to within the saturation range promptly.


\subsection{Stability and Asymptotic Analysis}

Let us verify that the AWPI controller constitutes a solution to our control problem. 
Since the anti-windup mechanism only functions to prevent integrator windup when the unsaturated input $\tilde{\bm{u}}(t)$ is saturated, we can analyze the nominal behavior by considering the situation where saturation is not reached. 
\RevJikuya{In this case, the input $\bm{u}(t)$ is identical to the unsaturated input $\tilde{\bm{u}}(t)$, \textit{i.e.}, $\bm{u}(t) = \mathrm{\mathbf{sat}} \left( \tilde{\bm{u}}(t) \right) = \tilde{\bm{u}}(t)$, and the anti-windup term in Eq.~\eqref{eqn:AWPI-01} vanishes. Therefore, we} perform the stability and asymptotic analysis of the pure PI controller: 
\begin{align}
	\hat{\bm{q}}(t) &= \bm{y}(t) - \bm{g}(\gamma(t)), \label{eqn:PI-00} \\
	\hat{\bm{\xi}}(t) &= \bm{M}^{\dagger} \hat{\bm{q}}(t), \label{eqn:PI-0} \\
	\dot{\bm{v}}(t) &= \hat{\bm{\xi}}(t), \label{eqn:PI-01} \\
	\bm{u}(t) &= - \bm{K}_{\mathrm{i}} \, \bm{v}(t) - \bm{K}_{\mathrm{p}} \, \hat{\bm{\xi}}(t). \label{eqn:PI-02} 
\end{align}

The {closed-loop system}, combining the plant in Eqs.~\eqref{eqn:plant-01-2} and \eqref{eqn:plant-02} and the PI controller in Eqs.~\eqref{eqn:PI-00} to \eqref{eqn:PI-02}, is described by the following state-space representation: 
\begin{align}
	\begin{bmatrix} \dot{\bm{v}}(t) \\ \dot{\bm{p}}(t) \end{bmatrix}
	= \begin{bmatrix} \bm{0}_{n \times n} & \bm{I}_n \\ - \bm{T}_{\mathrm{s}}^{-1} \bm{K}_{\mathrm{i}} & - \bm{T}_{\mathrm{s}}^{-1} \left( \bm{I}_n + \bm{K}_{\mathrm{p}} \right) \end{bmatrix} 
	\begin{bmatrix} \bm{v}(t) \\ \bm{p}(t) \end{bmatrix} 
	+ \begin{bmatrix} \bm{I}_n \\ - \bm{T}_{\mathrm{s}}^{-1} \bm{K}_{\mathrm{p}} \end{bmatrix} \bm{M}^{\dagger} \bm{w}, \label{eqn:CL-01}
\end{align}
where the disturbance $\bm{w}(t)$ has been simplified to a constant $\bm{w}$ for the purpose of asymptotic analysis. 
Here, we note that the process of the plant and the controller sharing the input $\bm{u}(t)$ and the output $\bm{y}(t)$ and forming a simultaneous system is called closing the feedback loop, and the resulting simultaneous equations are referred to as the closed-loop system.
We also note that a coupled set of first-order differential equations with inputs and outputs is called the state-space representation in control engineering, and this representation takes this form with the constant input $\bm{w}$ and without explicit outputs. 

\begin{figure}[th!]
	\begin{center}
		\includegraphics[width=0.75 \textwidth]{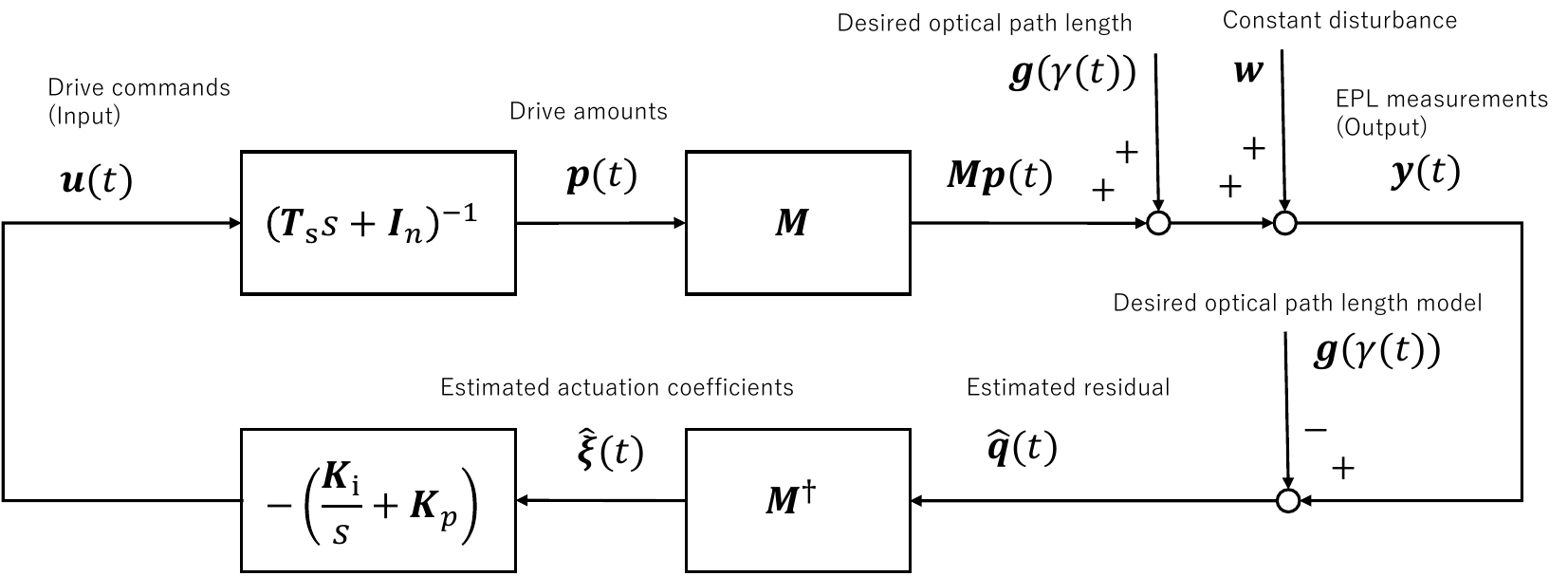}
		\caption{
			Block diagram of the closed-loop system consisting of the plant and the pure PI controller without the anti-windup mechanism. This simplified system is composed of the plant shown in Fig.~\ref{fig:FigPlant} and the PI controller derived from Fig.~\ref{fig:FigAWPI}. For analysis of the behavior near the operating point, the controller is simplified by omitting the saturation function $\mathrm{\mathbf{sat}}(\cdot)$ as well as anti-windup mechanism represented by the local feedback with the anti-windup gain $\bm{K}_{\mathrm{a}}$. 
		}\label{fig:FigCL}
	\end{center}
\end{figure}

The signal flow of the closed-loop system can be visualized using the block diagram shown in Fig.~\ref{fig:FigCL}. 
The symbol ``$- \left( \frac{\bm{K}_{\mathrm{i}}}{s} + \bm{K}_{\mathrm{p}} \right)$'' in Fig.~\ref{fig:FigCL} represents a parallel connection of proportional and integral actions in the frequency domain, serving as the core part of the PI controller without the anti-windup mechanism.

Let us investigate the closed-loop stability of the system described by Eq.~\eqref{eqn:CL-01}. 
The characteristic polynomial $\phi(s)$ of the closed-loop system is given by 
\begin{align}
	\phi(s) := \det \left( s \bm{I}_{2n} - \begin{bmatrix} \bm{0}_{n \times n} & \bm{I}_n \\ - \bm{T}_{\mathrm{s}}^{-1} \bm{K}_{\mathrm{i}} & - \bm{T}_{\mathrm{s}}^{-1} \left( \bm{I}_n + \bm{K}_{\mathrm{p}} \right) \end{bmatrix} \right), 
\end{align}
where $\det \bm{X}$ is the determinant of a square matrix $\bm{X}$. 
This polynomial is the key indicator of stability because its roots, known as the closed-loop poles, govern the temporal behavior of the state variables $\bm{v}(t)$ and $\bm{p}(t)$ of the closed-loop system. 
We note that a polynomial is defined as Hurwitz stable if and only if all of its roots lie strictly in the open left-half of the complex plane. 
This condition ensures that the corresponding system is asymptotically stable. 
Therefore, we can analyze the stability of the closed-loop system by examining the locations of the roots of $\phi(s)$ in the complex plane.

Because the matrices $\bm{T}_{\mathrm{s}}$, $\bm{K}_{\mathrm{i}}$, and $\bm{K}_{\mathrm{p}}$ are diagonal, the system matrix in Eq.~\eqref{eqn:CL-01} is decoupled into $n$ independent second-order systems. 
By pre- and post-multiplying the argument of the determinant by appropriate permutation matrices, the characteristic polynomial $\phi(s)$ is therefore factored as follows:
\begin{align}
	\phi(s) 
	= \prod_{i=1}^n \det \left( s \bm{I}_{2} - \begin{bmatrix} 0 & 1 \\ - \frac{k_{\mathrm{i},i}}{t_{\mathrm{s},i}} & - \frac{1 + k_{\mathrm{p},i}}{t_{\mathrm{s},i}} \end{bmatrix} \right) 
	= \prod_{i=1}^n \left( s^2 +  \frac{1 + k_{\mathrm{p},i}}{t_{\mathrm{s},i}} s + \frac{k_{\mathrm{i},i}}{t_{\mathrm{s},i}} \right), \label{eqn:phi-factor}
\end{align}
where $t_{\mathrm{s},i}, k_{\mathrm{i},i}, k_{\mathrm{p},i} > 0$ are the $i$-th diagonal elements of $\bm{T}_{\mathrm{s}}$, $\bm{K}_{\mathrm{i}}$, and $\bm{K}_{\mathrm{p}}$, respectively. 
Since all the coefficients of each quadratic factor are positive, the factor is Hurwitz stable. Because the product of Hurwitz polynomials is also Hurwitz stable, we can verify that the characteristic polynomial $\phi(s)$ is Hurwitz stable. 
Hence, the closed-loop system described by Eq.~\eqref{eqn:CL-01} is shown to be asymptotically stable. 
We note that the closed-loop stability is also guaranteed even for a bounded and time-varying $\bm{w}(t)$.

Let us also investigate the asymptotic convergence properties for Eq.~\eqref{eqn:CL-01}. 
Since the closed-loop system in Eq.~\eqref{eqn:CL-01} is asymptotically stable, both $\bm{v}(t)$ and $\bm{p}(t)$ are guaranteed to converge to constant steady-state values: 
\begin{align}
	\bm{v}(\infty) := \lim_{t \rightarrow \infty} \bm{v}(t), \\
	\bm{p}(\infty) := \lim_{t \rightarrow \infty} \bm{p}(t). 
\end{align}
The steady-state values can be found by setting the time derivatives to zero in Eq.~\eqref{eqn:CL-01}, which yields the following algebraic equations: 
\begin{align}
	\begin{bmatrix} \bm{0}_n \\ \bm{0}_n \end{bmatrix}
	= \begin{bmatrix} \bm{0}_{n \times n} & \bm{I}_n \\ - \bm{T}_{\mathrm{s}}^{-1} \bm{K}_{\mathrm{i}} & - \bm{T}_{\mathrm{s}}^{-1} \left( \bm{I}_n + \bm{K}_{\mathrm{p}} \right) \end{bmatrix} 
	\begin{bmatrix} \bm{v}(\infty) \\ \bm{p}(\infty) \end{bmatrix} 
	+ \begin{bmatrix} \bm{I}_n \\ - \bm{T}_{\mathrm{s}}^{-1} \bm{K}_{\mathrm{p}} \end{bmatrix} \bm{M}^{\dagger} \bm{w}. \label{eqn:CL-02}
\end{align}
Solving this system gives:
\begin{align}
	\bm{v}(\infty) &= \bm{K}_{\mathrm{i}}^{-1} \bm{M}^{\dagger} \bm{w}, \label{eqn:CL-03} \\
	\bm{p}(\infty) &= - \bm{M}^{\dagger} \bm{w}. \label{eqn:CL-04}
\end{align}
Furthermore, we can determine the steady-state values for the estimated residual, estimated actuation coefficients, and input:
\begin{align}
	\hat{\bm{q}}(\infty) &:= \lim_{t \rightarrow \infty} \hat{\bm{q}}(t) 
	= \bm{M} \bm{p}(\infty) + \bm{w} = \left( \bm{I}_m - \bm{M} \bm{M}^{\dagger} \right) \bm{w}, \label{eqn:CL-05} \\
	\hat{\bm{\xi}}(\infty) &:= \lim_{t \rightarrow \infty} \hat{\bm{\xi}}(t) 
	=  \bm{M}^{\dagger} \bm{q}(\infty) = \bm{0}_n, \label{eqn:CL-06} \\
	\bm{u}(\infty) &:= \lim_{t \rightarrow \infty} \bm{u}(t) 
	= - \bm{K}_{\mathrm{i}} \, \bm{v}(\infty) - \bm{K}_{\mathrm{p}} \, \bm{M}^{\dagger} \hat{\bm{q}}(\infty) 
	= - \bm{M}^{\dagger} \bm{w}. \label{eqn:CL-07}
\end{align}
Equations~\eqref{eqn:CL-04} and \eqref{eqn:CL-06} confirm that the asymptotic convergence properties required by Eq.~\eqref{eqn:ACP-01} are satisfied. 
Equation~\eqref{eqn:CL-05} is consistent with the residual decomposition form introduced in Eq.~\eqref{eqn:q-dec-01}. 
In other words, Eq.~\eqref{eqn:q-dec-01} states that the components belonging to $\left( \Image \bm{M} \right)^{\perp}$ cannot be suppressed by control action, and Eq.~\eqref{eqn:CL-05} explicitly shows that $\hat{\bm{q}}(\infty) \in \left( \Image \bm{M} \right)^{\perp}$, confirming the consistency. 


\subsection{Loop Shaping Design}
\label{sec:LoopShaping}

Let us consider the selection method for the integral gain $\bm{K}_{\mathrm{i}}$ and the proportional gain $\bm{K}_{\mathrm{p}}$ based on loop shaping technique. 
Since an overly large integral gain $\bm{K}_{\mathrm{i}}$ is known to destabilize the closed-loop system, we consider a design approach focused on securing appropriate stability margins. 
The {open-loop transfer function matrix} $\bm{L}(s)$ is derived by cutting the closed-loop system block diagram in Fig.~\ref{fig:FigCL} at the input $\bm{u}(t)$ as follows: 
\begin{align}
	\bm{L}(s) = \left( \frac{\bm{K}_{\mathrm{i}}}{s} + \bm{K}_{\mathrm{p}} \right) \bm{M}^{\dagger} \bm{M} \left( \bm{T}_{\mathrm{s}} \, s + \bm{I}_n \right)^{-1}
	= \left( \frac{\bm{K}_{\mathrm{i}}}{s} + \bm{K}_{\mathrm{p}} \right) \left( \bm{T}_{\mathrm{s}} \, s + \bm{I}_n \right)^{-1}. 
\end{align}
Here, the open-loop system is defined by the simultaneous equations obtained from the series connection of the plant and the controller, where the negative sign associated with the feedback path is conceptually separated and excluded from the connection. This connection is established by breaking the signal path at any point within the closed-loop system to obtain a system that only processes signals sequentially. The transfer representation of this system is called the open-loop transfer function.

Because the matrices $\bm{T}_{\mathrm{s}}$, $\bm{K}_{\mathrm{i}}$, and $\bm{K}_{\mathrm{p}}$ are diagonal, $\bm{L}(s)$ is decoupled and diagonal. 
The {decoupled open loop transfer function} $L(s)$, representing each diagonal element of $\bm{L}(s)$, is expressed in the following form:
\begin{align}
	L(s) = \frac{k_{\mathrm{p}} s + k_{\mathrm{i}}}{s(t_{\mathrm{s}} s + 1)},
\end{align}
where $k_{\mathrm{i}}, k_{\mathrm{p}}, t_{\mathrm{s}} > 0$ are the integral gain, the proportional gain, and the time constant, respectively, and the subscript indicating the order of the diagonal elements are omitted for simplicity. 
Note that the control period $t_{\mathrm{c}} > 0$, used for the Euler approximation in Eqs.~\eqref{eqn:AWPI-04f} to \eqref{eqn:AWPI-07f} or Eqs.~\eqref{eqn:AWPI-04b} to \eqref{eqn:AWPI-07b}, will be utilized in the subsequent discussion.

Loop shaping is a technique for adjusting control parameters such that the Bode plot of the open-loop transfer function achieves a desired shape, often to satisfy stability and performance criteria. 
Here, by assuming 
\begin{align}
	\frac{k_{\mathrm{i}}}{k_{\mathrm{p}}} < \frac{1}{t_{\mathrm{s}}}, \label{eqn:FreqRange-01} 
\end{align}
we derive a piecewise linear approximation of the gain diagram of $L(s)$ to obtain rough initial values for $k_{\mathrm{i}}$ and $k_{\mathrm{p}}$ for parameter tuning: 
\begin{align}
	20 \log_{10} | L (j \omega) | 
	= \left\{ \begin{array}{ll} - 20 \log_{10} \omega + 20 \log_{10} k_{\mathrm{i}} & \omega < \frac{k_{\mathrm{i}}}{k_{\mathrm{p}}} \\
	20 \log_{10} k_{\mathrm{p}} & \frac{k_{\mathrm{i}}}{k_{\mathrm{p}}} \leq \omega \leq \frac{1}{t_{\mathrm{s}}} \\
	- 20 \log_{10} \omega + 20 \log_{10} \frac{k_{\mathrm{p}}}{t_{\mathrm{s}}} & \frac{1}{t_{\mathrm{s}}} < \omega 
	\end{array} 	\right. , \label{eqn:Lgain-01}
\end{align}
where $j = \sqrt{-1}$ is the imaginary unit and $\omega$ represents the {angular frequency} in [rad/s]. 
Angular frequency is often used in control engineering and is related to frequency $f$ with units of [1/s] or [Hz] by the relation $\omega = 2 \pi f$. 

\begin{figure}[t!]
	\begin{center}
		\includegraphics[width=0.74 \textwidth]{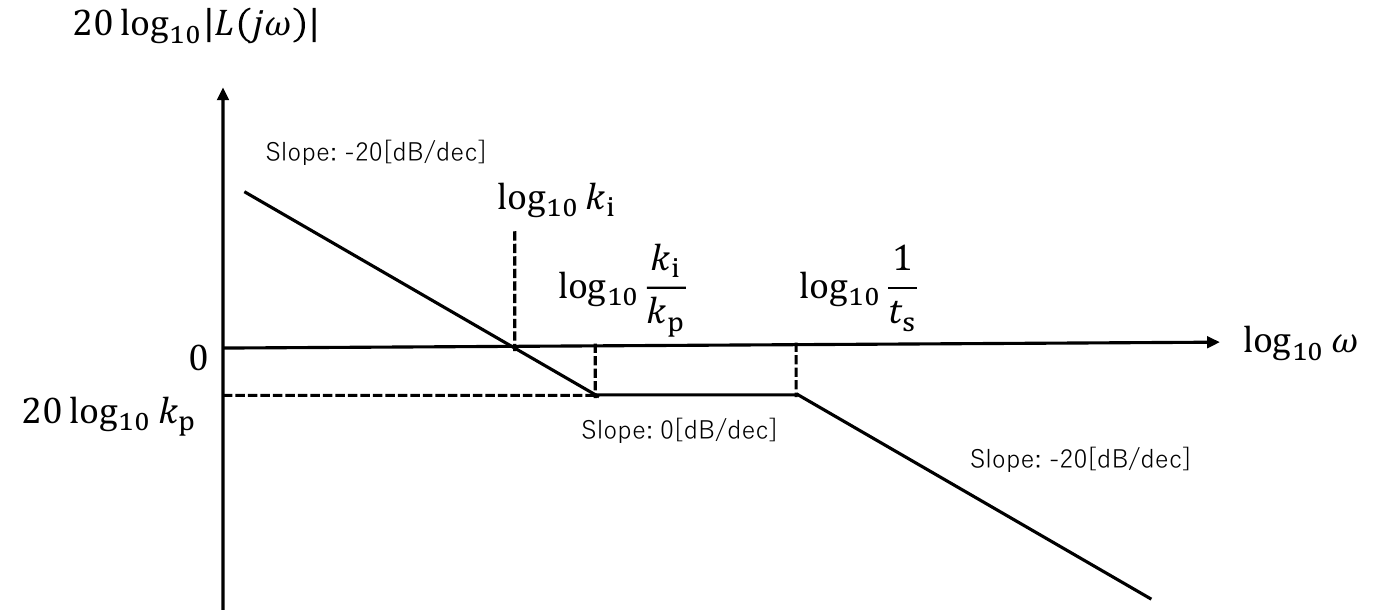}
		\caption{
			Piecewise linear approximation of the gain diagram of the open-loop transfer function $L(s)$ for loop-shaping design. This figure visualizes Eq.~\eqref{eqn:Lgain-01} on a log-log scale. The approximate value of the gain crossover frequency is shown to be \RevJikuya{coincident} with the integral gain $k_{\mathrm{i}}$ by finding the angular frequency of zero crossing. 
		}\label{fig:FigGainOP}
	\end{center}
\end{figure}

The piecewise linear approximation of the gain diagram of $L(s)$ is visualized in Fig.~\ref{fig:FigGainOP}. 
The slopes at both ends of the diagram are $-20$ [dB/dec], and the slope in the middle frequency range is $0$ [dB/dec], where dB stands for decibel and dec stands for decade, both representing a step on a logarithmic scale. 
Specifically, a slope of $-20$ [dB/dec] indicates that the gain decreases by $20$ [dB] for every tenfold increase in angular frequency.

The {gain crossover angular frequency} represents the angular frequency at which the open-loop gain is 0 [dB], 
and is a key index for evaluating the frequency range over which the control system is effective. 
From the piecewise approximation in Eq.~\eqref{eqn:Lgain-01}, the gain crossover frequency can be approximately derived as $k_{\mathrm{i}}$ [rad/s]. 
The gain crossover angular frequency must generally be set sufficiently lower than the Nyquist angular frequency $\pi/t_{\mathrm{c}}$ which is determined by Shannon's sampling theorem. Therefore, a conservative criterion is specified as
\begin{align}
	k_{\mathrm{i}} \leq \frac{\pi}{10 t_{\mathrm{c}}}. \label{eqn:FreqRange-02}
\end{align}
where the factor of 10 ensures a sufficient margin from the Nyquist frequency. 

The phase crossover angular frequency represents the angular frequency at which the phase delay is $-180$ [deg], and the gain margin can be evaluated at this angular frequency. 
If the Nyquist angular frequency is close to the gain crossover angular frequency, the phase crossover may occur in the angular frequency range where the gain slope is $0$ [dB/dec]. 
Therefore, it is crucial to ensure an appropriate gain margin in this critical angular frequency range. 
Servo systems typically require a gain margin of around 10 [dB] to 20 [dB]. This range is a common engineering rule of thumb, providing a necessary safety margin to ensure robust stability against modeling errors and external disturbances. This conservative criterion is specified as: 
\begin{align}
	-20 \leq 20 \log_{10} k_{\mathrm{p}} \leq -10 
	\quad \Leftrightarrow \quad 
	\frac{1}{10} \leq  k_{\mathrm{p}} \leq \frac{1}{\sqrt{10}}. \label{eqn:FreqRange-03}
\end{align}
\RevJikuya{The rationale for choosing $k_{\mathrm{p}} < 1$ is to provide a sufficient gain margin against high-frequency uncertainties and modeling errors that may not be fully captured in the simplified analysis in Eq.~\eqref{eqn:Lgain-01}.}

Combining the constraints from Eqs.~\eqref{eqn:FreqRange-01}, \eqref{eqn:FreqRange-02}, and \eqref{eqn:FreqRange-03} , the overall conditions for the integral and proportional gains are summarized as follows: 
\begin{align}
	k_{\mathrm{i}} \leq \min \left\{ \frac{\pi}{10 t_{\mathrm{c}}}, \frac{k_{\mathrm{p}}}{t_{\mathrm{s}}} 	\right\}, 
	\quad 
	\frac{1}{10} \leq k_{\mathrm{p}} \leq \frac{1}{\sqrt{10}}. \label{eqn:FreqRange-04}
\end{align}
After selecting the scalar gains for each diagonal element based on the above inequalities, $\bm{K}_{\mathrm{i}}$ and $\bm{K}_{\mathrm{p}}$ can be obtained by embedding them into the diagonal terms. 
When integral-only control is considered instead of PI control, it is sufficient to consider the constraint on the integral gain $k_{\mathrm{i}}$ in Eq.~\eqref{eqn:FreqRange-02} instead of Eq.~\eqref{eqn:FreqRange-04}. 

\begin{figure}[p] %
    \centering

    \begin{minipage}{1.0\textwidth}
        \centering
        \includegraphics[width=0.91\textwidth]{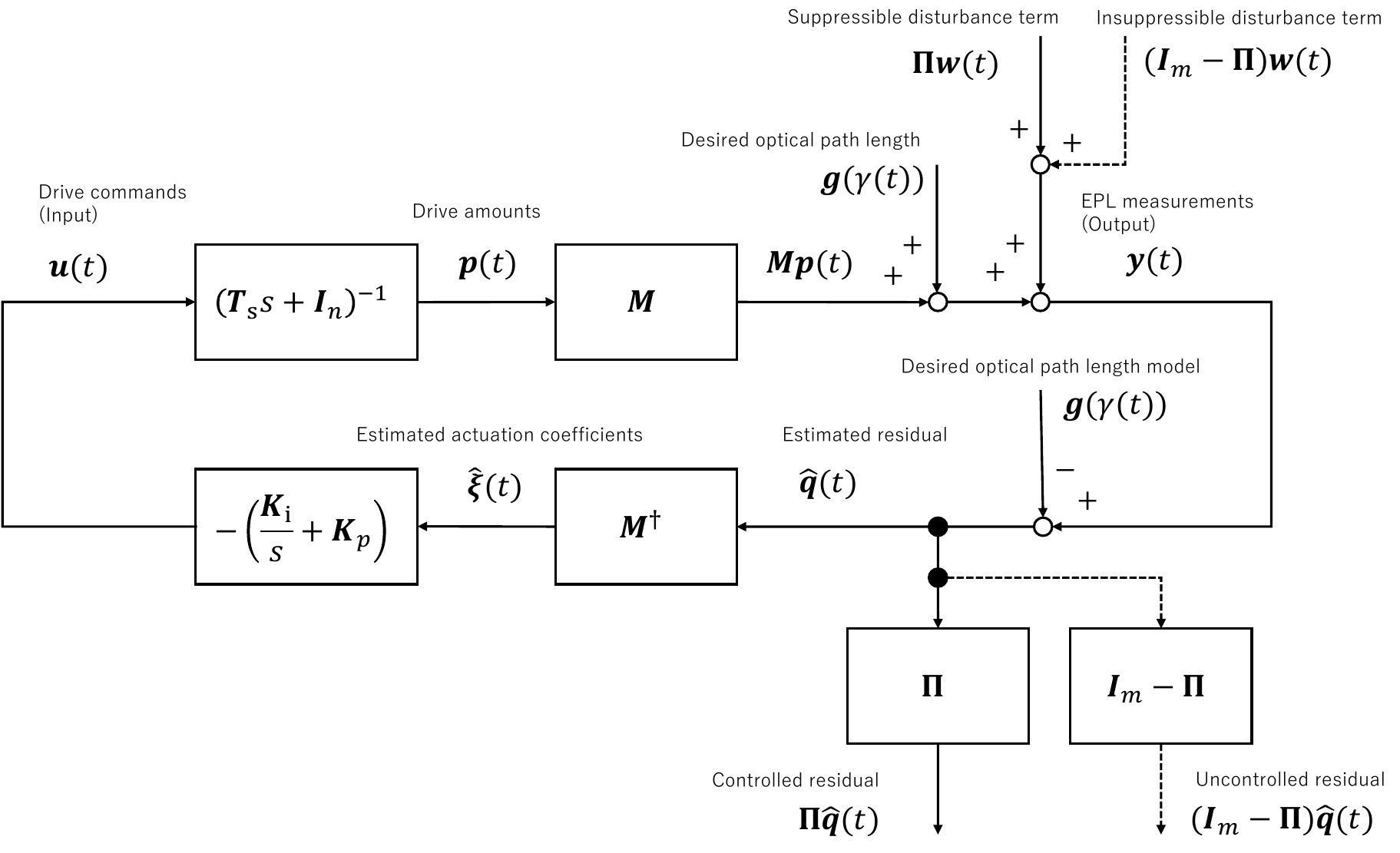}
        \caption{
            Decomposition of the disturbance $\bm{w}(t)$ and the estimated residual $\hat{\bm{q}}(t)$. The decomposition of $\bm{w}(s)$ in Eq.~\eqref{eqn:S-002} is interpreted as the partitioning into the suppressible and insuppressible terms. Correspondingly, the decomposition of $\hat{\bm{q}}(s)$ in Eq.~\eqref{eqn:S-003} is interpreted as the partitioning into the controlled and uncontrolled residuals. 
        }\label{fig:FigCLdecompose}
    \end{minipage}

    \vspace{3em}

    \begin{minipage}{1.0\textwidth}
        \centering
        \includegraphics[width=0.45\textwidth]{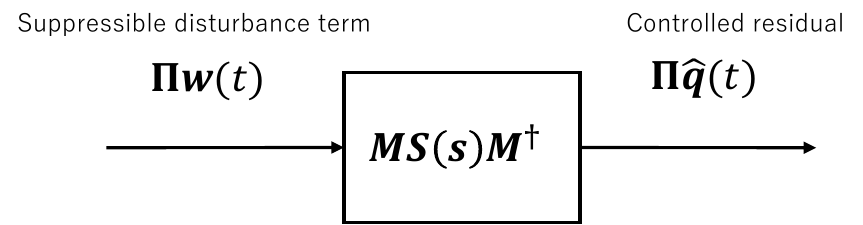}
        \caption{
            Effective disturbance suppression for the transfer from suppressible disturbance term to controlled residual. The block $\bm{M} \bm{S}(s) \bm{M}^{\dagger}$ indicates the effective disturbance suppression achieved by the control system. The sensitivity function $\bm{S}(s)$ defined in Eq.~\eqref{eqn:S-009} characterizes the frequency dependence, while $\bm{M}$ and $\bm{M}^{\dagger}$ characterize the directional dependence of the suppression performance.  
        }\label{fig:FigCLdecompose2}
    \end{minipage}

    \vspace{3em}

    \begin{minipage}{1.0\textwidth}
        \centering
        \includegraphics[width=0.92\textwidth]{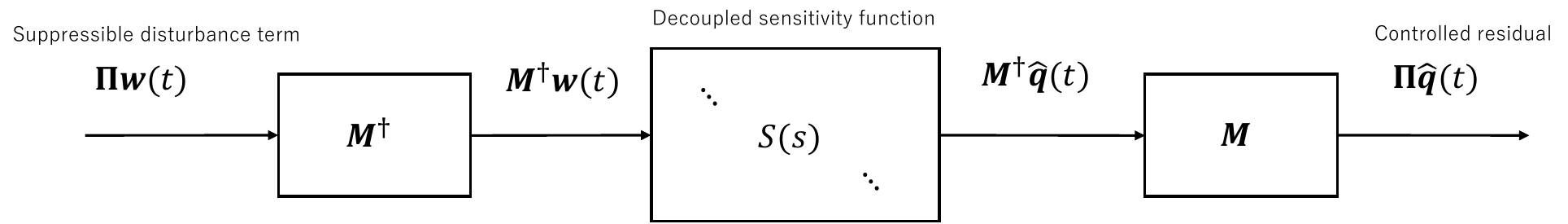}
        \caption{
            Decoupled structure of the transfer from suppressible disturbance term to controlled residual. The sensitivity function $\bm{S}(s)$ defined in Eq.~\eqref{eqn:S-009} is decoupled, resulting in a diagonal matrix. Its diagonal elements are referred to as the decoupled sensitivity functions $S(s)$, where subscripts indicating the order of the diagonal elements are omitted for simplicity. 
        }\label{fig:FigCLdecompose3}
    \end{minipage}
\end{figure}

For the analysis of the closed loop behavior, we investigate the input-output relation from the disturbance $\bm{w}(s)$ to the estimated residual $\hat{\bm{q}}(s)$ in the frequency domain. 
The fundamental relationship, derived considering the negative feedback loop, is given by:
\begin{align}
	\hat{\bm{q}}(s) = \bm{w}(s) - \bm{M} \bm{L}(s) \bm{M}^{\dagger} \hat{\bm{q}}(s)
	\quad \Leftrightarrow \quad 
	\left( \bm{I}_m + \bm{M} \bm{L}(s) \bm{M}^{\dagger} \right) \hat{\bm{q}}(s) = \bm{w}(s). \label{eqn:S-001}
\end{align}
To decompose the system into controllable and uncontrollable subspaces, we introduce an orthogonal projection matrix $\bm{\Pi} \in \mathbb{R}^{m \times m}$ onto $\Image \bm{M}$ defined by 
\begin{align}
	\bm{\Pi} := \bm{M} \bm{M}^{\dag}.
	\label{eqn:Pidef}
\end{align}
Then, its complement $\bm{I}_m - \bm{\Pi}$ is also an orthogonal projection matrix onto $\left( \Image \bm{M} \right)^{\perp}$. 
These orthogonal projection matrices form a complementary pair of orthogonal projection matrices and satisfy the following properties: 
\begin{align}
	\bm{\Pi} = \bm{\Pi}^{\T}, \quad 
    & \bm{\Pi}^2 = \bm{\Pi}, \quad 
    \bm{I}_m - \bm{\Pi} = \left( \bm{I}_m - \bm{\Pi} \right)^{\T}, \quad \nonumber \\
	& \left( \bm{I}_m - \bm{\Pi} \right)^2 = \left( \bm{I}_m - \bm{\Pi} \right), \quad 
	\bm{\Pi} \left( \bm{I}_m - \bm{\Pi} \right) = \bm{0}_{m \times m}.  
	\label{eqn:Piprop}
\end{align}
The disturbance $\bm{w}(s)$ and the estimated residuals $\hat{\bm{q}}(s)$ are projected onto $\Image \bm{M}$ and its orthogonal complement $\left( \Image \bm{M} \right)^{\perp}$ as follows:
\begin{align}
	\bm{w}(s) &= \bm{\Pi} \bm{w}(s) 
	+ \left( \bm{I}_m - \bm{\Pi} \right) \bm{w}(s), \label{eqn:S-002} \\
	\hat{\bm{q}}(s) &= \bm{\Pi} \hat{\bm{q}}(s) 
	+ \left( \bm{I}_m - \bm{\Pi} \right) \hat{\bm{q}}(s). \label{eqn:S-003}
\end{align}
Substituting these decompositions into Eq.~\eqref{eqn:S-001} and separating the components based on the orthogonality of the subspaces yields two decoupled equations for the suppressible component in $\Image \bm{M}$ and the insuppressible component in $\left( \Image \bm{M} \right)^{\perp}$:
\begin{align}
	\bm{M} \left( \bm{I}_n + \bm{L}(s) \right) \bm{M}^{\dagger} \hat{\bm{q}}(s) 
	&= \bm{\Pi} \bm{w}(s), \label{eqn:S-006} \\
	\left( \bm{I}_m - \bm{\Pi} \right) \hat{\bm{q}}(s) 
	&= \left( \bm{I}_m - \bm{\Pi} \right) \bm{w}(s). \label{eqn:S-007}
\end{align}
By multiplying $\bm{M}^{\dagger}$ from the left to Eq.~\eqref{eqn:S-006}, we isolate the transfer relation for the suppressible component:
\begin{align}
	& \left( \bm{I}_n + \bm{L}(s) \right) \bm{M}^{\dagger} \hat{\bm{q}}(s) 
	= \bm{M}^{\dagger} \bm{w}(s) 
	\quad \Leftrightarrow \quad
	\bm{M}^{\dagger} \hat{\bm{q}}(s) 
	= \bm{S}(s) \bm{M}^{\dagger} \bm{w}(s) \nonumber \\
	&\quad \Leftrightarrow \quad
	\bm{\Pi} \hat{\bm{q}}(s) 
	= \bm{M} \bm{S}(s) \bm{M}^{\dagger} \bm{w}(s) 
	\quad \Leftrightarrow \quad
	\bm{\Pi} \hat{\bm{q}}(s) 
	= \bm{M} \bm{S}(s) \bm{M}^{\dagger} \bm{\Pi} \bm{w}(s). 
	\label{eqn:S-008}
\end{align}
Here, $\bm{S}(s)$ is the {sensitivity function} defined by 
\begin{align}
	\bm{S}(s) := \left( \bm{I}_n + \bm{L}(s) \right)^{-1}. \label{eqn:S-009}
\end{align}
Equation~\eqref{eqn:S-008} indicates that the suppressible disturbance term $\bm{\Pi} \bm{w}(s)$ is mapped to the controlled residual $\bm{\Pi} \hat{\bm{q}}(s)$ via the multiplication of the transfer function  $\bm{M} \bm{S}(s) \bm{M}^{\dagger}$. 
In contrast, Eq.~\eqref{eqn:S-007} shows that the insuppressible disturbance term $\left( \bm{I}_m - \bm{\Pi} \right) \bm{w}(s)$ is directly transferred to the uncontrolled residual $\left( \bm{I}_m - \bm{\Pi} \right) \hat{\bm{q}}(s)$. 
Hence, the decomposition of $\bm{w}(s)$ in Eq.~\eqref{eqn:S-002} is interpreted as the decomposition into the suppressible and insuppressible terms, and the decomposition of $\hat{\bm{q}}(s)$ in Eq.~\eqref{eqn:S-003} is interpreted as the decomposition into the controlled and uncontrolled terms, as indicated in Fig.~\ref{fig:FigCLdecompose}. 
Feedback control operates solely on the suppressible disturbance term $\bm{\Pi} \bm{w}(s)$, mapping it to the controlled residual $\bm{\Pi} \hat{\bm{q}}(s)$ as shown in Fig.~\ref{fig:FigCLdecompose2}, but the control mechanism cannot affect the insuppressible disturbance term $\left( \bm{I}_m - \bm{\Pi} \right) \bm{w}(s)$. 

\begin{figure}[t!]
	\begin{center}
		\includegraphics[width=0.74 \textwidth]{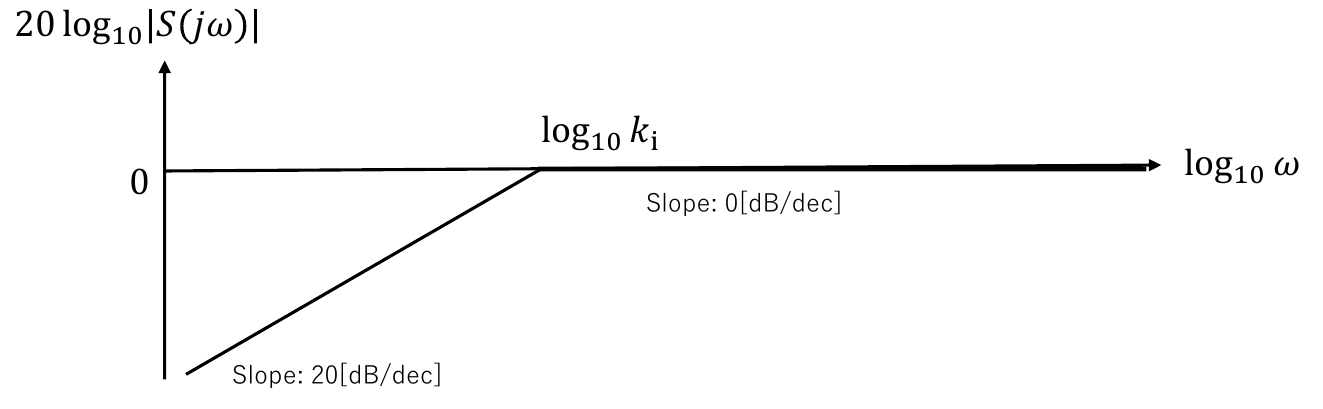}
		\caption{
			Piecewise linear approximation of the gain diagram of the sensitivity funciton $S(s)$. This figure visualizes Eq.~\eqref{eqn:S-05} on a log-log scale. The control bandwidth $\omega_{\mathrm{b}}$ is defined by the gain crossover frequency of the open-loop transfer function $L(s)$, which is approximated by the integral gain $k_{\mathrm{i}}$. Then, the disturbance suppression is shown to be effective at angular frequencies lower than $k_{\mathrm{i}}$.  
		}\label{fig:FigGainCL}
	\end{center}
\end{figure}

Let us further investigate the structure of $\bm{M} \bm{S}(s) \bm{M}^{\dagger}$ in Fig.~\ref{fig:FigCLdecompose2} by utilizing the non-interacting property of the sensitivity function $\bm{S}(s)$. The analysis is simplified by focusing on the {decoupled sensitivity function} $S(s)$, which is defined as the diagonal element of the transfer function matrix $\bm{S}(s)$ from $\bm{M}^{\dagger} \bm{w}(s)$ to $\bm{M}^{\dagger} \bm{\hat{q}}(s)$ as shown in Fig.~\ref{fig:FigCLdecompose3}. This diagonal element is given by: 
\begin{align}
	S(s) = \frac{1}{1+L(s)} 
	= \frac{1}{1 + \dfrac{k_{\mathrm{p}} s + k_{\mathrm{i}}}{s(t_{\mathrm{s}} s + 1)}}
	= \frac{s(t_{\mathrm{s}} s + 1)}{t_s s^2 + (k_{\mathrm{p}}+1) s + k_{\mathrm{i}}}, \label{eqn:S-03}
\end{align}
where the subscript indicating the order of the diagonal elements are omitted, analogous to $L(s)$ for simplicity. 
By approximating $1 + L(j \omega)$ to be $k_{\mathrm{i}} / j \omega$ for $| L(j \omega)|$ to be sufficiently large and by approximating 
it to be $1$ for $| L(j \omega)|$ to be sufficiently small, the rough approximation for $S(j \omega)$ is given by 
\begin{align}
	S(j \omega) \simeq \left\{ \begin{array}{ll} \dfrac{j \omega}{k_{\mathrm{i}}} & \omega \ll k_{\mathrm{i}} \\
	\dfrac{1}{2} & \omega = k_{\mathrm{i}} \\
	1 & k_{\mathrm{i}} \ll \omega \end{array} \right. . \label{eqn:S-04}
\end{align}
Then, the piecewise linear approximation of the gain diagram of $S(s)$ is given by 
\begin{align}
	20 \log_{10} | S(j \omega) |
	&\simeq \left\{ \begin{array}{ll} 20 \log_{10} \omega - 20 \log_{10} k_{\mathrm{i}} &  \omega \ll k_{\mathrm{i}} \\
	- 20 \log_{10} 2 & \omega = k_{\mathrm{i}} \nonumber \\
	0  & k_{\mathrm{i}} \ll \omega \end{array} \right.  \\
	&\simeq \left\{ \begin{array}{ll} 20 \log_{10} \omega - 20 \log_{10} k_{\mathrm{i}} &  \omega < k_{\mathrm{i}} \\
	0  & k_{\mathrm{i}} < \omega \end{array} \right. ,  \label{eqn:S-05}
\end{align}
as shown in Fig.~\ref{fig:FigGainCL}. 
\RevJikuya{The first line of Eq.~\eqref{eqn:S-05} highlights the gain reduction at the corner frequency $\omega = k_{\mathrm{i}}$, whereas the second line provides a further simplified piecewise linear approximation using only two straight lines to capture the overall asymptotic trends.} 
The {control bandwidth} $\omega_{\mathrm{b}}$ is defined by the gain crossover frequency of the open-loop transfer function $L(s)$ where its gain is $0$ [dB], and is approximated by the integral gain $k_{\mathrm{i}}$, i.e., $\omega_{\mathrm{b}} = k_{\mathrm{i}}$, as shown in Fig.~\ref{fig:FigGainOP}. 
This bandwidth $\omega_{\mathrm{b}}$ characterizes the frequency range where disturbance attenuation is effective for $\omega < \omega_{\mathrm{b}} = k_{\mathrm{i}}$.
The admissible range of $\omega_{\mathrm{b}}$ restricted by the control period $t_{\mathrm{c}}$ is obtained as follows: 
\begin{align}
	\omega_{\mathrm{b}} &= k_{\mathrm{i}} \leq \frac{\pi}{10 t_{\mathrm{c}}}, \label{eqn:S-06}
\end{align}
based on Eq.~\eqref{eqn:FreqRange-02}. 
This inequality shows the fundamental limitation on control performance, which can be relaxed as a shorter control period $t_{\mathrm{c}}$ allows a larger integral gain $k_{\mathrm{i}}$. 
This concludes our discussion on the control system design and stability analysis of the AWPI controller.

%


\section{Performance Evaluation and Practical Application}\label{sect:performance_evaluation}


\subsection{Performance Comparison with Proportional Control}

Proportional (P) control, which relies on negative feedback without an integral term, is often the initial choice for constructing a feedback control system. Even when Proportional-Integral (PI) control is the primary candidate, P control may be selected to perform initial stability tests. Therefore, we provide a self-contained introduction to P control and perform an asymptotic analysis of its behavior. Comparing the asymptotic analyses of P and PI control will help explain the necessity of incorporating an integrator, which is particularly useful for astronomers who are unfamiliar with servo systems in control engineering.

\begin{figure}[th!]
	\begin{center}
		\includegraphics[width=0.89 \textwidth]{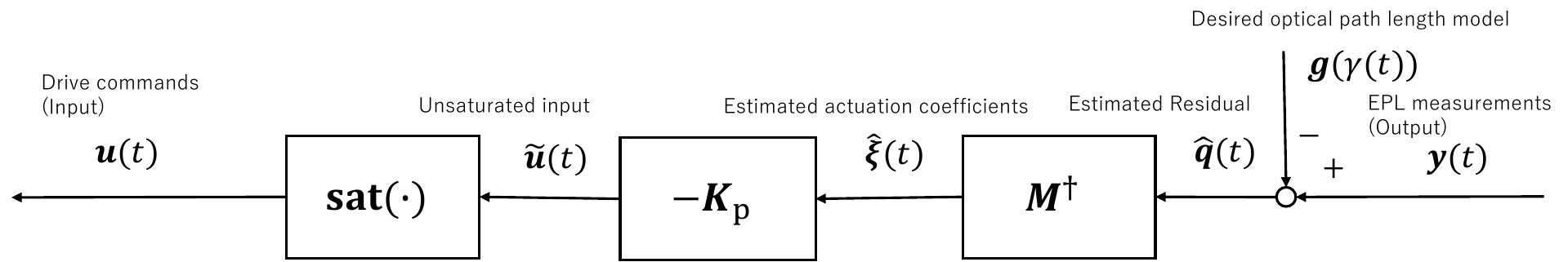}
		\caption{
			Block diagram of the P controller. This controller corresponds to Eqs.~\eqref{eqn:P-01}-\eqref{eqn:P-05}. It is equivalent to the AWPI controller shown in Fig.~\ref{fig:FigAWPI} by setting the integral gain $\bm{K}_{\mathrm{i}}$ and the anti-windup gain $\bm{K}_{\mathrm{a}}$ to zero matrices. Since the integrator becomes unnecessary due to this configuration, the resulting controller is static, and the anti-windup mechanism represented by the local feedback with the anti-windup gain $\bm{K}_{\mathrm{a}}$ is also eliminated. 
		}\label{fig:FigP}
	\end{center}
\end{figure}

The P controller compatible with the AWPI controller in Eqs.~\eqref{eqn:AWPI-00} to \eqref{eqn:AWPI-03} is defined by:
\begin{align}
	\hat{\bm{q}}(t) &= \bm{y}(t) - \bm{g}(\gamma(t)), \label{eqn:P-01} \\
	\hat{\bm{\xi}}(t) &= \bm{M}^{\dagger} \hat{\bm{q}}(t), \label{eqn:P-02} \\
	\tilde{\bm{u}}(t) &=  - \bm{K}_{\mathrm{p}} \, \hat{\bm{\xi}}(t), \label{eqn:P-04} \\
	\bm{u}(t) &= \mathrm{\mathbf{sat}} \left( \tilde{\bm{u}}(t) \right), \label{eqn:P-05}
\end{align}
where $\bm{K}_{\mathrm{p}} \in \mathbb{R}^{n \times n}$ is the diagonal matrix with positive elements representing the proportional gain and $\mathrm{\mathbf{sat}}( \cdot )$ is the saturation function defined in Eq.~\eqref{eqn:satdef}. 
This P controller is the necessary simplification of the AWPI controller when the integral gain is set to zero, $\bm{K}_{\mathrm{i}} = \bm{0}$, and the integral state is removed. The block diagram of the P controller is illustrated in Fig.~\ref{fig:FigP}. 

\begin{figure}[th!]
	\begin{center}
		\includegraphics[width=0.84 \textwidth]{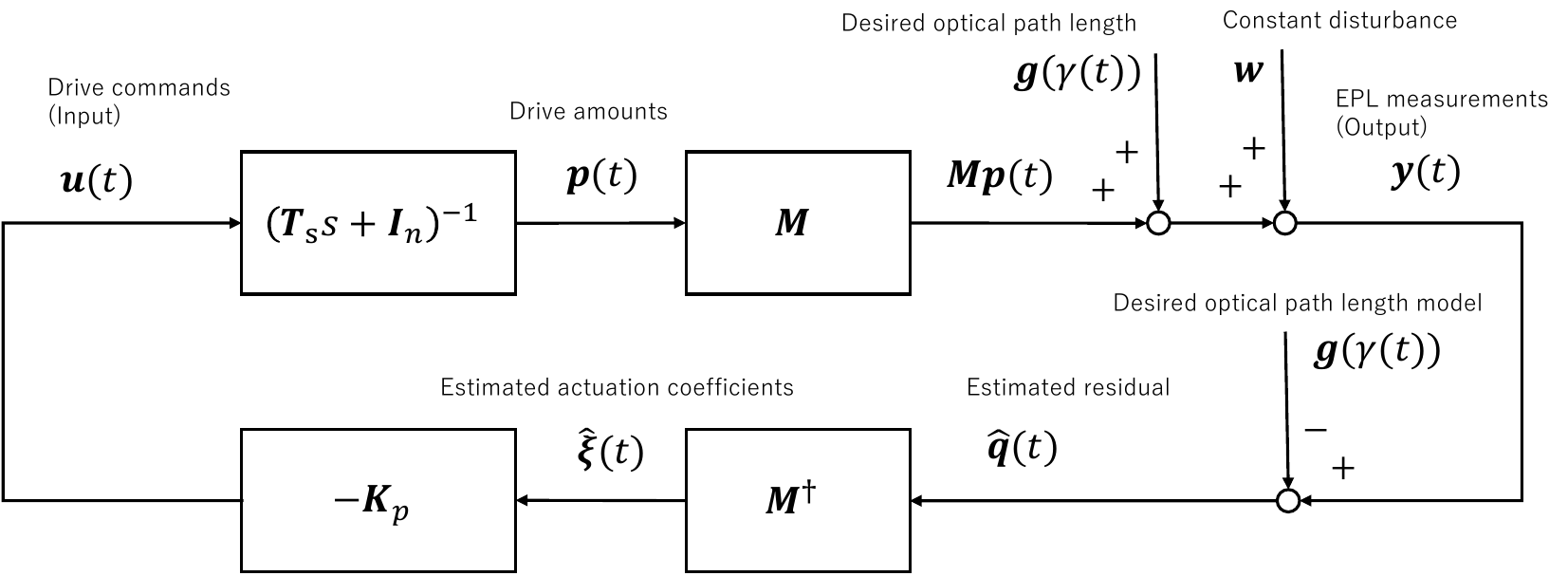}
		\caption{
			Block diagram of the closed-loop system consisting of the plant and the pure P controller. Similar to Fig.~\ref{fig:FigCL}, the saturation function $\mathrm{\mathbf{sat}}(\cdot)$ is omitted for analysis of the behavior near the operating point.  
		}\label{fig:FigCLP}
	\end{center}
\end{figure}

Analogous to the AWPI controller analysis, we examine the nominal behavior by considering the situation when saturation is not reached. The closed-loop system of the plant and the pure P controller is given by:
\begin{align}
	\dot{\bm{p}}(t) = - \bm{T}_{\mathrm{s}}^{-1} \left( \bm{I}_n + \bm{K}_{\mathrm{p}} \right) \bm{p}(t) 
	- \bm{T}_{\mathrm{s}}^{-1} \RevSakibara{\bm{K}_{\mathrm{p}}} \bm{M}^{\dagger} \bm{w} \label{eqn:P-06}
\end{align}
which is compatible with Eq.~\eqref{eqn:CL-01} for the nominal PI controller. 
Figure~\ref{fig:FigCLP} shows the block diagram of the closed-loop system. 
This closed-loop system is stable because $\bm{K}_{\mathrm{p}}$ is diagonal with positive diagonal elements by assumption. 
The asymptotic convergence is evaluated as follows:
\begin{align}
	\bm{p}(\infty) &= 
	- \left( \bm{I}_n + \bm{K}_{\mathrm{p}} \right)^{-1} \RevSakibara{\bm{K}_{\mathrm{p}}} \bm{M}^{\dagger} \bm{w} \label{eqn:P-07} \\ 
	\hat{\bm{q}}(\infty) &= \bm{M} \bm{p}(\infty) + \bm{w} = \left( \bm{I}_m - \bm{M} \left( \bm{I}_n + \bm{K}_{\mathrm{p}} \right)^{-1} \RevSakibara{\bm{K}_{\mathrm{p}}} \bm{M}^{\dagger} \right) \bm{w}, \label{eqn:P-08} \\
	\hat{\bm{\xi}}(\infty) &= \bm{M}^{\dagger} \RevSakibara{\hat{\bm{q}}}(\infty) 
	=  \left( \bm{I}_n + \bm{K}_{\mathrm{p}} \right)^{-1} \bm{M}^{\dagger} \bm{w} 
	\neq \bm{0}_n, \label{eqn:P-09} \\
	\bm{u}(\infty) &= - \bm{K}_{\mathrm{p}} \, \bm{M}^{\dagger} \hat{\bm{q}}(\infty) 
	= - \bm{K}_{\mathrm{p}}  \left( \bm{I}_n + \bm{K}_{\mathrm{p}} \right)^{-1} \bm{M}^{\dagger} \bm{w}. \label{eqn:P-10}
\end{align}
Equations~\eqref{eqn:P-07} and \eqref{eqn:P-09} confirm that the asymptotic convergence properties required by Eq.~\eqref{eqn:ACP-01} are not satisfied. 
Equation~\eqref{eqn:P-08} is not consistent with the residual decomposition form introduced in Eq.~\eqref{eqn:q-dec-01}. 
In other words, Eq.~\eqref{eqn:q-dec-01} states that the components belonging to $\left( \Image \bm{M} \right)^{\perp}$ cannot be suppressed by control action, and Eq.~\eqref{eqn:P-08} explicitly shows that $\hat{\bm{q}}(\infty) \not\in \left( \Image \bm{M} \right)^{\perp}$, confirming the inconsistency. 
Hence, the integrator inside the controller is shown to be necessary for the asymptotic disturbance attenuation. 


\subsection{Manual Focus Adjustment in Astronomical Observation}

Let us consider the operational requirements that may arise for the control system during actual field operation. 
When the AWPI control law in Eqs.~\eqref{eqn:AWPI-00} to \eqref{eqn:AWPI-03} is applied, the optical drive system is automatically adjusted to absorb the source of suppressible disturbance $\bm{M}^{\dagger} \bm{w}$, as confirmed by Eq.~\eqref{eqn:CL-04}. 
Because the disturbance $\bm{w}$ may include unwanted components, such as modeling errors in $\bm{g}(\gamma(t))$, fine-tuning of the optical drive system may be required during astronomical observations. 

We therefore consider the issue of manual focus adjustment during feedback control. 
One solution is to adjust the drive amount of the optical system by adding an external offset, $\bm{u}_{\mathrm{o}}$. 
However, simply adding an input offset $\bm{u}_{\mathrm{o}}$ to the input, as shown in Eq.~\eqref{eqn:MFC-03} below, would result in the offset being integrated out by the PI controller, effectively canceling the intended manual adjustment. 
Therefore, by subtracting an output offset $\bm{M} \bm{u}_{\mathrm{o}}$ from the output, as shown in Eq.~\eqref{eqn:MFC-00} below, the presence of the offset has been successfully hidden from the PI controller. 

The AWPI controller with manual focus adjustment is now summarized below: 
\begin{align}
	\hat{\bm{q}}(t) &= \bm{y}(t) - \bm{g}(\gamma(t)) - \bm{M} \bm{u}_{\mathrm{o}}, \label{eqn:MFC-00} \\
	\hat{\bm{\xi}}(t) &= \bm{M}^{\dagger} \hat{\bm{q}}(t), \label{eqn:MFC-0} \\
	\dot{\bm{v}}(t) &= \hat{\bm{\xi}}(t) + \bm{K}_{\mathrm{a}} \left( \tilde{\bm{u}}(t) - \bm{u}(t)   \right), \label{eqn:MFC-01} \\
	\tilde{\bm{u}}(t) &= - \bm{K}_{\mathrm{i}} \bm{v}(t) - \bm{K}_{\mathrm{p}} \hat{\bm{\xi}}(t), \label{eqn:MFC-02} \\
	\bm{u}(t) &= \mathrm{\mathbf{sat}} \left( \tilde{\bm{u}}(t) \right) + \bm{u}_{\mathrm{o}}, \label{eqn:MFC-03}
\end{align}
where the focus position can be adjusted by specifying the amount of change $\bm{u}_{\mathrm{o}}$ in the drive amount. 

\begin{figure}[th!]
	\begin{center}
		\includegraphics[width=0.8 \textwidth]{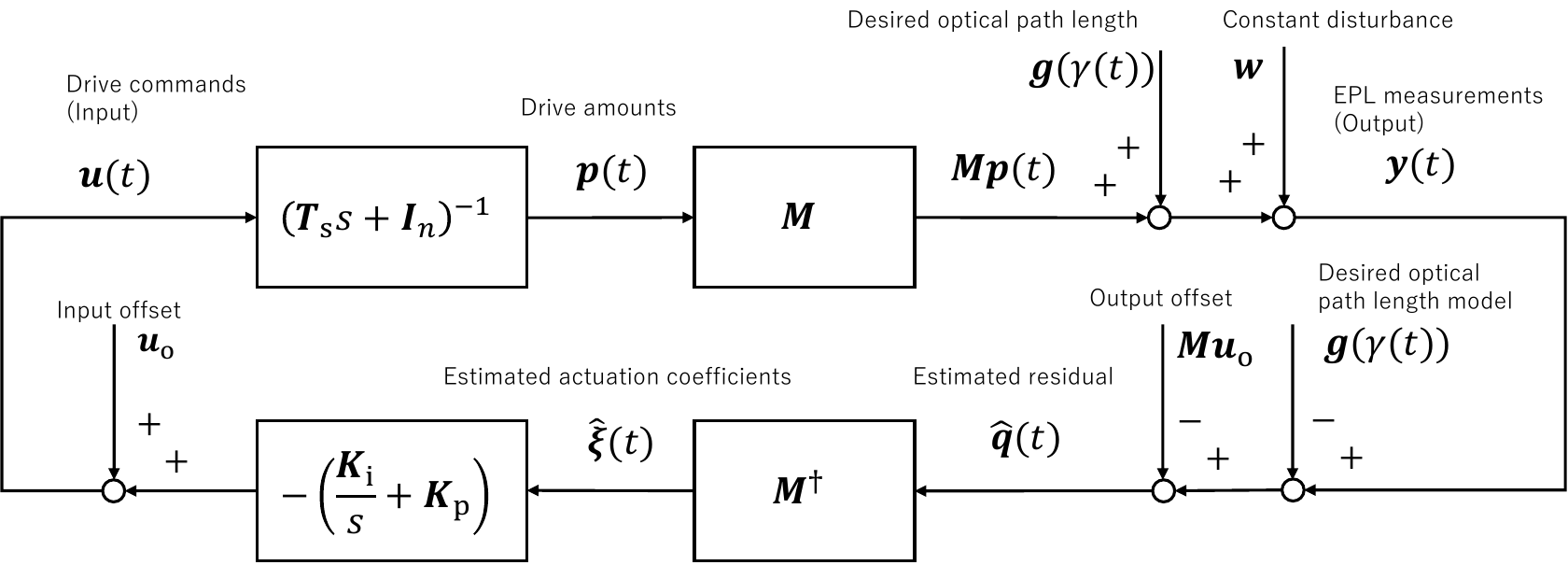}
		\caption{
			Manual focus adjustment by adding input offset $\bm{u}_{\mathrm{o}}$ and subtracting output offset $\bm{M} \bm{u}_{\mathrm{o}}$, where the anti-windup mechanism is excluded from the AWPI controller in asymptotic analysis. This manual focus adjustment is assumed to be required when the operator desires to perform a manual correction based on astronomical observation results during the observation process. The objective of manual focus is to introduce an offset $\bm{u}_{\mathrm{o}}$ to the drive amount $\bm{p}(t)$. Simply adding the offset $\bm{u}_{\mathrm{o}}$ to the input would cause the offset to be canceled by the integrator inside the PI controller. Therefore, by subtracting the output offset $\bm{M} \bm{u}_{\mathrm{o}}$, the existence of the input offset is hidden from the integrator. 
		}\label{fig:FigOffset}
	\end{center}
\end{figure}

To perform asymptotic analysis, we consider the PI controller without the anti-windup mechanism. By adding the input offset $\bm{u}_{\mathrm{o}}$ and subtracting the output offset $\bm{M} \bm{u}_{\mathrm{o}}$, as shown in Fig.~\ref{fig:FigOffset}, and performing the asymptotic analysis in the same way as in Eq.~\eqref{eqn:CL-02}, the asymptotic convergence is evaluated as follows:
\begin{align}
	\bm{v}(\infty) &= \bm{K}_{\mathrm{i}}^{-1} \bm{M}^{\dagger} \bm{w}, \label{eqn:MFC-04} \\
	\bm{p}(\infty) &= \bm{u}_{\mathrm{o}} - \bm{M}^{\dagger} \bm{w}, \label{eqn:MFC-05} \\
	\hat{\bm{q}}(\infty) &= \left( \bm{I}_m - \bm{M} \bm{M}^{\dagger} \right) \bm{w}, \label{eqn:MFC-06} \\
	\hat{\bm{\xi}}(\infty) &=  \bm{0}_n, \label{eqn:MFC-07} \\
	\bm{u}(\infty) &= \bm{u}_{\mathrm{o}} - \bm{M}^{\dagger} \bm{w}, \label{eqn:MFC-08}
\end{align}
where only $\bm{p}(\infty)$ and $\bm{u}(\infty)$ are affected by the input offset $\bm{u}_{\mathrm{o}}$ while the others remain unaffected. 
Thus, we can see that subtracting the output offset hides the presence of the offset from the integrator in the controller, as intended. 

In Fig.~\ref{fig:FigOffset}, the signal line between $-\bm{g}\left(\gamma(t)\right)$ and $-\bm{M} \bm{u}_{\mathrm{o}}$ represents the displacements at the radiator positions; specifically, $\bm{M} \bm{u}_{\mathrm{o}}$ corresponds to the resulting changes in the radiator positions. In practice, $\bm{u}_{\mathrm{o}}$ can be fine-tuned during astronomical observations by back-calculating from real-time results to achieve a desired effect, such as improving the focus. 

%


\subsection{Performance Verification by Artificial Disturbance}
\label{sec:ArtificialDisturbance}

Let us consider the requirements that may arise for the control system during its initial operation. 
When implementing a control law and verifying its operation, we will probably first want to confirm that the control system operates stably. 
We may further want to intentionally generate a disturbance and evaluate its performance and verify whether the AWPI controller can really suppress the disturbance. 
Obviously, it would be difficult to intentionally deform the telescope or induce actual wind disturbances in a real environment.
Instead, we can consider simulating disturbances that may occur in reality by artificially adding a disturbance inside the controller.
This is very simple; just subtract $\bm{g}(\gamma(t))$ from the output $\bm{y}(t)$ and add an artificial disturbance $\bm{w}_{\mathrm{o}}$.

The AWPI controller with artificial disturbance is now summarized below: 
\begin{align}
	\hat{\bm{q}}(t) &= \bm{y}(t) - \bm{g}(\gamma(t)) + \bm{w}_{\mathrm{o}}, \label{eqn:EFC-00} \\
	\hat{\bm{\xi}}(t) &= \bm{M}^{\dagger} \hat{\bm{q}}(t), \label{eqn:EFC-0} \\
	\dot{\bm{v}}(t) &= \hat{\bm{\xi}}(t) + \bm{K}_{\mathrm{a}} \left( \tilde{\bm{u}}(t) - \bm{u}(t)   \right), \label{eqn:EFC-01} \\
	\tilde{\bm{u}}(t) &= - \bm{K}_{\mathrm{i}} \, \bm{v}(t) - \bm{K}_{\mathrm{p}} \, \hat{\bm{\xi}}(t), \label{eqn:EFC-02} \\
	\bm{u}(t) &= \bm{\mathrm{sat}} \left( \tilde{\bm{u}}(t) \right), \label{eqn:EFC-03}
\end{align}
where the artificial disturbance can be emulated by specifying the amount of change in measurement. 

\begin{figure}[th!]
	\begin{center}
		\includegraphics[width=0.84 \textwidth]{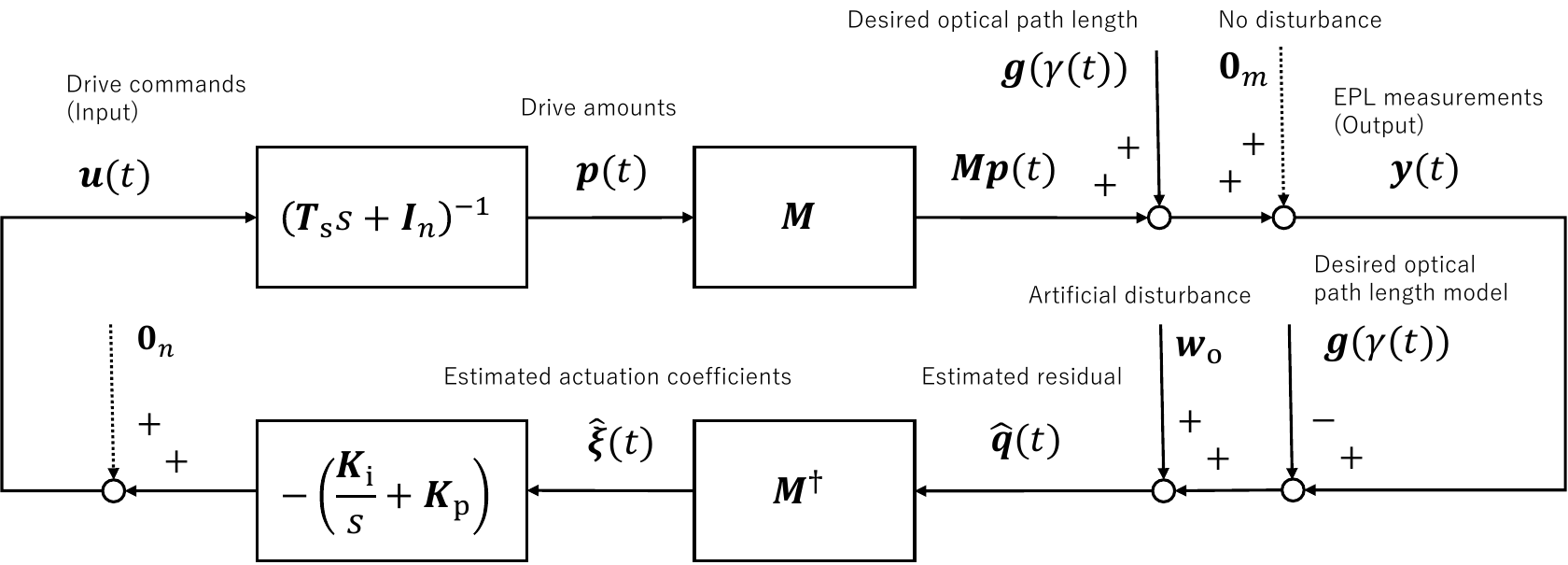}
		\caption{
			Transient evaluation test by adding artificial disturbance \RevSakibara{$\bm{w}_{\mathrm{o}}$}, where the anti-windup mechanism is excluded from the AWPI controller in asymptotic analysis. This test is motivated by the desire to evaluate the transient response of the control law under controlled conditions, simulating the effects of an unmodeled external disturbance by adding the artificial disturbance $\bm{w}_{\mathrm{o}}$. While this procedure resembles the input offset addition shown in Fig.~\ref{fig:FigOffset}, the objective here is not to introduce an offset to the drive amount $\bm{p}(t)$, but rather to deliberately introduce an offset into the measured value $\bm{y}(t)$ for the purpose of control performance evaluation.  
		}\label{fig:FigArtificialError}
	\end{center}
\end{figure}

Comparing the block diagrams in Figs.~\ref{fig:FigCL} and \ref{fig:FigArtificialError}, the only difference is whether the output $\bm{y}(t)$ contains the disturbance $\bm{w}$ or not, but the other signals remain completely identical. 
The asymptotic analysis results are therefore consistent with Eqs.~\eqref{eqn:CL-03} to \eqref{eqn:CL-07}.

To better understand the effect of the AWPI controller, let us consider two complementary scenarios related to the artificially created disturbance $\bm{w}_{\mathrm{o}}$.
The first scenario is to consider the artificially created suppressible disturbance term $\bm{w}_{\mathrm{o}} = \bm{M} \bm{\xi}_{\mathrm{o}} \in \Image \bm{M}$. 
The asymptotic analysis for this scenario shows that the steady-state values are:
\begin{align}
	\bm{v}(\infty) &= \bm{K}_{\mathrm{i}}^{-1} \bm{M}^{\dagger} \bm{w}_{\mathrm{o}} = \bm{K}_{\mathrm{i}}^{-1} \bm{\xi}_{\mathrm{o}}, \label{eqn:EFC-04} \\
	\bm{p}(\infty) &= - \bm{M}^{\dagger} \bm{w}_{\mathrm{o}} = - \bm{\xi}_{\mathrm{o}},  \label{eqn:EFC-05} \\
	\hat{\bm{q}}(\infty) &= \left( \bm{I}_m - \bm{M} \bm{M}^{\dagger} \right) \bm{w}_{\mathrm{o}} = \bm{0}_m, \label{eqn:EFC-06} \\
	\hat{\bm{\xi}}(\infty) &=  \bm{0}_n, \label{eqn:EFC-07} \\
	\bm{u}(\infty) &= - \bm{M}^{\dagger} \bm{w}_{\mathrm{o}} = - \bm{\xi}_{\mathrm{o}}. \label{eqn:EFC-08}
\end{align}
Thus, it is clear that the final drive amount $\bm{p}(\infty)$ is directly affected by the artificially created suppressible disturbance term $\bm{w}_{\mathrm{o}} \in \Image \bm{M}$, resulting in $\bm{p}(\infty) = -\bm{\xi}_{\mathrm{o}}$. 
If a suppressible disturbance term corresponding to Fig.~\ref{fig:FigCL} occurs in a real environment, it will be canceled by the drive amount and focus will be achieved. However, if an artificially created suppressible disturbance term corresponding to Fig.~\ref{fig:FigArtificialError} is added, the drive amount will attempt to cancel the artificial disturbance, leading to a temporary loss of focus.  
Therefore, while this operation is meaningless from the perspective of continuous astronomical observation, it constitutes a meaningful control test, as its effects can be confirmed and quantified in the context of the astronomical observation. 
Furthermore, since a transient response occurs when an artificial disturbance is added, the control performance can be fully verified by comparing this measured response with the predicted response from the simulation model.

The second scenario is to consider the artificially created insuppressible disturbance term $\bm{w}_{\mathrm{o}} = \bm{\eta}_{\mathrm{o}} \in \left( \Image \bm{M} \right)^{\perp}$. 
The asymptotic analysis for this scenario shows that the steady-state values are: 
\begin{align}
	\bm{v}(\infty) &= \bm{K}_{\mathrm{i}}^{-1} \bm{M}^{\dagger} \bm{w}_{\mathrm{o}} = \bm{0}_n, \label{eqn:EFC-09} \\
	\bm{p}(\infty) &= - \bm{M}^{\dagger} \bm{w}_{\mathrm{o}} = \bm{0}_n,  \label{eqn:EFC-10} \\
	\hat{\bm{q}}(\infty) &= \left( \bm{I}_m - \bm{M} \bm{M}^{\dagger} \right) \bm{w}_{\mathrm{o}} = \RevSakibara{\bm{w}_{\mathrm{o}}}, \label{eqn:EFC-11} \\
	\hat{\bm{\xi}}(\infty) &=  \bm{0}_n, \label{eqn:EFC-12} \\
	\bm{u}(\infty) &= - \bm{M}^{\dagger} \bm{w}_{\mathrm{o}} = \bm{0}_n. \label{eqn:EFC-13}
\end{align}
Thus, we can see that the drive amount $\bm{p}(\infty)$ is not affected by the artificially created insuppressible disturbance term $\bm{w}_{\mathrm{o}} \in \left( \Image \bm{M} \right)^{\perp}$, resulting in $\bm{p}(\infty) = \bm{0}_n$. 
If an insuppressible disturbance term corresponding to Fig.~\ref{fig:FigCL} occurs in a real environment, the drive amount cannot cancel it, resulting in a loss of focus. 
However, if an artificially created insuppressible disturbance term corresponding to Fig.~\ref{fig:FigArtificialError} is added, the drive amount will not attempt to cancel the insuppressible artificial disturbance term, and focus will be kept. 
Therefore, this operation is also meaningless from the perspective of continuous astronomical observation, yet it remains a meaningful control experiment to confirm the characteristic of ignoring insuppressible disturbance term.  
Furthermore, it is a remarkable observation that no transient response occurs even when this insuppressible artificial disturbance is introduced!

At first glance, the manual focus adjustment concept in Fig.~\ref{fig:FigOffset} and the transient response test in Fig.~\ref{fig:FigArtificialError} appears similar and might be confused. 
We will briefly explain the relationship between these two operational modes. 
In the manual adjustment framework in Fig.~\ref{fig:FigOffset}, if we set the input offset $\bm{u}_{\mathrm{o}} = - \bm{\xi}_{\mathrm{o}}$ and assume the disturbance to be $\bm{w} = \bm{0}$, the asymptotic variables in Eqs.~\eqref{eqn:MFC-04} to \eqref{eqn:MFC-08} are simplified as follows:
\begin{align}
	\bm{v}(\infty) = \hat{\bm{\xi}}(\infty) = \bm{0}_n, \quad \hat{\bm{q}}(\infty) = \bm{0}_m, \quad \bm{p}(\infty) = \bm{u}(\infty) = - \bm{\xi}_{\mathrm{o}} 
\end{align}
Comparing this result with the equations for the transient response test in Eqs.~\eqref{eqn:EFC-04} to \eqref{eqn:EFC-08}, we confirm that while the asymptotic integrator variable $\bm{v}(\infty)$ is different, all other asymptotic variables are identical. 
This demonstrates that the steady-state control outcome achieved by the manual focus adjustment in Fig.~\ref{fig:FigOffset} can replicate the effect of controlling suppressible errors in the transient response test. 
Conversely, the manual focus adjustment in Fig.~\ref{fig:FigOffset} cannot be used to intentionally introduce the insuppressible disturbance into the feedback control system. 
Therefore, manual focus adjustment can be considered a special case of the transient response test framework. The primary advantage of manual focus adjustment is its intuitive nature, as the focus amount can be specified directly. 


\subsection{Cosine Similarity Index} 

When evaluating suppressible and insuppressible modes, it is useful to express the wavefront distortion with an orthogonal system. 
Zernike modes are often employed in describing the wavefront distortions for a circular aperture of a telescope, and the suppression of specific Zernike modes is crucial for high-quality astronomical observations. 
Zernike modes are characterized by Zernike polynomials $Z_n^m(\rho,\phi)$ defined by 
\begin{align}
	Z_n^m(\rho,\phi) 
	= \left\{ \begin{array}{ll} R_n^m(\rho) \cos \left( m \phi \right) & m \geq 0 \\
	R_n^{|m|}(\rho) \sin \left( |m| \phi \right) & m < 0 
	\end{array} 	\right. , 
\end{align}
where the nonnegative integer $n$ is radial order,  the integer $m$ is the azimuthal frequency satisfying $n \geq | m |$, $\rho$ is the radius, $\phi$ is the argument, and the radial polynomials $R_n^m(\rho)$ are defined by 
\begin{align}
	R_n^m(\rho) = \left\{ \begin{array}{ll}
	\sum_{k=0}^{\frac{n-m}{2}} 
	\frac{(-1)^k (n-k)!}{k! \left( \frac{(n+m)}{2} - \RevSakibara{k} \right) ! \left( \frac{(n-m)}{2} - \RevSakibara{k} \right) !} \rho^{n-2k} 
	& \mbox{$n-m$: even} \\
	0 & \mbox{$n-m$: odd}
	\end{array}	\right. .
\end{align}
We note that the symbols $n$ and $m$ used in the control problem overlap with the Zernike polynomial indices $n$ and $m$, but they must be distinguished contextually.
Zernike polynomials $Z_n^m(\rho,\phi)$ form an infinite-dimensional orthogonal system of functions defined on the unit circle. 
The output \RevSakibara{$\bm{y}(t)$} is measured by placing radiators at a finite number of points on the M1 reflector, but by mapping the M1 reflector to a unit circle and determining the radius $\rho$ and argument $\phi$ of each point, we can define the Zernike mode vector $\bm{z} \in \mathbb{R}^m$ by substituting the pairs of $\rho$ and $\phi$ to $Z_n^m(\rho,\phi)$ of interest. 
We can also define the Zernike mode matrix $\bm{Z} \in \mathbb{R}^{m \times p}$ by arranging $p$ Zernike mode vectors in the column direction. 

Although the Zernike mode vector $\bm{z}$ is merely a finite-dimensional approximation of the infinite-dimensional Zernike polynomials $Z_n^m(\rho,\phi)$, the fundamental assumption is that if the truncated Zernike mode vector $\bm{z}$ can be successfully suppressed by the controller, the corresponding infinite-dimensional Zernike modes will also be suppressed to a large extent across the continuous surface.

The Zernike mode vector $\bm{z}$ can be completely suppressed by the controller if and only if it lies within the image space of the measurement matrix, $\bm{z} \in \Image \bm{M}$. 
If this condition does not hold, complete suppression is not possible. However, the degree of suppression can be estimated by orthogonally decomposing $\bm{z}$ into a suppressible component $\bm{\Pi} \bm{z} \in \Image \bm{M}$ and an insuppressible component $\left( \bm{I}_m - \bm{\Pi} \right) \bm{z} \in \left( \Image \bm{M} \right)^{\perp}$ as follows:
\begin{align}
	\bm{z} = \bm{\Pi} \bm{z} + \left( \bm{I}_m - \bm{\Pi} \right) \bm{z}
\end{align}
\RevJikuya{analogous to} Eqs.~\eqref{eqn:S-002} and \eqref{eqn:S-003}. 
Here, $\bm{\Pi} \in \mathbb{R}^{m \times m}$ is the orthogonal projection matrix onto $\Image \bm{M}$ defined by Eq.~\eqref{eqn:Pidef} and $\bm{I}_m - \bm{\Pi}$ is also the orthogonal projection onto $\left( \Image \bm{M} \right)^{\perp}$ satisfying the standard properties in Eq.~\eqref{eqn:Piprop}. 

Now, we propose the concept of a {cosine similarity index} $\cos_{\mathrm{I}}  \left( \bm{z}, \bm{M} \right) $, which serves as a quantitative criterion for evaluating the degree of suppressibility of the mode vector $\bm{z}$, defined as:
\begin{align}
	\cos_{\mathrm{I}} \left( \bm{z}, \bm{M} \right) := \frac{\| \bm{\Pi} \bm{z} \|}{\| \bm{z} \|}, 
\end{align}
where $\| \cdot \|$ is the Euclidean norm. 
As is clear from the definition and the properties of orthogonal projection, the cosine similarity index $\cos_{\mathrm{I}}  \left( \bm{z}, \bm{M} \right)$ is bounded by $0 \leq \cos_{\mathrm{I}}  \left( \bm{z}, \bm{M} \right)  \leq 1$. 
Specifically, $\bm{z}$ is completely suppressible if and only if $\cos_{\mathrm{I}}  \left( \bm{z}, \bm{M} \right)  = 1$, and $\bm{z}$ is completely insuppressible if and only if $\cos_{\mathrm{I}}  \left( \bm{z}, \bm{M} \right)  = 0$.  
We note that $\cos_{\mathrm{I}} \left( \bm{z}, \bm{M} \right)$ is mathematically undefined when the norm of $\bm{z}$ is zero, as this results in an indeterminate form $0/0$. In such cases, and when the norm is negligibly small in numerical computations, $\cos_{\mathrm{I}} \left( \bm{z}, \bm{M} \right)$ should be treated as NaN (Not a Number).

\begin{figure}[th!]
	\begin{center}
		\includegraphics[width=0.43 \textwidth]{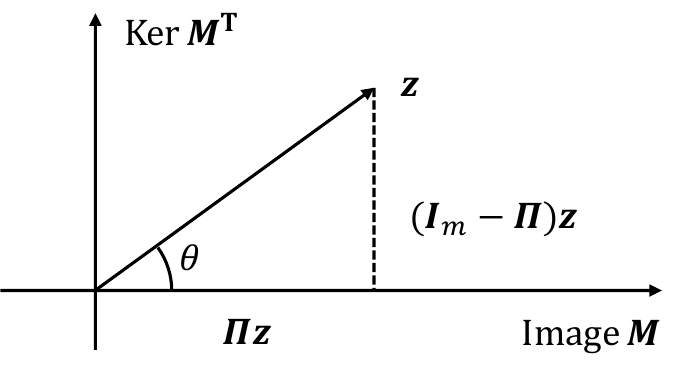}
		\caption{
			Geometrical interpretation of cosine similarity index $\cos_{\mathrm{I}} \left( \bm{z}, \bm{M} \right)$. In the space of the estimated residuals, the projection of the vector $\bm{z}$ onto the image space of the matrix $\bm{M}$ yields the projected vector $\bm{\Pi} \bm{z}$. The cosine similarity index $\cos_{\mathrm{I}} \left( \bm{z}, \bm{M} \right)$ is equivalent to the directional cosine between $\bm{z}$ and $\bm{\Pi} \bm{z}$. This value represents the extent to which $\bm{z}$ lies within the subspace $\Image \bm{M}$, effectively quantifying how closely the direction of $\bm{z}$ aligns with the best possible direction within the subspace. Furthermore, this value serves as an indicator of control efficiency, representing the ratio of the suppressible component contained in $\bm{z}$, which can be minimized by appropriately selecting the input $\bm{u}(t)$. 
		}\label{fig:FigCosine}
	\end{center}
\end{figure}

As conceptually illustrated in Fig.~\ref{fig:FigCosine}, $\bm{\Pi} \bm{z}$ is the orthogonal projection of $\bm{z}$ onto the subspace $\Image \bm{M}$. The cosine similarity index $\cos_{\mathrm{I}} \left( \bm{z}, \bm{M} \right)$ is equivalent to $\cos \theta$, where $\theta$ represents the angle of separation between the vector $\bm{z}$ and the subspace $\Image \bm{M}$.

Formally, the cosine similarity index can be extended to evaluate multiple Zernike modes simultaneously, considering the Zernike mode matrix $\bm{Z}$. Two primary methods for this extension are discussed below. 
The {Frobenius norm extension} $\cos_{\mathrm{F}} \left( \cdot , \cdot \right)$ of cosine similarity index is defined by vertically stacking the column components of $\bm{\Pi} \bm{Z}$ and $\bm{Z}$:
\begin{align}
	\cos_{\mathrm{F}} \left( \bm{Z}, \bm{M} \right) := \frac{\| \bm{\Pi} \bm{Z} \|_{\mathrm{F}}}{\| \bm{Z} \|_{\mathrm{F}}} 
	= \frac{\|  \mathrm{vec} ( \bm{\Pi} \bm{Z} ) \|}{\|  \mathrm{vec} ( \bm{Z} ) \|}, 
\end{align}
where $\| \cdot \|_{\mathrm{F}}$ is the Frobenius norm and $\mathrm{vec}( \cdot )$ is the vectorized operation;  
namely, $\mathrm{vec}(\bm{Z})$ corresponds to a vector of size $mp$ formed by vertically stacking the column components of $\bm{Z}$. 
Alternatively, the {weighted arithmetic mean extension} $\cos_{\mathrm{A}} \left( \cdot, \cdot \right)$ of cosine similarity index is defined by by the weighted arithmetic mean of the column-wise cosine similarity indices:
\begin{align}
	\cos_{\mathrm{A}} \left( \bm{Z}, \bm{M} \right) := \sum_{i = 1}^p w_i \cos_{\mathrm{I}} \left( \bm{Z}_{\RevSakibara{i}}, \bm{M} \right),
\end{align}
where $w_i> 0$ are weighting factors satisfying $\sum_{i=1}^p w_i =1$ and $\bm{Z}_i$ is the column components of $\bm{Z}$ representing the $i$-th Zernike mode for each $i = 1, \ldots, p$. 
As is clear from the definitions, $\cos_{\mathrm{I}} \left( \bm{z}, \bm{M} \right) = \cos_{\mathrm{F}} \left( \bm{z}, \bm{M} \right) = \cos_{\mathrm{A}} \left( \bm{z}, \bm{M} \right)$ for a single Zernike vector, \text{i.e.}, $p=1$ and $\bm{z} = \bm{Z}$,  provided $w_1 = 1$ for $\cos_{\mathrm{A}} \left( \bm{z}, \bm{M} \right)$. However, $\cos_{\mathrm{F}} \left( \bm{Z}, \bm{M} \right) \neq \cos_{\mathrm{A}} \left( \bm{Z}, \bm{M} \right)$ generally holds for multiple Zernike vector, \text{i.e.}, $p > 1$. 
The extended cosine similarity index $\cos_{\#}  \left( \bm{Z}, \bm{M} \right)$ maintains the property $0 \leq \cos_{\#}  \left( \bm{Z}, \bm{M} \right)  \leq 1$ for both ${\#} = {\mathrm{F}}$ and ${\#} = {\mathrm{A}}$.  
All column components of $\bm{Z}$ are completely suppressible if and only if $\cos_{\#} \left( \bm{Z}, \bm{M} \right)  = 1$ and all column components of $\bm{Z}$ is completely insuppressible if and only if $\cos_{\#} \left( \bm{Z}, \bm{M} \right)  = 0$, for both ${\#} = {\mathrm{F}}$ and ${\#} = {\mathrm{A}}$.   

As a practical application example of cosine similarity index, let us consider the {vector selection problem} to select the most suppressible and insuppressible Zernike vector from the set of Zernike vectors $\{ \bm{z}_1, \cdots, \bm{z}_p \}$. 
The most suppressible Zernike vector is determined by solving the following maximization problem: 
\begin{align}
	\bm{z}_{\mathrm{suppressible}} = 
	\mathrm{arg} \max_{\bm{z} \in \{ \bm{z}_1, \cdots, \bm{z}_p \}} \cos_{\mathrm{I}} \left( \bm{z}, \bm{M} \right),
    \label{eqn:z_suppressible}
\end{align}
and the most insuppressible Zernike vector is determined by solving the following minimization problem: 
\begin{align}
	\bm{z}_{\mathrm{insuppressible}} = 
	\mathrm{arg} \min_{\bm{z} \in \{ \bm{z}_1, \cdots, \bm{z}_p \}} \cos_{\mathrm{I}} \left( \bm{z}, \bm{M} \right). 
    \label{eqn:z_insuppressible}
\end{align}
Note that, although the subscripts ``suppressible'' and ``insuppressible'' for $\bm{z}$ are used for convenience, these vectors are actually quasi-suppressible and quasi-insuppressible in nature, as they do not strictly belong to $\Image \bm{M}$ or $\left( \Image \bm{M} \right)^{\perp}$.

As a practical application example of extended cosine similarity index, let us consider the {optimal radiator placement problem} that can suppress a particular combination of multiple Zernike modes. 
The size of $\bm{M} \in \mathbb{R}^{m \times n}$ indicates $m$-radiators and $n$-optical drivers. 
We define the placement vector $\bm{\chi} \in X \subset \mathbb{R}^{2m}$, which consists of $m$-radii, $\rho_1, \cdots, \rho_m$, and $m$-arguments, $\phi_1, \cdots, \phi_m$ of $m$-radiators in the admissible set $X$.  
In the radiator placement problem, it is appropriate to interpret the measurement matrix and the Zernike mode matrix as functions of the placement vector, $\bm{M}(\bm{\chi})$ and $\bm{Z}(\bm{\chi})$, where $\bm{M}(\bm{\chi})$ is typically obtained via numerical electromagnetic analysis or simulation and $\bm{Z}(\bm{\chi})$ is obtained by substituting the paired elements $\rho$ and $\phi$ of $\bm{\chi}$ into the Zernike polynomials. 
The optimal radiator placement  is then determined by solving the following maximization problem: 
\begin{align}
	\bm{\chi}_{\mathrm{optimal}} = \mathrm{arg} \max_{\bm{\chi} \in X} \cos_{\#} \left( \bm{Z}(\bm{\chi}), \bm{M}(\bm{\chi}) \right),  
	\label{eqn:PlacementProblem}
\end{align}
where the extended index $\cos_{\#}( \cdot, \cdot )$ is selected as either the Frobenius norm extension ${\#} = {\mathrm{F}}$ or the weighted arithmetic mean extension ${\#} = {\mathrm{A}}$. 
Given the lack of a rational basis for prioritizing either ${\#} = {\mathrm{F}}$ or ${\#} = {\mathrm{A}}$, we recommend calculating both metrics and evaluating the results empirically. 
The optimal radiator placement problem is the important next step in realizing an actual metrology system, which is beyond the scope of the present paper. We will leave this for future studies.


\subsection{Artificial Disturbance Selection via Cosine Similarity Index}

Let us reconsider the selection of disturbances in performance verification by introducing artificial disturbances discussed in Section \ref{sec:ArtificialDisturbance}, leveraging the knowledge gained from the discussion of cosine similarity index.  
Here, we consider a situation where the most suppressible Zernike vector $\bm{z}_{\mathrm{suppressible}}$ and the most insuppressible Zernike vector $\bm{z}_{\mathrm{insuppressible}}$ have been pre-selected for testing; these may be defined by Eqs.~\eqref{eqn:z_suppressible} and \eqref{eqn:z_insuppressible}, or alternatively, chosen based on heuristic knowledge of the suppression capabilities. 
In this context, it is desirable that the suppressible disturbance term $\bm{w}_{\mathrm{suppressible}} \in \Image \bm{M}$ is in the direction closest to $\bm{z}_{\mathrm{suppressible}}$, and the insuppressible disturbance term $\bm{w}_{\mathrm{insuppressible}} \in \left( \Image \bm{M} \right)^{\perp}$ is in the direction closest to $\bm{z}_{\mathrm{insuppressible}}$. 
It is also desirable that both $\bm{w}_{\mathrm{suppressible}}$ and $\bm{w}_{\mathrm{insuppressible}}$ be scaled to be large enough to induce a measurable effect on the astronomical observation for demonstration purposes, yet small enough to avoid violating drive saturation constraints.  
We note that if only a subset of $n-l$ actuations is permissible for the transient response test while the remaining $l$ are impermissible, a simplified plant model can be utilized. This is achieved by defining a selection matrix $\bm{S} \in \mathbb{R}^{n \times (n-l)}$ that maps the permissible drive inputs to the full $n$-dimensional space, and then replacing the original measurement matrix $\bm{M}$ with the reduced measurement matrix $\bm{M} \bm{S}$. 
However, detailed explanations of this practical implementation are omitted here to avoid unnecessary complexity in the discussion. 

Here, we suppose that $\bm{z}_{\mathrm{suppressible}} \notin \left( \Image \bm{M} \right)^{\perp}$ and $\bm{z}_{\mathrm{insuppressible}} \notin \Image \bm{M}$ and that they are normalized without loss of generality. 
for $\bm{w}_{\mathrm{suppressible}}$ and $\bm{w}_{\mathrm{insuppressible}}$ is determined from the specifications of the optical drive system and preliminary experiments. 
To find the vector $\bm{M} \bm{\xi} \in \Image \bm{M}$ that is closest to $\bm{z}_{\mathrm{suppressible}}$ in the least squares sense, we compute the gradient of the quadratic error %
\begin{align}
	\frac{\partial}{\partial \bm{\xi}} \left( \bm{z}_{\mathrm{suppressible}} - \bm{M} \bm{\xi} \right)^{\T} \left( \bm{z}_{\mathrm{suppressible}} - \bm{M} \bm{\xi} \right)
	= - \RevSakibara{2} \left( \bm{z}_{\mathrm{suppressible}} - \bm{M} \bm{\xi} \right)^{\T} \bm{M}.  
\end{align}
By setting the gradient to zero, the optimal $\bm{\xi}^{\ast}$ is given by
\begin{align}
	\bm{\xi}^{\ast} = \bm{M}^{\dagger} \bm{z}_{\mathrm{suppressible}}.  
\end{align}
Therefore, the optimal direction $\bm{M} \bm{\xi}^{\ast}$ in the image space is given by
\begin{align}
	\bm{M} \bm{\xi}^{\ast} = \bm{\Pi} \bm{z}_{\mathrm{suppressible}}. 
\end{align}
The suppressible disturbance term $\bm{w}_{\mathrm{suppressible}}$ is then defined by normalizing this optimal direction to the admissible magnitude $\sigma$:
\begin{align}
	\bm{w}_{\mathrm{suppressible}} := \frac{\sigma}{\left\|  \bm{M} \bm{\xi}^{\ast} \right\|} \bm{M} \bm{\xi}^{\ast} 
	= \frac{\sigma}{\left\|  \bm{\Pi} \bm{z}_{\mathrm{suppressible}} \right\|} \bm{\Pi} \bm{z}_{\mathrm{suppressible}}, 
\end{align}
where the denominator is nonzero because of the assumption $\bm{z}_{\mathrm{suppressible}} \notin \left( \Image \bm{M} \right)^{\perp}$. 

The vector $\bm{\eta} \in \left( \Image \bm{M} \right)^{\perp}$ is parametrized by $\bm{\eta} = \bm{U}_2 \bm{e}$, where $\bm{U}_2 \in \mathbb{R}^{m \times (m-n)}$ is the orthogonal basis for $\left( \Image \bm{M} \right)^{\perp}$ defined by the SVD in Eq.~\eqref{eqn:Msvd}, and $\bm{e} \in \mathbb{R}^{m-n}$ is the free parameter vector.
To find the vector $\bm{\eta}$ that is closest to $\bm{z}_{\mathrm{insuppressible}}$ in the least squares sense, we compute the gradient of the quadratic error 
\begin{align}
	\frac{\partial}{\partial \bm{e}} \left( \bm{z}_{\mathrm{insuppressible}} - \bm{U}_2 \bm{e} \right)^{\T} \left( \bm{z}_{\mathrm{insuppressible}} - \bm{U}_2 \bm{e} \right)
	= - \RevSakibara{2} \left( \bm{z}_{\mathrm{insuppressible}} - \bm{U}_2 \bm{e} \right)^{\T} \bm{U}_2. 
\end{align}
By setting the gradient to zero, the optimal $\bm{e}^{\ast}$ is given by 
\begin{align}
	\bm{e}^{\ast} = \left( \bm{U}_2^{\T} \bm{U}_2 \right)^{-1} \bm{U}_2^{\T} \bm{z}_{\mathrm{insuppressible}} = \bm{U}_2^{\T} \bm{z}_{\mathrm{insuppressible}}.  
\end{align}
Therefore, the optimal direction $\bm{\eta}^{\ast}$ in the orthogonal complementary space is given by
\begin{align}
	\bm{\eta}^{\ast} := \bm{U}_2 \bm{e}^{\ast} = \left( \bm{I}_m - \bm{\Pi} \right) \bm{z}_{\mathrm{insuppressible}}. 
\end{align}
The insuppressible disturbance term $\bm{w}_{\mathrm{insuppressible}}$ is then defined by normalizing this optimal direction to the admissible magnitude $\sigma$:
\begin{align}
	\bm{w}_{\mathrm{insuppressible}} := \frac{\sigma}{\left\|  \bm{\eta}^{\ast} \right\|} \bm{\eta}^{\ast}  
	= \frac{\sigma}{\left\|  \left( \bm{I}_m - \bm{\Pi} \right) \bm{z}_{\mathrm{insuppressible}} \right\|} \left( \bm{I}_m - \bm{\Pi} \right) \bm{z}_{\mathrm{insuppressible}},  
\end{align}
where the denominator is nonzero because of the assumption $\bm{z}_{\mathrm{insuppressible}} \notin \Image \bm{M}$ . 

Once the disturbance is obtained, a simple yet effective test scenario is achieved by applying the following artificial disturbance:
\begin{align}
	\bm{w}(t) := \left\{ \begin{array}{ll} \bm{0}_m & 0 \leq  t < \frac{t_{\mathrm{e}}}{4}, \; \frac{t_{\mathrm{e}}}{2} \leq t < \frac{3 t_{\mathrm{e}}}{4} \\
	\bm{w}_{\mathrm{suppressible}} & \frac{t_{\mathrm{e}}}{4} \leq t < \frac{t_{\mathrm{e}}}{2} \\
	\bm{w}_{\mathrm{insuppressible}} & \frac{3 t_{\mathrm{e}}}{4} \leq t \leq t_{\mathrm{e}} 
	\end{array}
	\right.
\end{align}
where $t_{\mathrm{e}}$ is the transient test time interval. This scenario allows for the clear evaluation of the controller's performance in absorbing the suppressible component $\bm{w}_{\mathrm{suppressible}}$ followed by the system's inherent inability to fully remove the insuppressible component $\bm{w}_{\mathrm{insuppressible}}$.


\section{Numerical Examples}\label{sect:numerical_examples}


\subsection{Plant Description}

\begin{figure}[p]
    \centering

    \begin{minipage}{1.0\textwidth}
        \centering
        \includegraphics[width=0.48\textwidth]{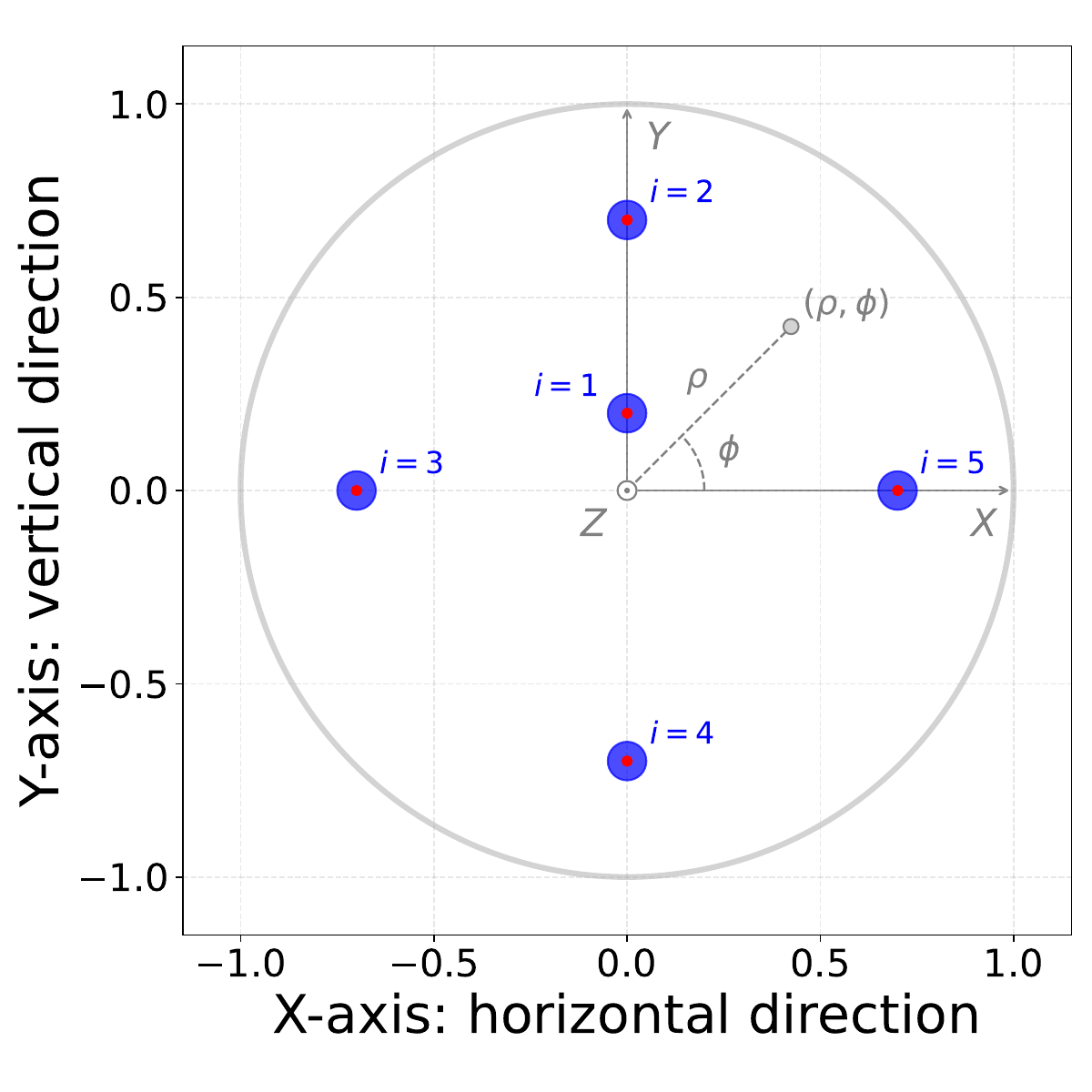}
        \caption{
            Locations of radiators on the M1-reflector. The M1-reflector is represented as a unit circle, and the relative positions of the radiators are indicated by blue circles within this unit circle. 
            The figure corresponds to the view looking down at the M1-reflector along the optical axis. The X-axis indicates the horizontal rightward direction of the reflector, and the Y-axis is defined to form a right-handed coordinate system with the X and Z axes. The Z-axis points toward the sky, representing the direction opposite to the incoming plane waves. The symbols $(\rho,\phi)$ represent the polar coordinates used to define the radiator positions. 
        }\label{fig:FigRadiators}
    \end{minipage}

    \vspace{3em} 

    \begin{minipage}{1.0\textwidth}
        \centering
        \includegraphics[width=0.42\textwidth]{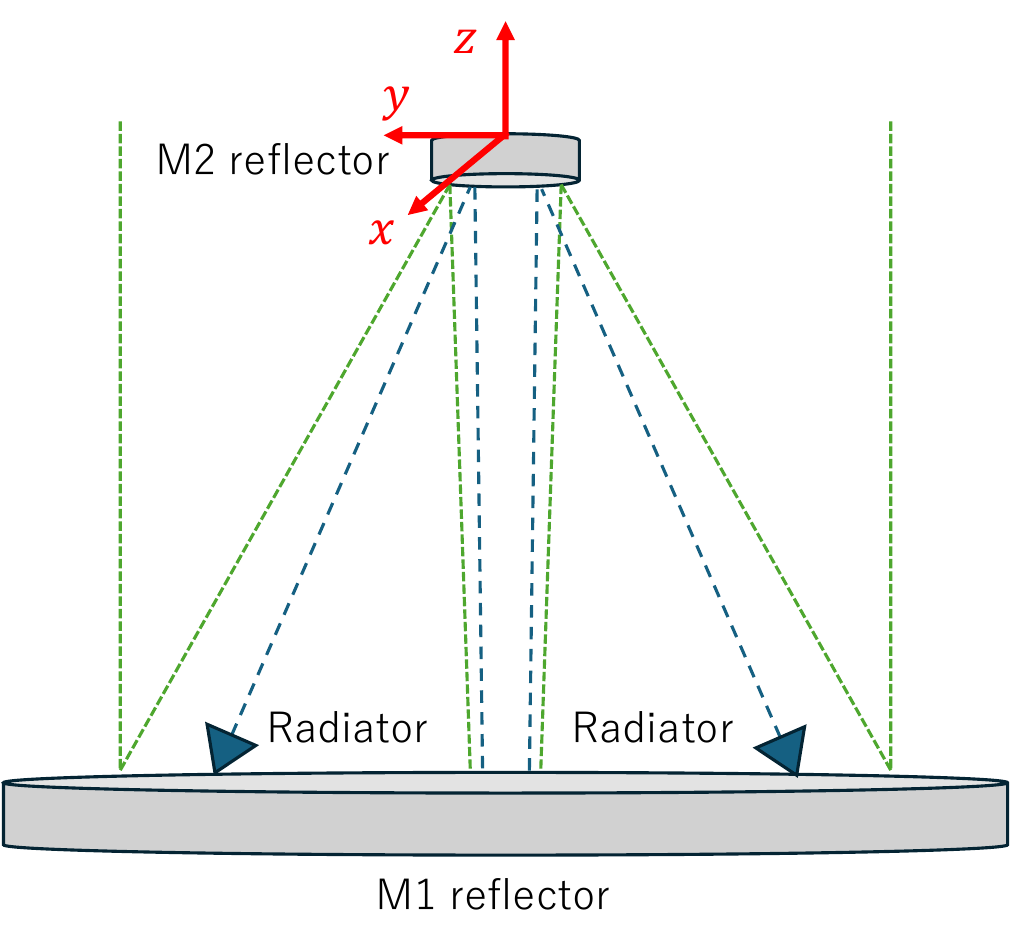}
        \caption{
            Framework of MAO system by EPL measurements and M2-reflector actuation, viewed from an off-zenith direction looking down at the telescope. The system simultaneously receives astronomical and emitted radio waves, which are schematically shown as green and blue dashed lines. The real-time EPL measurement is achieved by determining the cross-power spectrum between the radio waves emitted from the M1-reflector's radiators and those collected by the measurement system after reflection from the M2-reflector. For adaptive optics compensation, the M2-reflector is driven using an optical drive system, which enables three-axis translational motion in XYZ directions to actively change the EPLs, serving as one example of optical drive systems. 
        }\label{fig:M1M2}
    \end{minipage}
\end{figure}

In this section, we demonstrate the effectiveness of the proposed AWPI control using numerical examples.
Figure~\ref{fig:FigRadiators} illustrates the M1-reflector as viewed from the optical axis. 
This radiator arrangement can be easily implemented in a real telescope and is expected to be effective in compensating tip-tilt and defocus components. 
We assume five radiators are arranged concentrically on this reflector. 
The coordinates of these radiators, given as pairs of radius $\rho_i$ and argument $\phi_i$ [deg] for $i = 1, \ldots, 5$, are specified below:
\begin{align}
	( \rho_1, \phi_1) &= ( 0.2, 90 ), \label{eqn:rhophi-1-1} \\
	( \rho_2, \phi_2) &= ( 0.7, 90 ), \label{eqn:rhophi-1-2} \\
	( \rho_3, \phi_3) &= ( 0.7, 180 ), \label{eqn:rhophi-1-3} \\ 
	( \rho_4, \phi_4) &= ( 0.7, 270 ), \label{eqn:rhophi-1-4} \\
	( \rho_5, \phi_5) &= ( 0.7, 0 ), \label{eqn:rhophi-1-5}
\end{align}
where the argument $\phi_i$ is measured counterclockwise from the positive horizontal axis.  
This arrangement replicates the radiator configuration used in the 2025 experiment at the Nobeyama 45-m Telescope. While $\rho_1$ would be ideally zero for a point-symmetric arrangement of radiators, it is set to $\rho_1 \neq 0$ to simulate a practical layout that avoids M2 blocking.

Figure~\ref{fig:M1M2} illustrates the M1 and M2 reflectors as viewed perpendicular to the optical axis. 
We assume that incident radio waves from celestial bodies are parallel to the optical axis. 
The celestial waves are reflected by the M1 reflector and subsequently by the M2 reflector, converging to form an image at the focal point. 
Conversely, reference radio waves, utilizing a different wavelength \RevSakibara{from} the observation waves, are radiated from the on-reflector radiators. 
These reference waves are also reflected by the M2 reflector and converge at the focal point. 
Subsequently, signal processing is performed on the combined waves to measure the EPLs $\bm{y}(t) \in \mathbb{R}^5$, \textit{i.e.}, $m=5$.

Furthermore, we model the M2 reflector \RevSakibara{as a three-axis optical drive system}. 
The coordinate system for the M2 drive is defined as follows: 
the Z-axis points from the M1 reflector toward the M2 reflector along the optical axis; 
\RevJikuya{the X-axis is perpendicular to the Z-axis and points in the elevation direction;} 
and the Y-axis completes the left-handed orthogonal system.
The drive amounts in the Z, X, and Y directions are denoted by $\bm{p}(t) \in \mathbb{R}^3$. 
The input $\bm{u}(t) \in \mathbb{R}^3$, \textit{i.e.}, $n=3$,  is the drive command to the secondary reflector drive system. 
We assume that the time constant for each axis is identical and that the matrix $\bm{T}_{\mathrm{s}}$ is defined by:
\begin{align}
	\bm{T}_{\mathrm{s}} = t_{\mathrm{s}} \bm{I}_3, \quad t_{\mathrm{s}} = 0.2 \, \mathrm{[s]}. \label{eqn:ex-01} 
\end{align}
We introduce the measurement matrix $\bm{M} \in \mathbb{R}^{5 \times 3}$, which is obtained via electromagnetic field analysis, as follows:
\begin{align}
	\bm{M} = \begin{bmatrix} 1.960 & -0.259 & 0.0 \\ 1.607 & -0.731 & 0.0 \\ 1.607 & 0.0 &  -0.731 \\ 1.607 & 0.731 & 0.0 \\ 1.607 & 0.0 & 0.731 \end{bmatrix}. \label{eqn:ex-02} 
\end{align}
This matrix represents the interaction between the subreflector drive amounts and the EPL measurements. The columns of $\bm{M}$ correspond to the drive amounts in the Z, X, and Y directions, in that order, while its rows correspond to the sequence of the five radiators. Note that if the direction of the Y-axis is reversed to adopt a right-handed coordinate system, the signs of the third column should be inverted. Similarly, changing the drive order or the radiator sequence would require a corresponding permutation of the columns or rows of $\bm{M}$, respectively. 

To formulate the control problem for MAO, the plant model is constructed by substituting Eqs.~\eqref{eqn:ex-01} and \eqref{eqn:ex-02} into Eqs.~\eqref{eqn:plant-01-2} and \eqref{eqn:plant-02}. For simplicity in this numerical example, the desired optical path length $\bm{g}(\gamma(t))$ is neglected \RevJikuya{assuming it is ideally pre-compensated as presented in Eq.~\eqref{eqn:AWPI-00}}, and the external disturbance term $\bm{w}(t)$ will be determined subsequently. 
Driving the secondary reflector changes the optical path length from the radiators to the focal point, thus causing $\bm{y}(t)$ to vary according to the drive command $\bm{u}(t)$. 
Since the \RevSakibara{measurement} $\bm{y}(t)$ corresponds to variations in the optical path length of \RevJikuya{the radiators, which can be identified with the variations in the optical path length of celestial radio waves}, the core task of MAO is actually realized by determining the appropriate drive command $\bm{u}(t)$ based on the measurement $\bm{y}(t)$.

\subsection{Controller Design}

The MAO is formulated as an asymptotic disturbance attenuation problem, and the AWPI controller defined by Eqs.~\eqref{eqn:AWPI-00} to \eqref{eqn:AWPI-03} is applied. 
Similar to the plant model modeling process, the term $\bm{g}(\gamma(t))$ is neglected.
The upper and lower saturation limits for each $i = 1, 2, 3$ are set to $\pm \bar{u}_i = \pm 3$ [mm]. 
The integral gain $\bm{K}_{\mathrm{i}}$ and proportional gain $\bm{K}_{\mathrm{p}}$ are selected using the loop shaping technique. 
They are assumed to be diagonal matrices:
\begin{align}
	\bm{K}_{\mathrm{i}} = k_{\mathrm{i}} \bm{I}_3, \quad \bm{K}_{\mathrm{p}} = k_{\mathrm{p}} \bm{I}_3,
\end{align}
and the scalar gains $k_{\mathrm{i}}$ and $k_{\mathrm{p}}$ are chosen to satisfy the inequalities in Eq.~\eqref{eqn:FreqRange-04} as follows:
\begin{align}
	k_{\mathrm{i}} = 1.57 \, \mathrm{[1/s]}, \quad k_{\mathrm{p}} = 0.316 \, \mathrm{[-]},
\end{align}
where the control period is supposed to be $t_{\mathrm{c}} = 0.1$ [s]. 
The anti-windup gain is set to $\bm{K}_{\mathrm{a}} = \bm{K}_{\mathrm{p}}^{-1}$. 

\begin{figure}[p] %
    \centering

    \begin{minipage}{1.0\textwidth}
        \centering
        \includegraphics[width=0.83\textwidth]{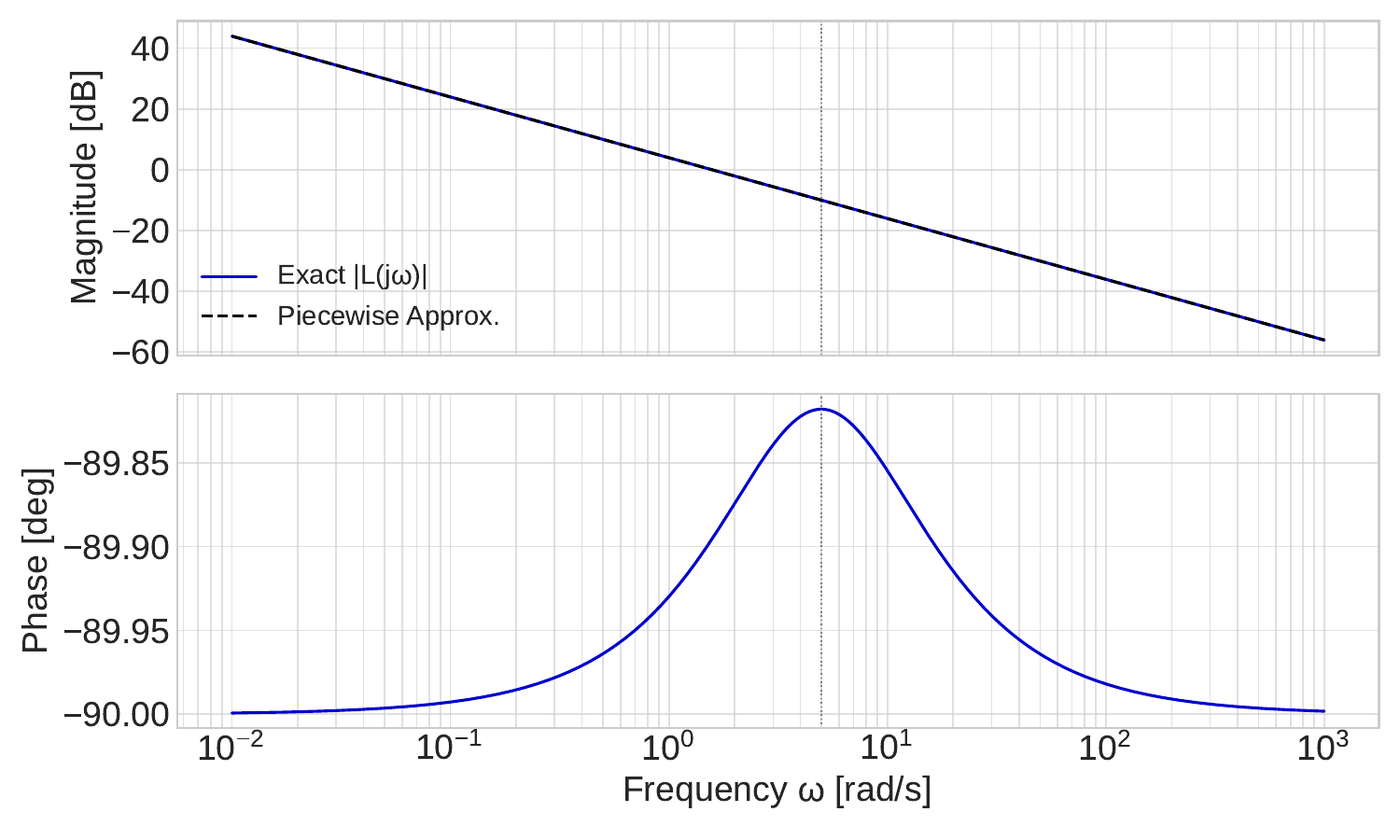}
        \caption{
            Bode diagram of the open loop transfer function $L(s)$, where the piecewise linear approximation is overlapped with the gain diagram. Although the range corresponding to the $0$~[dB/dec] slope is very narrow, making it difficult to clearly distinguish this segment in the gain diagram, its existence is indicated by the phase lead observed in the phase diagram. 
        }\label{fig:FigLbode}
    \end{minipage}

    \vspace{3em} %

    \begin{minipage}{1.0\textwidth}
        \centering
        \includegraphics[width=0.83\textwidth]{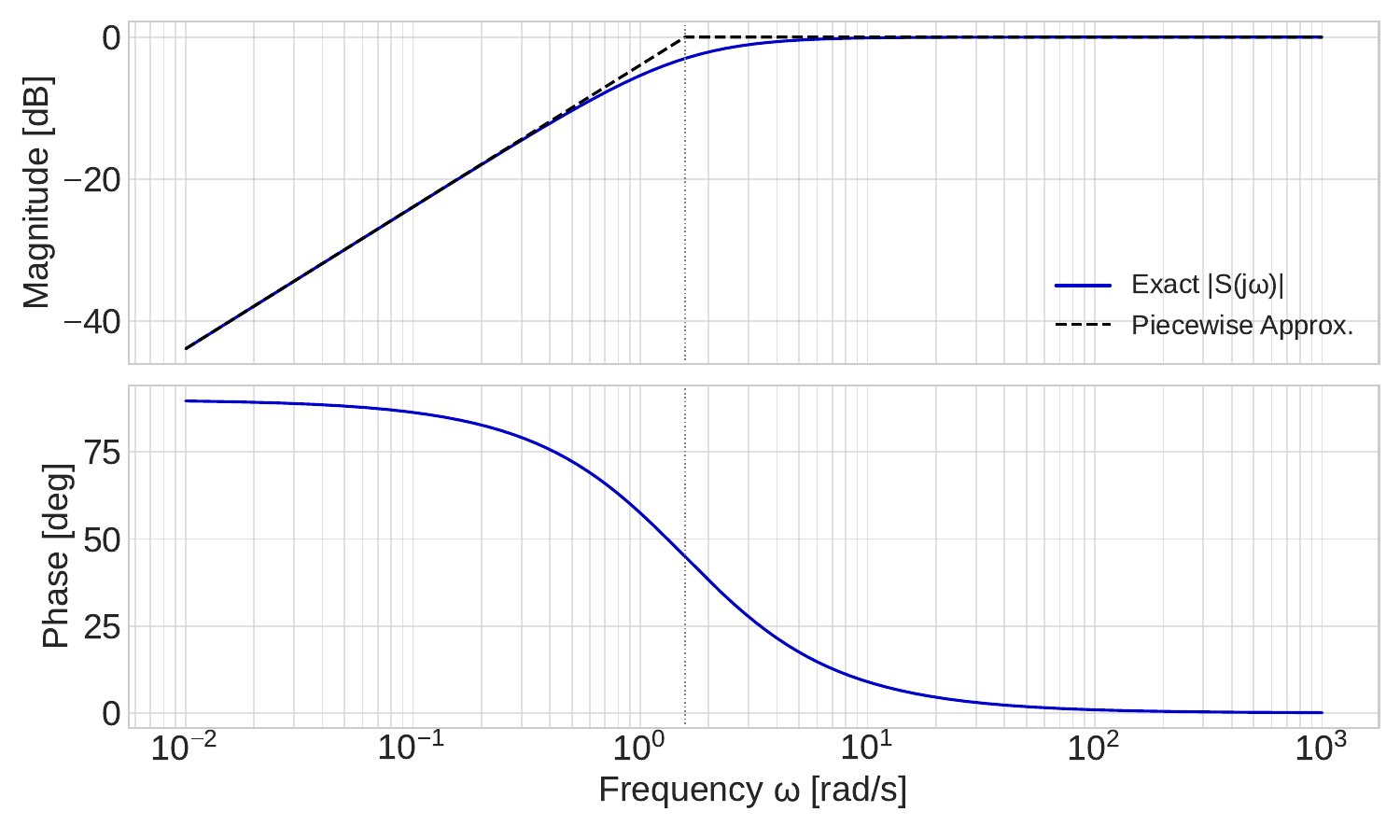}
        \caption{
            Bode diagram of the sensitivity function $S(s)$, where the piecewise linear approximation is overlapped with the gain diagram. The piecewise linear approximation provides a good approximation of the gain diagram. Furthermore, the gain crossover frequency calculated from the open-loop transfer function $L(s)$ provides a good approximation of the corner frequency in the gain diagram of $S(s)$.
        }\label{fig:FigSbode}
    \end{minipage}
\end{figure}

To confirm the effect of loop shaping, Fig.~\ref{fig:FigLbode} shows the Bode plot of the open-loop transfer function $L(s)$. 
Since $k_{\mathrm{i}} / k_{\mathrm{p}} = 4.97$ [rad/s] and $1/t_{\mathrm{s}} = 5.0$ [rad/s], the section with a slope of $0$ [dB/dec], as represented in Eq.~\eqref{eqn:Lgain-01} and Fig.~\ref{fig:FigGainOP}, hardly appears in the gain diagram. 
The gain crossover frequency is 1.57 [rad/s], which is consistent with the selected integral gain $k_{\mathrm{i}}$.
 The gain margin is 10 [dB] to 20 [dB] around 5 [rad/s], confirming the characteristics predicted by Eq.~\eqref{eqn:FreqRange-03}. 
The phase margin is greater than 90 [deg]. 
Even after converting the control period $t_{\mathrm{c}} = 0.1$ [s] into a phase delay of $t_{\mathrm{c}} \omega$ [rad], which corresponds to $28.6$ [deg], still provides a sufficient phase margin of greater than $60$ [deg]. 
This demonstrates the inherent robustness of the system, which can tolerate some unmodeled response delay in the M2 drive system and unmodeled processing delay in the EPL measurement system.

To confirm the asymptotic disturbance suppression performance, Fig.~\ref{fig:FigSbode} also shows a Bode plot of the sensitivity function $S(s)$. 
The piecewise linear approximation shown in Eq.~\eqref{eqn:S-05} and Fig.~\ref{fig:FigGainCL} provides a good approximation of the gain diagram of $S(s)$. 
At frequencies lower than the gain crossover frequency of $1.57$ [rad/s], the gain decreases and the condition $S(0) = 0$ is satisfied, confirming the asymptotic disturbance suppression performance. 

\subsection{Numerical Simulation: Asymptotic Disturbance Suppression}

Numerical simulation has been carried out to demonstrate the effectiveness of the proposed AWPI controller. 
The disturbance $\bm{w}(t)$ is supposed to be piecewise constant and is defined by 
\begin{align}
	\bm{w}(t) := \left\{ \begin{array}{ll}
	10 \bm{w}_1 & 10 < t < 20  \\
	10 \bm{w}_2 & 30 < t < 40 \\
	20 \bm{w}_1 & 50 < t < 60 \\
	20 \bm{w}_2 & 70 < t < 80 \\ 
	\bm{0} & \mathrm{else} 
	\end{array} 	\right., \quad 
	\bm{w}_1 := \begin{bmatrix} 0.435 \\ 0.302 \\ 0.302 \\ 0.376 \\ 0.698 \end{bmatrix} \, \mathrm{[mm]}, \quad 
	\bm{w}_2 := \begin{bmatrix} 0.434 \\ 0.224 \\ 0.224 \\ 0.597 \\ 0.597 \end{bmatrix} \, \mathrm{[mm]}. \label{eqn:w-01}
\end{align}
The disturbance $\bm{w}_1$ is neither suppressible nor insuppressible 
\begin{align}
	\bm{w}_1 \not\in \Image \bm{M} \; \& \; \bm{w}_1 \not\in \left( \Image \bm{M} \right)^{\perp}
	\quad \Leftrightarrow \quad 
	0 < \cos_{\mathrm{I}} \left( \bm{w}_1, \bm{M} \right) < 1, 
\end{align}
where $\cos_{\mathrm{I}} \left( \bm{w}_1, \bm{M} \right) = 0.986$, 
and therefore, the estimated residual $\hat{\bm{q}}(t)$ is not expected to converge to zero. 
In contrast, the disturbance $\bm{w}_2$ is suppressible 
\begin{align}
	\bm{w}_2 \in \Image \bm{M} 
	\quad \Leftrightarrow \quad 
	\cos_{\mathrm{I}} \left( \bm{w}_2, \bm{M} \right) = 1, 
\end{align}
and therefore, the estimated residual $\hat{\bm{q}}(t)$ should theoretically converge to zero.  

\begin{figure}[p]
	\begin{center}
		\includegraphics[width=0.83\textwidth]{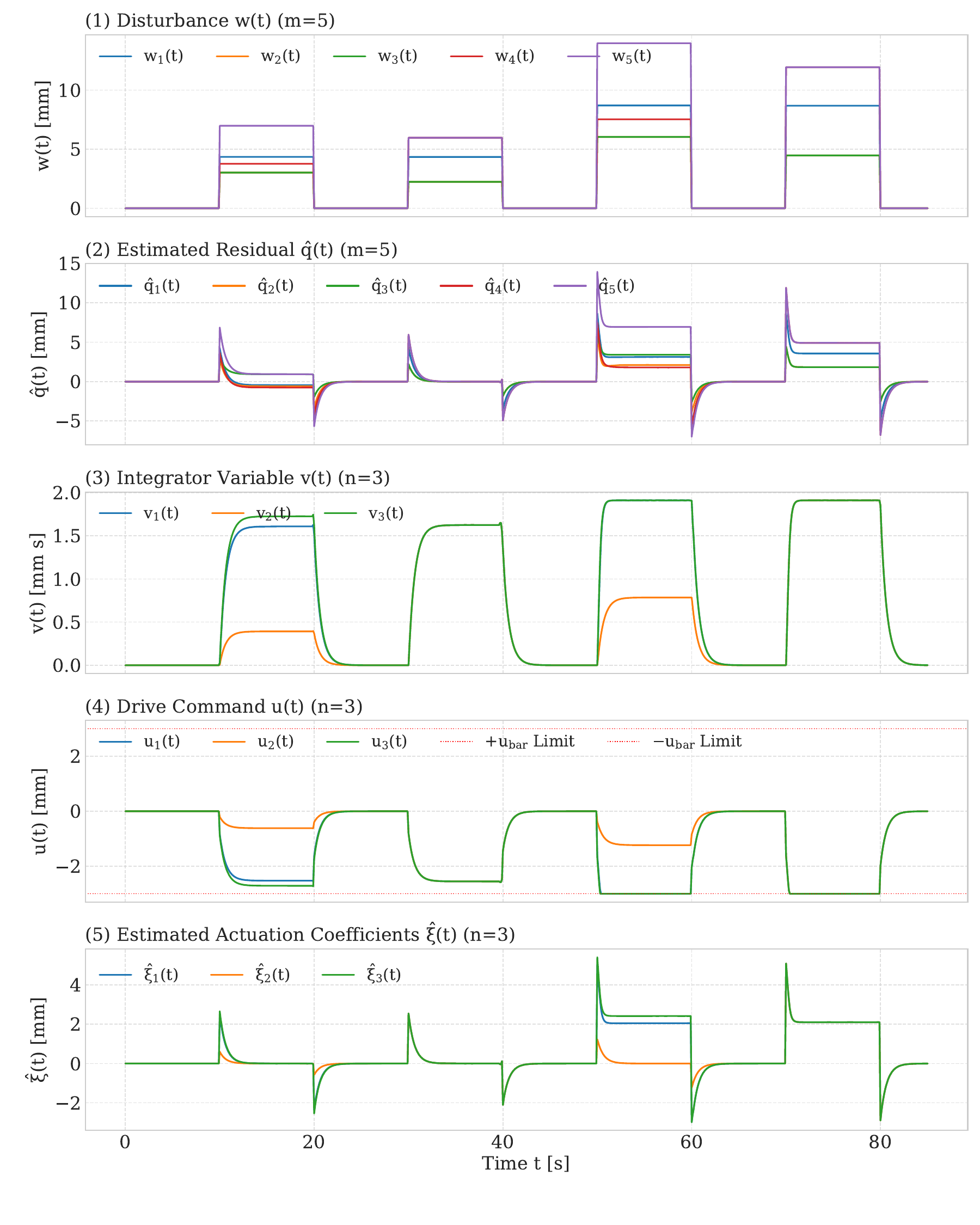}
		\caption{
			Simulation results verifying the direction-dependent disturbance suppression performance and the effectiveness of the anti-windup mechanism are presented, showing the time response of the estimated residual, integrator variable, drive command, and estimated actuation coefficient. The closed-loop system consists of the plant in Fig.~\ref{fig:FigPlant} and the AWPI controller in Fig.~\ref{fig:FigAWPI}. To evaluate the control performance, four combinations of rectangular wave signals are applied as the disturbance $\bm{w}(t)$, varying across two directions and two amplitudes. The estimated residual confirms the response characteristics consistent with the classification of suppressible and insuppressible disturbance terms, demonstrating that the response was obtained as designed. Furthermore, the drive command monitors the response characteristics when saturation occurs under large amplitude inputs, confirming the effective operation of the anti-windup mechanism. 
		}\label{fig:FigAWPIsimulation}
	\end{center}
\end{figure}

Figure~\ref{fig:FigAWPIsimulation} shows the simulation results. These results represent the forced response of the closed-loop system composed of the plant in Eqs.~\eqref{eqn:plant-01-2} and \eqref{eqn:plant-02} and the AWPI controller defined by Eqs.~\eqref{eqn:AWPI-00} to \eqref{eqn:AWPI-03}, when subjected to the disturbance $\bm{w}(t)$ in Eq.~\eqref{eqn:w-01}. 
During the simulation, the estimated residual $\hat{\bm{q}}(t)$, the integrator variable \RevSakibara{$\bm{v}(t)$}, the input $\bm{u}(t)$, and the estimated actuation coefficients  $\hat{\bm{\xi}}(t)$ remain bounded, confirming the stability of the closed-loop system. 

During the interval $10 < t < 20$, the plant is subjected to the disturbance $\bm{w}_1$, which is neither suppressible nor insuppressible. 
As a result, $\hat{\bm{q}}(t)$ does not converge to zero, while $\hat{\bm{\xi}}(t)$ converges to zero, as expected. 
Conversely, on the interval $30 < t < 40$, the plant is affected by the suppressible disturbance term $\bm{w}_2$. 
In this case, both $\hat{\bm{q}}(t)$ and $\hat{\bm{\xi}}(t)$ converge to zero, as expected. 
Furthermore, in both intervals, $\bm{u}(t)$ does not reach the saturation limits $\pm \bar{u}_i = \pm 3$ [mm] for $i = 1, \ldots 3$. 
Consequently, the controller's behavior is functionally identical to that of a pure PI controller.

In contrast, \RevJikuya{to counteract the large positive disturbances,} $\bm{u}(t)$ reaches the lower limit $- \bar{u}_i = -3$ [mm] during the intervals $50 < t < 60$ and $70 < t < 80$. 
As a result, both $\hat{\bm{\xi}}(t)$ and $\hat{\bm{q}}(t)$ fail to converge to zero. 
The integrator variable $\bm{v}(t)$ remains constant while $\bm{u}(t)$ is constrained by the lower limit. 
Although it may initially appear that control has been lost, the anti-windup mechanism is actively working to suppress divergence of $\bm{v}(t)$, clearly demonstrating its intended function.

In summary, the AWPI controller is verified to be a stabilizing controller that achieves asymptotic disturbance suppression performance. 
Furthermore, the anti-windup mechanism has been thoroughly confirmed to be highly effective in suppressing the divergence of the integral variable when the input is saturated, demonstrating that it functions exactly as intended.

%


\subsection{Numerical Simulation: Wind Disturbance Suppression}

Numerical simulation has been also carried out to demonstrate the effectiveness of the proposed AWPI controller. 
The closed-loop analysis based on Eq.~\eqref{eqn:S-008} indicates that the frequency dependence of the transfer from the disturbance $\bm{w}(s)$ to the estimated residual $\hat{\bm{q}}(s)$ is primarily dominated by the diagonal elements of the transfer function matrix $\bm{S}(s)$ from $\bm{M}^\dagger \bm{w}(s)$ to $\bm{M}^{\dagger} \hat{\bm{q}}(s)$, which can be identified with $\hat{\bm{\xi}}(t)$ as defined in Eq.~\eqref{eqn:AWPI-0}. 

To simplify the problem setting and isolate the effects of the wind disturbance spectrum, we neglect the directional dependency contained in $\bm{M}^\dagger \bm{w}(s)$ by neglecting the influence of the matrix $\bm{M}^\dagger$ and focus solely on the input-output relation governed by the decoupled sensitivity function $S(s)$. Specifically, we define a scalar disturbance $w(s)$ as an element of $\bm{M}^\dagger \bm{w}(s)$ where its spectral properties are assumed to be described by a standard wind gust model. 
This allows us to consider the scalar time-domain counterpart $w(t)$ of the scalar frequency-domain signal $w(s)$ and subsequently to investigate the scalar time-domain output $\hat{\xi}(t) := \mathcal{L}^{-1} \left[ S(s) w(s) \right](t)$ under a realistic, yet simplified, wind disturbance scenario.

The Power Spectral Density (PSD) $\Phi_u(f)$ [m$^2$/s] of the continuous wind gust component $u$ along the horizontal direction is commonly modeled using the von Karman wind turbulence spectrum, a mathematical model that describes isotropic turbulence. 
It accurately models the inertial subrange, where the spectrum follows Kolmogorov's $-5/3$ power law slope, while also providing a realistic representation of the energy-containing range at low frequencies. 
This spectrum is given by the following equation: 
\begin{align}
	\Phi_u(f) &= \frac{4 \sigma_u^2 L_u }{\bar{U} \left( 1 + 70.8 \, f_L^2 \right)^{\frac{5}{6}}}, \label{eqn:vonKarman-01} 
\end{align}
where the non-dimensional frequency $f_L$ is defined by 
\begin{align}
	f_L &= \frac{f  L_u}{\bar{U}}. \label{eqn:vonKarman-02} 
\end{align}
In these expressions, $\sigma_u$ [m/s] represents the root mean square (RMS) value or standard deviation of the longitudinal wind velocity fluctuation, 
$L_u$  [m] is the longitudinal integral length scale which characterizes the average size of the largest energy-containing eddies within the wind turbulence, 
$\bar{U}$ [m/s]  is the mean wind speed at the reference height, and $f$ [Hz] is the \RevJikuya{temporal} frequency. 
By applying Taylor's frozen turbulence hypothesis $\Omega = 2 \pi f / \bar{U}$ and the spectral conversion rule $\Phi_u(f) = \Phi_u(\Omega) \frac{d \Omega}{df}$, 
this spectral form $\Phi_u(f)$ in Eq.~\eqref{eqn:vonKarman-01} based on the temporal frequency $f$ 
is strictly equivalent to the form $\Phi_u(\Omega)$ based on the spatial wavenumber $\Omega$ [rad/m] as follows: 
\begin{align}
	\Phi_u(\Omega) = \frac{2 \sigma_u^2 L_u}{\pi} \frac{1}{\left( 1 + \left( 1.339 \, L_u \Omega \right)^2 \right)^{\frac{5}{6}}}. \label{eqn:vonKarman-03} 
\end{align}

The scalar time-domain disturbance $w(t)$ is constructed based on the derived PSD $\Phi_u(f)$. 
Two typical approaches for constructing $w(t)$ are: approximating $\Phi_u(f)$ with a linear filter, such as a second order filter, and driving it with white noise; or applying the Inverse Fourier Transform (IFT) to $\Phi_u(f)$ with a random phase. 
We adopt the latter approach because the PSD of the resulting time-domain signal $w(t)$ is nearly identical to the theoretical spectrum $\Phi_u(f)$, which is essential for accurate simulation.

The numerical simulation employed a specific set of parameters for $\Phi_u(f)$ \RevJikuya{as a representative case to evaluate the spectral characteristics and the disturbance suppression performance of the control system}. 
The average wind speed was set to  $\bar{U} =10.0$ [m/s], 
the target standard deviation of the turbulence was set to $\sigma_u = 2.0$ [m/s], 
and the turbulence length scale was set to $L_u = 20.0$ [m].  
\RevJikuya{While $L_u$ varies depending on the telescope aperture and site conditions, these specific values} leads to a characteristic corner frequency $f_{\Phi} \simeq 0.080$ [Hz] for the spectrum, marking the transition point where turbulence energy begins to decay rapidly. 
\RevJikuya{The choice of $L_u$ primarily shifts this corner frequency and does not qualitatively alter the fundamental mechanism of disturbance attenuation within the control bandwidth.} 

The simulation was conducted in the discrete-time domain using a control period $t_c = 0.1$  [s], corresponding to a sampling frequency of $f_s = 10.0$ [Hz]. 
A total simulation time was chosen to $T_{\mathrm{sim}} = 10,000$ [s] to ensure sufficient spectral resolution, resulting in $N_{\mathrm{points}} = 100,000$ data points.

The frequency vector $\bm{f}$ is \RevSakibara{defined} by an array of $N_{\mathrm{points}}$ discrete frequency components, generated to correspond precisely to the time-series data points. 
This \RevSakibara{vector} spans the two-sided frequency range from $-f_s/2$ to $f_s/2$ ensuring all frequency content up to the Nyquist limit is captured for accurate spectral representation.
The complex amplitude spectrum, $\bm{w}(\bm{f})$, is constructed by taking the square root of $\Phi_u(f)$ in Eq.~\eqref{eqn:vonKarman-01} at each frequency $f$ in the generated frequency vector $\bm{f}$ and incorporating a random phase that is uniformly distributed between $0$ and $2 \pi$ [rad]. 
The magnitudes are appropriately scaled to account for the discrete Fourier transform properties and the simulation length. 
Namely, $\bm{w}(\bm{f})$ is the bank of components $w(f)$ given by
\begin{align}
	w(f) &= A(f) e^{i \phi(f)}, \\
	A(f) &= \sqrt{ \left| \Phi_u(f) \right| N_{\mathrm{points}} \Delta f / 2},
\end{align}
where $\phi (f)\in [0,2\pi)$ [rad] is the uniformly distributed random phase and $f$ is an element of $\bm{f}$,  
$\Delta f$ is the frequency resolution $f_s / N_{\mathrm{points}}$, and the factor of $1/2$ accounts for the transformation from the one-sided PSD $\Phi_u(f)$ to the required two-sided spectrum for IFT input. 

The time vector $\bm{k}$ is defined by an array of $N_{\mathrm{points}}$ discrete time components, generated to correspond precisely to the time-series data points. 
Applying the Inverse Fast Fourier Transform (IFFT) to the complex vector $\bm{w}(\bm{f})$ yields the unscaled, discrete time-domain vector $\bm{w}_{\mathrm{unscaled}}(\bm{k})$. 
Hermitian symmetry is strictly enforced on the complex vector $\bm{w}(\bm{f})$ to guarantee that the resulting time-domain vector $\bm{w}_{\mathrm{unscaled}}(\bm{k})$ obtained after applying the IFFT is purely real.  
The $k$-th element of this vector $w_{\mathrm{unscaled}}[k]$ represents the scalar wind disturbance at discrete time index $k = 0, 1, \ldots, N_{\mathrm{points}}-1$.   
Finally, to ensure strict adherence to the defined turbulence intensity $\sigma_u$ of the von Karman model, the scaled IFFT output $w[k]$ is obtained by multiplying the unscaled IFFT output $w_{\mathrm{unscaled}}[k]$ by the constant scaling factor \RevSakibara{$C$}: 
\begin{align}
	w[k] &= C w_{\mathrm{unscaled}}[k], \\
	C &= \frac{\sigma_{u}}{\sigma(\bm{w}_{\mathrm{unscaled}}(\bm{k}))},
\end{align}
where the constant scaling factor $C$ is derived from the ratio of the standard deviation $\sigma_{u}$ in the von Karman model to the numerically evaluated standard deviation $\sigma(\bm{w}_{\mathrm{unscaled}}(\bm{k}))$ of the generated series $\bm{w}_{\mathrm{unscaled}}(\bm{k})$. 
Since $\sigma(\bm{w}_{\mathrm{unscaled}}(\bm{k}))$ is a single statistical measure calculated across the entire time series and does not depend on time $k$, $C$ is a constant value applied uniformly across the signal. 
This correction ensures that the final wind disturbance $w[k]$ precisely matches the intended turbulence intensity. 

The resulting scaled signal $\bm{w}[k]$ is then utilized as the input to the discrete-time decoupled sensitivity function $S(z)$. This function was obtained by discretizing the continuous-time decoupled sensitivity function $S(s)$ using a Zero-Order Hold (ZOH) method with an ideal sampler at the control period $t_c = 0.1$ [s]. Subsequently, the output $\hat{\bm{\xi}}[k]$ is obtained. 
The input spectrum $P_w(f)$ and output spectrum $P_{\xi}(f)$ are calculated using Welch's method\cite{Welch1967}.

\begin{figure}[th!]
	\begin{center}
		\includegraphics[width=\textwidth]{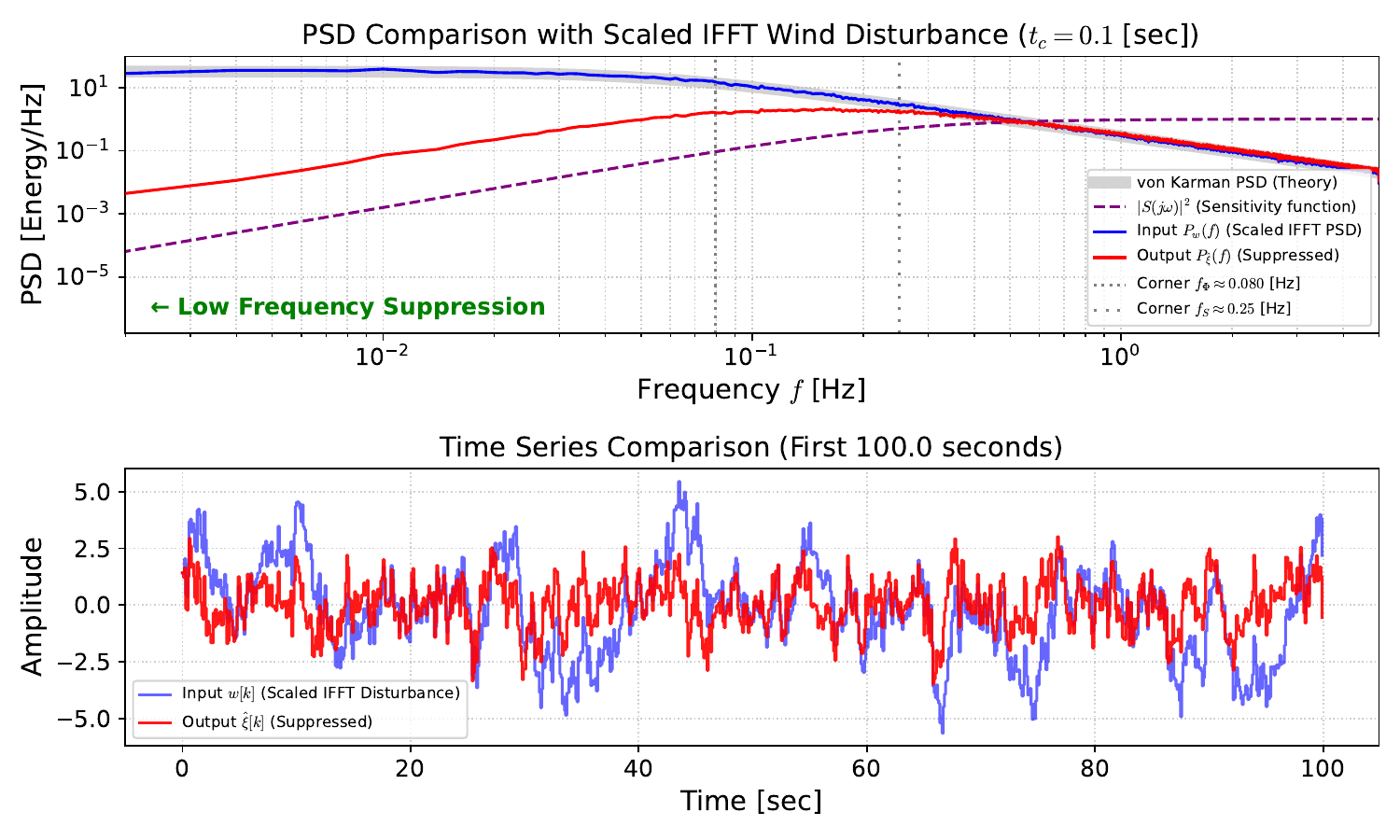}
		\caption{
			Simulation results demonstrating the frequency-dependent disturbance suppression performance using the decoupled sensitivity function $S(s)$ in Eq.~\eqref{eqn:S-03} utilized in the control system design. Wind disturbance is modeled using the von Karman spectrum in Eq.~\eqref{eqn:vonKarman-01}. The performance is evaluated by numerically generating a disturbance that faithfully reproduces the von Karman PSD, applying it as the input to $S(s)$, and comparing the input and output PSDs. The comparison confirms that low-frequency disturbance suppression is achieved, corresponding to the gain characteristics of $S(s)$, which corresponds to the gain diagram in Fig.~\ref{fig:FigSbode}. Furthermore, the time-domain sample process shows that slowly fluctuating components of the disturbance are effectively suppressed.
		}\label{fig:FigWindDisturbance}
	\end{center}
\end{figure}

The upper subplot of Fig.~\ref{fig:FigWindDisturbance} presents the PSD comparison, featuring four key curves. 
\RevJikuya{The thick gray solid line} represents the theoretical von Karman PSD $\Phi_u(f)$, establishing the benchmark for turbulence energy distribution\cite{Simiu2019}. 
The blue solid line shows the numerically estimated PSD of the input wind disturbance $P_w(f)$, calculated from the IFFT-generated time series. 
\RevJikuya{As shown in the figure, the blue curve is centrally aligned with the thick gray line; this excellent agreement} validates the disturbance generation methodology. 
The purple dashed line plots the squared magnitude of the decoupled continuous-time sensitivity function $\left| S(j \omega) \right|^2$, where $\omega = 2 \pi f$, which represents the theoretical frequency-domain attenuation of the control system. 
Finally, the red solid line displays the numerically estimated PSD of the system output $P_{\xi}(f)$ , which is the energy of the estimated coefficients.

The primary point of focus in this subfigure is the low-frequency suppression. 
The red output PSD $P_{\xi}(f)$ is significantly lower than the blue input PSD $P_w(f)$ at low frequencies, demonstrating that the control system effectively attenuates the high-energy, slow-varying components of the wind disturbance. 
The suppression effect is clearly dominant for frequencies less than the corner frequency $f_S = 0.25$ [Hz] of $S(s)$, which provides substantial attenuation.
This is visually confirmed by the output PSD closely tracking the theoretical attenuation provided by the purple $\left| S(j \omega) \right|^2$ curve, thereby proving the robust performance of the decoupled sensitivity function in rejecting wind turbulence.

The lower subplot of Fig.~\ref{fig:FigWindDisturbance} presents the time-domain comparison, visually validating the suppression effect observed in the PSD analysis. This plot displays a representative 100-second segment of the simulation for visual clarity, showing the input wind disturbance $w[k]$ (blue solid line) alongside the estimated actuation coefficients $\hat{\xi}[k]$ (red solid line). The primary observation is the reduction in the amplitude and overall variance of the output signal compared to the input. While both remain random signals, the estimated coefficients $\hat{\xi}[k]$ lacks the large, slow fluctuations that characterize the high-energy, low-frequency components of $w[k]$. This demonstrates the control performance characterized by the decoupled sensitivity function $S(s)$ in reducing the overall displacement or motion caused by wind turbulence.

%


\section{Conclusions}\label{sect:conclusions}

This study has established a unified control-theoretic formalism for the management of Excess Path Length (EPL) in Millimeter-wave Adaptive Optics (MAO). By modeling the optical drive system as a dynamical system and formulating the control task as an asymptotic disturbance suppression problem, we have successfully bridged the gap between abstract control theory and the practical requirements of submillimeter telescope instrumentation.

The proposed Anti-Windup Proportional-Integral (AWPI) controller provides a robust solution for suppressing low-frequency environmental disturbances, such as thermal and wind-induced deformations. Through decoupling control and loop-shaping design, we showed that asymptotic disturbance suppression can be achieved with guaranteed stability margins. 
A significant contribution of our asymptotic analysis is the revelation that disturbance suppression performance possesses a distinct directionality within the residual space; specifically, only the disturbance term lying within the image space of the measurement matrix can be fundamentally suppressed. To characterize this inherent constraint, we introduced the cosine similarity index, providing engineers and astronomers with a rigorous geometric metric to evaluate suppressibility of disturbance.
Furthermore, the inclusion of an intuitive manual focus adjustment scheme and a robust anti-windup mechanism ensures that the system is both high-performing and operationally reliable even under actuator saturation. Performance verification by artificial disturbance also provides a practical tool for testing the control efficiency during initial operation.

Numerical simulations confirmed that the controller performs exactly as theoretically predicted: it effectively attenuates disturbances in suppressible directions within the designed bandwidth and maintains system integrity during large-amplitude fluctuations. These results validate the AWPI framework as a viable architecture for next-generation large-aperture telescopes. While the numerical examples in this study assumed a three-axis sub-reflector drive and five-point EPL measurements, the proposed formalism is inherently scalable and can be readily extended to more complex optical drive systems and high-density, multi-point wavefront sensing configurations. Future work will focus on extending this formalism to handle non-stationary disturbances and multi-modal coupling, culminating in on-site control experiments integrated with live astronomical observations to further verify the enhancement of imaging quality in real-world environments. Additionally, the investigation of optimal sensor allocation based on the cosine similarity index remains an important future direction.





\subsection*{Disclosures}

The authors declare there are no financial interests, commercial affiliations, or other potential conflicts of interest that have influenced the objectivity of this research or the writing of this paper.

\subsection* {Code, Data, and Materials Availability} 

The code and data are available from the corresponding author upon request.


\subsection* {Acknowledgments}

The authors would like to thank Mr.~Akinobu Miyake of Nagoya University for providing the measurement matrices, as well as Dr.~Akio Taniguchi of Kitami Institute of Technology and Mr.~Masaki Sakakibara of Nagoya University for their fruitful discussions.  
The first author expresses his sincere gratitude to Professor Emeritus Katsuhiko Yamada of Osaka University for his insightful guidance and invaluable discussions on loop-shaping design. 
This study is financially supported by JSPS KAKENHI (Nos.\ 17H06206, 22H04939).


\bibliography{report}   
\bibliographystyle{spiejour}   


\vspace{2ex}\noindent\textbf{Ichiro Jikuya} is Associate Professor at Kanazawa University. He received his Ph.~D.\ in Engineering from the University of Tokyo (UTokyo) in 2001 and worked at Twente University and Nagoya University before joining Kanazawa University in 2016. He brings over 20 years of experience in the education and research of control engineering. His main research interests include control theory, particularly for linear time-varying systems, as well as control applications such as satellite attitude control and the development of telescope instrumentation.

\vspace{2ex}\noindent\textbf{Yoichi Tamura} is Professor of Astronomy at Nagoya University. He received his Ph.~D.\ from UTokyo and has worked for NAOJ and UTokyo before joining Nagoya University. He brings a 20-year experience in submillimeter astronomy. His main interests include a future submillimeter telescope, instrumentation, signal processing and commissioning, which facilitate his astronomical interests in galaxy formation in the early Universe.



\end{document}